\documentclass[journal]{IEEEtran}

\usepackage{cite}

\usepackage{amsmath}
\usepackage{mathrsfs}

\usepackage{algorithmic}

\usepackage{array}

\ifCLASSOPTIONcompsoc
\usepackage[caption=false,font=normalsize,labelfont=sf,textfont=sf]{subfig}
\else
\usepackage[caption=false,font=footnotesize]{subfig}
\fi

\usepackage{url}
\usepackage{graphicx,amsmath,amssymb,amsfonts}
\usepackage{algorithmic,algorithm}

\usepackage[bookmarks=true,
					 colorlinks  =true,
					 breaklinks =true,
					 citecolor   =black,
					 linkcolor   =black,
					 urlcolor     =black,
					]{hyperref}

\usepackage{setspace}

\usepackage{xcolor}

\newtheorem{theorem}{\textbf{Theorem}}

\newtheorem{lemma}{\textbf{Lemma}}

\newtheorem{corollary}{\textbf{Corollary}}

\newtheorem{assumption}{\textbf{Assumption}}

\makeatletter
\newcommand{\rmnum}[1]{\romannumeral #1}
\newcommand{\Rmnum}[1]{\expandafter\@slowromancap\romannumeral #1@}
\makeatother

\usepackage{framed} 

\usepackage{cancel}

\usepackage{booktabs}
\usepackage{tabularx,makecell,multirow}

\begin{document}
% reference control
\bstctlcite{ref:BSTcontrol}

\title{STAR-RIS Integrated Non-Orthogonal Multiple Access and Over-the-Air Federated Learning: Framework, Analysis, and Optimization}

\author{Wanli~Ni,~\IEEEmembership{Graduate~Student~Member,~IEEE,}
		Yuanwei~Liu,~\IEEEmembership{Senior~Member,~IEEE,}
		Yonina~C.~Eldar,~\IEEEmembership{Fellow,~IEEE,}
		Zhaohui~Yang,~\IEEEmembership{Member,~IEEE,}
		and~Hui~Tian,~\IEEEmembership{Senior~Member,~IEEE}
		\vspace{-3 mm}
		\thanks{This work was supported in part by the National Natural Science Foundation of China under Grant 62071068 and in part by the Beijing University of Posts and Telecommunications (BUPT) Excellent Ph.D. Students Foundation under Grant CX2022301.
		An earlier version of this paper has been presented at the IEEE Global Communications Conference (GLOBECOM), Madrid, Spain, December 2021, DOI: 10.1109/GLOBECOM46510.2021.9685556. \nocite{Ni2021Enabling} \textit{(Corresponding author: Hui Tian.)}}
		\thanks{Wanli Ni and Hui Tian are with the State Key Laboratory of Networking and Switching Technology, Beijing University of Posts and Telecommunications, Beijing 100876, China (e-mail: charleswall@bupt.edu.cn; tianhui@bupt.edu.cn).}
		\thanks{Yuanwei Liu is with the School of Electronic Engineering and Computer Science, Queen Mary University of London, London E1 4NS, UK (e-mail: yuanwei.liu@qmul.ac.uk).}
		\thanks{Yonina C. Eldar is with the Faculty of Mathematics and Computer Science, Weizmann Institute of Science, Rehovot 7610001, Israel (e-mail: yonina.eldar@weizmann.ac.il).}
		\thanks{Zhaohui Yang is with Zhejiang Lab, Hangzhou 31121, China, and also with the College of Information Science and Electronic Engineering, Zhejiang University, Hangzhou, Zhejiang 310027, China, and Zhejiang Provincial Key Lab of Information Processing, Communication and Networking (IPCAN), Hangzhou, Zhejiang 310007, China (e-mail: yang\_zhaohui@zju.edu.cn).}
}

\maketitle

\begin{abstract}
	This paper integrates non-orthogonal multiple access (NOMA) and over-the-air federated learning (AirFL) into a unified framework using one simultaneous transmitting and reflecting reconfigurable intelligent surface (STAR-RIS).
	The STAR-RIS plays an important role in adjusting the decoding order of hybrid users for efficient interference mitigation and omni-directional coverage extension.
	To capture the impact of non-ideal wireless channels on AirFL, a closed-form expression for the optimality gap (a.k.a. convergence upper bound) between the actual loss and the optimal loss is derived.
	This analysis reveals that the learning performance is significantly affected by the active and passive beamforming schemes as well as wireless noise.
	Furthermore, when the learning rate diminishes as the training proceeds, the optimality gap is explicitly shown to converge with linear rate.
	To accelerate convergence while satisfying quality-of-service requirements, a mixed-integer non-linear programming (MINLP) problem is formulated by jointly designing the transmit power at users and the configuration mode of STAR-RIS.
	Next, a trust region-based successive convex approximation method and a penalty-based semidefinite relaxation approach are proposed to handle the decoupled non-convex subproblems iteratively.
	An alternating optimization algorithm is then developed to find a suboptimal solution for the original MINLP problem.
	Extensive simulation results show that
	\rmnum{1}) the proposed framework can efficiently support NOMA and AirFL users via concurrent uplink communications,
	\rmnum{2}) our algorithms achieve faster convergence rate on IID and non-IID settings compared to existing baselines,
	and \rmnum{3}) both the spectrum efficiency and learning performance is significantly improved with the aid of the well-tuned STAR-RIS.
\end{abstract}

\begin{IEEEkeywords}
	 Non-orthogonal multiple access, over-the-air federated learning, reconfigurable intelligent surface, convergence analysis, resource allocation.
\end{IEEEkeywords}

\section{Introduction}
Beyond transmitting bits, the next-generation wireless networks are expected to support diverse services (from sensing and communication to edge intelligence) in different application scenarios, such as smart cities, metaverse, intelligent manufacturing, and autonomous driving \cite{Walid2020A, Zhang20196G}.
In order to support ubiquitous connectivity and pervasive intelligence at the network edge \cite{Park2019Wireless, Shi2020Communication}, future sixth generation (6G) wireless networks require a new multiple access paradigm to provide guaranteed communication quality and learning performance for stable and customized services by effectively coordinating limited wireless resources in distributed facilities.
One major challenge is that of supporting data-intensive applications and privacy-sensitive services with non-ideal noisy links and limited communication resources, so that the diverse requirements of future intelligent 6G networks can be satisfied \cite{Yang2020Federated, Deniz2020Communicate}.

The rapid development of Internet of Things (IoT) has led to an exponential growth in the number of terminals (e.g., sensors, cameras, and smart phones), some of which are sensing-oriented monitoring devices while others are likely to be computing-oriented learning devices.
For the former, non-orthogonal multiple access (NOMA) can be used to improve spectral efficiency by serving multiple devices over the same time-frequency resource \cite{Liu2017NOMA, Liu2022Evolution}.
Such a spectrum sharing scheme has attracted significant interest due to its ability to break orthogonality and meet stringent networking requirements of future 6G networks, especially for massive connectivity \cite{Liu2022Evolution} and low latency communication \cite{Kotaba2021How}.
For the latter, a promising distributed machine learning method, called federated learning (FL), has be conceived to making full use of decentralized data while protecting user privacy.
Although FL can effectively exploit the substantial computational resources of geo-distributed smart devices, the communication bottleneck problem has attracted a lot of attention due to the ever-increasing high-dimensional model parameters of deep neural networks \cite{Yang2021Federated, Chen2020A}.
Compared with most existing FL proposals based on conventional orthogonal multiple access (OMA) \cite{Chen2020A, Zhong2021Parallelizing, Nguyen2021Efficient, Chen2021Convergence, Yang2021Energy, Chen2021PNAS}, over-the-air FL (AirFL) allows all users to aggregate a shared model by utilizing the superposition property of the uplink wireless channel \cite{Yang2020TWC, Zhu2020Broadband, Guo2021Analog, Zhu2021One, Zhang2021Gradient, Amiri2020Machine}.
Different from the interference-combating OMA schemes, AirFL is a customized scheme that enjoys the improved throughput and reduced latency by harnessing the interference of multiple access channels.
Despite recent growing efforts on joint design of sensing- and computing-oriented applications, there are still many open issues yet to be addressed in this entirely new area exploring the interplay between communication and learning \cite{Park2019Wireless, Shi2020Communication}.
In particular, network architectures fulfilling 6G’s vision of agile sensing and ubiquitous intelligence remain largely undefined, which motivates the exploration of practical techniques that can efficiently integrate wireless communication and distributed learning into a unified framework.
For example, an potential solution is to combine the state-of-the-art networking and learning technologies (e.g., NOMA and AirFL) for satisfying the diverse demands in 6G networks.
However, this combination may make the design, optimization, and analysis of such an integrated network totally distinct from that of conventional networks consisting of only NOMA or AirFL users.

Distinctly different from conventional wireless networks where the environment is uncontrollable, reconfigurable intelligent surface (RIS, a.k.a., intelligent reflecting surface) suggest a new communication paradigm that can tune the phase shifts of incident signals flexibly to create favorable propagation conditions \cite{Wu2020Towards, Liu2021Reconfigurable, Huang2020Holographic}.
However, due to its limited physical implementation, a key issue is that both transmitters and receivers have to be at the same side of the RIS, which inherently results in incomplete wireless coverage \cite{Liu2021STAR, Zhang2020Beyond, Huang2020Holographic}.
In contrast to existing reflecting-only RIS \cite{Huang2019EE, Pan2020Multicell, Di2020Hybrid, Renzo2020JSAC, Ni2021Integrating}, a simultaneous transmitting and reflecting (STAR) RIS was recently developed in \cite{Liu2021STAR} to realize a full-space smart radio environment, where the source and the destination can be located in either side of the STAR-RIS \cite{Xu2021STAR, Mu2021Simultaneously, Wu2021Coverage}.
By reaping its benefits, introducing STAR-RIS into existing wireless networks is beneficial to adjust the multiple access channel on demand.
Furthermore, compared to traditional AirFL schemes that only optimize the transceiver, the STAR-RIS introduces additional control dimensions to better match the function computation requirements of AirFL, which can combat the aggregation error caused by signal distortion and further improve the learning performance of AirFL.
In order to support heterogeneous services (e.g., ubiquitous connectivity and pervasive intelligence) and guarantee their respective performance, we suggest to integrate STAR-RIS into the considered network for flexible signal processing and enhanced coverage.
However, the introduction of STAR-RIS increases the complexity of channel estimation and resource optimization due to the newly established reflective/transmissive links.
Besides, such an integration also raises more stringent signal processing requirements for the receiver to decode individual information and aggregate model updates in an efficient manner.

\subsection{Motivations and Challenges}
The goal of this work is to consider how to efficiently integrate wireless communication and federated learning with limited wireless resources at the network edge to provide ubiquitous connectivity and pervasive intelligence for future 6G networks, while still guaranteeing convergence and quality-of-service (QoS) requirements.
Toward this end, the following critical issues and potential challenges in an integrated network should be addressed:
\begin{itemize}
	\item \textbf{Constrained resources and limited coverage:}
	Simultaneously supporting communication and computation/learning services at the network edge suffer from shortage in the available resources \cite{Deniz2020Communicate, Yonina2021Federated}.
	Dedicated bandwidth allocation will inevitably decrease the spectrum and energy efficiency \cite{Nguyen2021Efficient}, so that properly sharing spectrum resource among users is essential.
	Furthermore, the battery capacity is usually insufficient for these low-cost end devices \cite{Shi2020Communication}.
	Therefore, it is important to perform power control at energy-constrained users for long-term utility maximization \cite{Huang2019EE}.
	Finally, wireless channels may be blocked by unfavorable propagation conditions and thus the effective coverage area of the base station (BS) is limited. Though the channel gains can be enhanced by the carefully designed STAR-RIS, the number of reflecting/transmitting elements is finite, namely its ability to adjust the wireless environment is limited as well, in particular when the phase shifts and amplitudes are selected from a finite number of discrete values in practice \cite{Wu2020Towards}.
	\item \textbf{Different effects of interference:}
	Although both NOMA users and AirFL users can be supported simultaneously by sharing spectrum skillfully and orchestrating resource elaborately, the co-channel interference between communication signals and learning signals has totally different effects on the performance of these heterogeneous users with different transmission goals \cite{Qi2020Integrated,Ni2021Integrating}.
	Concretely, for communication-centric NOMA users, the individual signal is expected to be separated from the received superposition signal.
	Thus, co-channel interference is an obstacle to guarantee QoS requirements for communication signals \cite{Ni2021Resource}.
	In contrast, for learning-centric AirFL users, the co-channel interference can be well exploited to complete the gradient/model aggregation via concurrent uplink communications, which is proved to be beneficial for bandwidth saving and latency reduction \cite{Zhu2020Broadband}.
	However, most prior works only consider the interference from single-type users, which results in existing solutions not being applicable to  integrated networks of communication and learning.
\end{itemize}

\subsection{Contributions and Organization}
This paper advocates a unified framework serving NOMA and AirFL users simultaneously via STAR-RIS aided concurrent communications, which aims to overcome the scarcity of system bandwidth and to support various on-demand applications.
We address the problem of interference mitigation by developing a successive interference cancellation (SIC) based signal processing scheme with the aid of the STAR-RIS.
It is worth pointing out that the STAR-RIS in this framework plays an indispensable role to coordinate the decoding order between two types of users and is also an essential component to extend the efficient coverage area of wireless networks as compared to conventional reflecting-only RIS.
Accordingly, theoretical analysis is performed to confirm the convergence guarantee of our framework.
Then, subject to the power budget, binary mode switching and unit modulus constraints, performance optimization is applied to achieve a faster convergence speed while satisfying the QoS requirements with limited resources.
To the best of our knowledge, this is the first effort to provide a compatible uplink framework by integrating NOMA and AirFL seamlessly with the aid of the STAR-RIS.
The main contributions of this work can be summarized as follows:
\begin{itemize}
	\item \textbf{Framework design and interference cancellation:}
	We design a STAR-RIS assisted heterogeneous fusion framework by supporting communication-centric NOMA users and learning-centric AirFL users in a non-orthogonal manner.
	Since the complete temporal and spectral resources are utilized by all users simultaneously, an efficient co-channel interference cancellation technique is exploited to guarantee the quality of the communication signal while harnessing the superposition property of multiple-access channels for the computation signal.
	The key advantage of the developed signal processing scheme is that it can skillfully separate the individual signal of NOMA users while preserving the aggregated signal of AirFL users, which provides the best of both worlds.
	\item \textbf{Convergence analysis and problem formulation:}
	We derive a closed-form expression for the optimality gap between the actual loss and the optimal loss to quantify the impact of constrained resources and wireless noise on the learning performance of AirFL users.
	We prove that, using a diminishing learning rate, the optimality gap is guaranteed to converge linearly with a rate of $\mathcal{O} \left( 1/t \right)$, where $t$ is the index of communication rounds.
	Based on the theoretical analysis results, we formulate a resource allocation problem in the heterogeneous fusion network to improve the learning performance of AirFL users while satisfying the QoS requirements of NOMA users.
	The problem captures a joint design of active and passive beamforming over training rounds, which is a mixed-integer non-linear programming (MINLP) problem.
	\item \textbf{Performance optimization and experimental validation:}
	To tackle the resulting NP-hard problem, we first decompose it into two tractable subproblems, and then an alternating optimization technique is adopted to find a suboptimal solution.
	Specifically, for the power allocation (active beamforming) problem, we develop a trust region-based successive convex approximation (SCA) method.
	For the STAR-RIS configuration (passive beamforming) problem, we propose a penalty-based semidefinite relaxation (SDR) approach to address the issues with rank-one constraints and binary variables.
	Finally, we conduct comprehensive experiments to validate the effectiveness of our theoretical analysis and algorithm design by training linear regression models on a synthetic dataset as well as building image classification models on real datasets.
	Compared to benchmarks, simulation results demonstrate that our solution yields better performance in terms of communication capacity and learning behavior, even if the data is not identically distributed.
\end{itemize}

The remainder of this paper is organized as follows.
The system model of the STAR-RIS integrated NOMA and AirFL framework is given in Section \ref{section_system_model} where the concurrent communication mechanism is designed and an efficient signal processing scheme is developed to mitigate the interference among hybrid users.
Convergence analysis is provided in Section \ref{section_convergence_and_problem} where a non-convex resource allocation problem is formulated.
Next, Section \ref{section_optimization} proposes an alternating optimization algorithm to control uplink transmit power and STAR-RIS configuration.
Finally, simulation results are presented in Section \ref{section_simulation}, followed by conclusions in Section \ref{section_conclusion}.
The key symbols and main notations are listed in Table \ref{table_notations}.

\begin{table}[t]
	\color{black}
	\caption{Summary of Key Symbols and Main Notations}
	\label{table_notations}
	\vspace{-4 mm}
	\begin{tabular}[t]{p{.25 \columnwidth}| p{.65 \columnwidth}}
		\toprule
		$\mathcal{N}$, $\mathcal{K}$, $\mathcal{U}$, $\mathcal{M}$ & Sets of NOMA users, AirFL users, hybrid users, and STAR-RIS elements \\
		$\mathbf{\Theta}_u$ $\&$ $\boldsymbol{\beta}_u$ & Diagonal reflection/transmission matrix and mode switching vector of STAR-RIS \\
		$h_u$, $\boldsymbol{r}_u$, $\boldsymbol{\bar{r}}$, $\bar{h}_u$ & BS-User link, RIS-User link, BS-RIS link, and the combined channel \\
		$\mathcal{T}$ $\&$ $\lambda$ & Set of training rounds, and the learning rate adopted by the BS \\
		$F_k ( \cdot )$ $\&$ $F ( \cdot )$ & Local loss function at the $k$-th user and global loss function at the BS \\
		$\boldsymbol{g}_k^{(t)}$ $\&$ $\boldsymbol{g}^{(t)}$ & Local gradient at the $k$-th user and global gradient in the $t$-th round \\
		$\boldsymbol{w}^{(t)}$ $\&$ $\boldsymbol{w}^{*}$ & Global model in the $t$-th round, and the optimal model that minimizes $F ( \boldsymbol{w} )$\\
		$s_n$ $\&$ $s_k$ & Transmit symbol at the $n$-th ($k$-th) user \\
		$p_n$ $\&$ $p_k$ & Power scalar at the $n$-th ($k$-th) user \\
		\midrule
		${\rm diag} \{ \cdot \}$ &  A diagonal matrix with each diagonal element being the element in a vector \\
		${\rm Diag} \{ \cdot \}$ & A vector with each element being the main diagonal elements in a matrix \\
		$( \cdot )^{\rm H}$ $\&$ $( \cdot )^{\top}$ & The conjugate transpose and transpose of a vector or matrix \\
		$| \cdot |$ $\&$ $\| \cdot \|_2$ & The magnitude of a complex scalar and the Euclidean norm of a vector \\
		$\nabla$ $\&$ $E[ \cdot ]$ & The gradient of a function and the expectation of a random variable \\
		${\rm tr}( \cdot )$ $\&$ ${\rm rank}( \cdot )$ & The trace and the rank of a matrix \\
		\bottomrule
	\end{tabular}
\end{table}

\section{System Model} \label{section_system_model}

\begin{figure*} [ht]
	\centering
	\includegraphics[width=6.8 in]{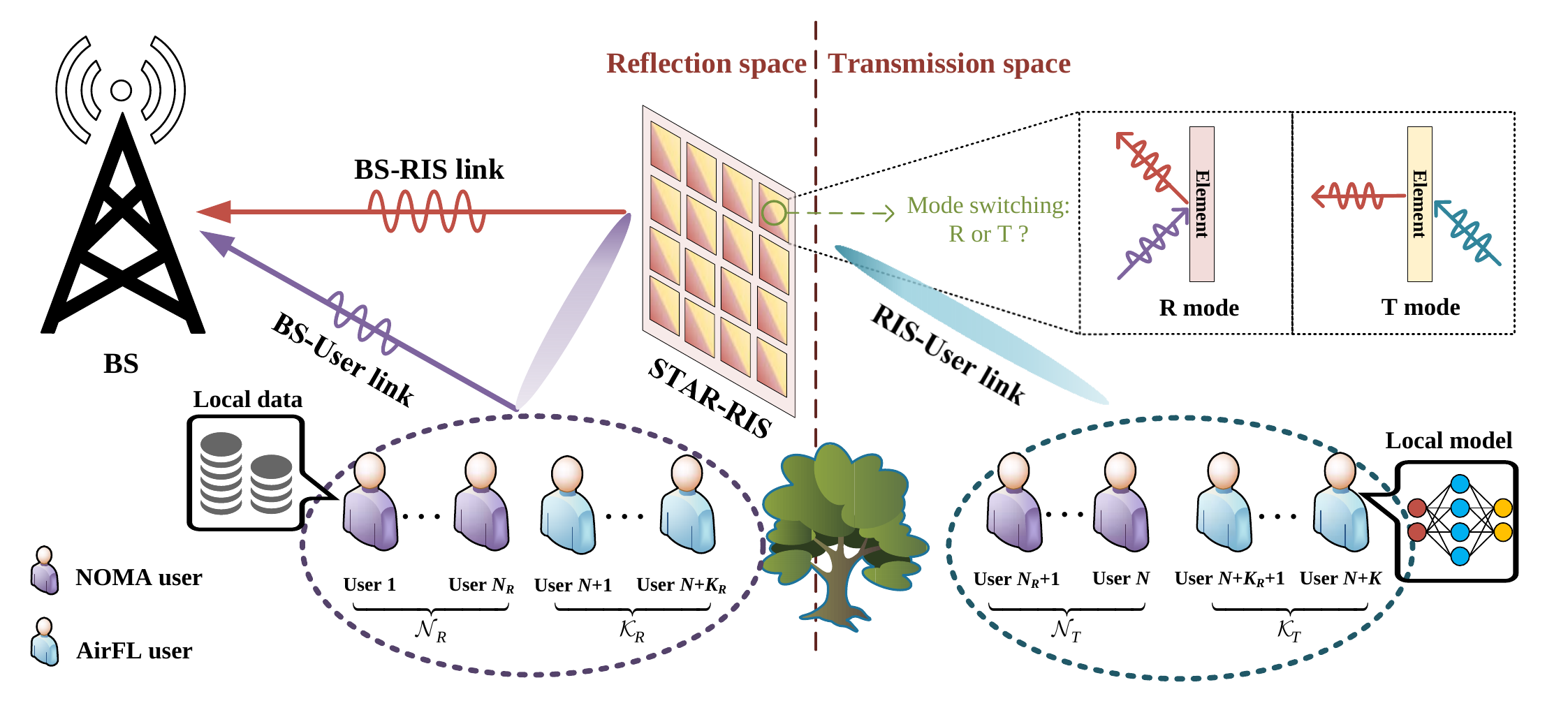}
	\caption{An illustration of the STAR-RIS integrated NOMA and AirFL framework.}
	\label{system_model}
\end{figure*}

\subsection{Network and Channel Model}
We consider a STAR-RIS assisted heterogeneous network, consisting of one BS and $N+K$ hybrid users, as illustrated in Fig. \ref{system_model}.
Specifically, both the BS and users are equipped with one single antenna each.
The STAR-RIS is assumed to be equipped with $M$ passive reflecting/transmitting elements.
The randomly distributed users are classified into two types: NOMA users (communication-centric) and AirFL users (learning-centric).
The set of hybrid users is denoted by $\mathcal{U} = \mathcal{N} \cup \mathcal{K}$, where the sets of NOMA users and AirFL users are indexed by $\mathcal{N} = \{ 1,2, \ldots, N\}$ and $\mathcal{K} = \{ N+1, N+2, \ldots, N+K\}$, respectively.
The full space of signal propagation is divided into two parts by the STAR-RIS, namely the reflection space (left half space) and the transmission space (right half space) \cite{Liu2021STAR}.
The location of each user determines that its uplink signal is reflected or transmitted by the STAR-RIS \cite{Xu2021STAR, Mu2021Simultaneously, Wu2021Coverage}.
More precisely, the set of NOMA users located in the reflection space is denoted by $\mathcal{N}_R = \{ 1,2, \ldots, N_R\}$, and other NOMA users located in the transmission space are indexed by $\mathcal{N}_T = \{ N_R+1, N_R+2, \ldots, N\}$, while $\mathcal{N} = \mathcal{N}_R \cup \mathcal{N}_T$ and $\mathcal{N}_R \cap \mathcal{N}_T = \varnothing$.
Similarly, the set of AirFL users located in the reflection space is denoted by $\mathcal{K}_R = \{ N+1, N+2, \ldots, N+K_R\}$, while $\mathcal{K}_T = \mathcal{K} \backslash \mathcal{K}_R$ denotes AirFL users located in the transmission space.
In practice, the sets of $\mathcal{N}_R$, $\mathcal{N}_T$, $\mathcal{K}_R$, and $\mathcal{K}_T$ can be easily obtained, as long as the number of users is given and the locations of all users are fixed.

Different from the conventional reflection-only RIS, the STAR-RIS enables omni-directional radiation by introducing the equivalent electric and magnetic currents into the hardware implementation of each element \cite{Xu2021STAR}.
On the whole, STAR-RIS has three protocols for operating in wireless networks: energy splitting, mode switching, and time switching \cite{Liu2021STAR, Mu2021Simultaneously}.
In this paper, we focus on the mode switching protocol\footnote{Although the other two protocols (i.e., energy splitting and time switching) have their respective advantages, the mode switching protocol is more applicable to the implementation of concurrent uplink communications in our designed framework.} where each element of the STAR-RIS can operate in reflection mode (referred to as R mode) or transmission mode (referred to as T mode).
The set of passive elements is indexed by $\mathcal{M} = \{1, 2, \ldots, M\}$.
We denote $\beta_m \in \{0, 1\}, \forall m \in \mathcal{M}$ as the mode indicator of the $m$-th element, where $\beta_m = 1$ for R mode and $\beta_m = 0$ for T mode.
The users in reflection space (i.e., $\mathcal{N}_R \cup \mathcal{K}_R$) are served by these R-mode elements, while users in transmission space (i.e., $\mathcal{N}_T \cup \mathcal{K}_T$) are served by these T-mode elements.
Then, the $M \times M$ diagonal reflection matrix is denoted by $\mathbf{\Theta}_u = {\rm diag} \lbrace  \beta_1 e^{j\theta_1}, \beta_2 e^{j\theta_2}, \dots, \beta_M e^{j\theta_M} \rbrace, \forall u \in \mathcal{N}_R \cup \mathcal{K}_R$, where $\theta_m \in [0, 2 \pi], \forall m \in \mathcal{M}$ represents the reflection phase shift of the $m$-th element operating in R mode.
Likewise, the diagonal transmission matrix is denoted by $\mathbf{\Theta}_u = {\rm diag} \{  (1-\beta_1) e^{j \phi_1}, (1-\beta_2) e^{j \phi_2}, \dots, (1-\beta_M) e^{j \phi_M} \}, \forall u \in \mathcal{N}_T \cup \mathcal{K}_T$, where $\phi_m \in [0, 2 \pi], \forall m \in \mathcal{M}$ represents the transmission phase shift of the $m$-th element operating in T mode.

Let $h_u \in \mathbb{C}^{1 \times 1}$, $\boldsymbol{r}_u \in \mathbb{C}^{M \times 1}$, and $\boldsymbol{\bar{r}} \in \mathbb{C}^{M \times 1}$ denote the channel from the $u$-th user to the BS, from the $u$-th user to the STAR-RIS, and from the STAR-RIS to the BS, respectively.
The large-scale path loss is modeled as $L(d) = \varsigma_0 (d)^{-\alpha}$, where $\varsigma_0$ is the path loss at the reference distance of one meter,  $d$ is the individual link distance, and $\alpha$ is the path loss exponent.
Similar to \cite{Ni2021Resource} and \cite{Liu2020RIS}, we assume that all user-related links follow Rayleigh fading due to the extensive scattering, while the BS-RIS link obeys Rician fading due to the high altitude of the BS and STAR-RIS.
Similar to \cite{Ni2021Resource} and \cite{Liu2020RIS}, we assume that all user-related links follow Rayleigh fading due to the extensive scattering, while the BS-RIS link obeys Rician fading due to the high altitude of the BS and STAR-RIS.
Thus, the channel coefficient of the BS-RIS link is given by
\begin{align}
\boldsymbol{\bar{r}} = \sqrt{\frac{L(d_0)}{\kappa+1}} \left( \sqrt{\kappa} \boldsymbol{r}^{\rm LoS} + \boldsymbol{r}^{\rm NLoS} \right),
\end{align}
where $d_0$ is the distance between the BS and the STAR-RIS, $\kappa$ is Rician factor, $\boldsymbol{r}^{\rm LoS}$ and $\boldsymbol{r}^{\rm NLoS}$ are the deterministic line-of-sight (LoS) and Rayleigh fading components, respectively.
In specific, the LoS component $\boldsymbol{r}^{\rm LoS}$ is given by $\boldsymbol{r}^{\rm LoS} =[1, e^{j \frac{2 \pi \bar{d}}{\bar{\lambda}} \sin \vartheta^{\rm AoD}}, \cdots, e^{j \frac{2 \pi \bar{d}}{\bar{\lambda}}\left(M-1\right) \sin \vartheta^{\rm AoD}}]^{\rm T}$,
	where $\bar{d}$ is the element separation distance, $\bar{\lambda}$ is the wavelength, $\vartheta^{\rm AoD}$ is the angle of departure that follows equal distribution within $[0,2 \pi]$. For simplicity, we set $\bar{d} / \bar{\lambda}=1 / 2$ \cite{Pan2020Multicell}.

We assume that all channels follow the quasi-static flat-fading model, and that the perfect channel state information (CSI) of all channels is available at the BS \cite{Ni2021Resource, Liu2020RIS, Wu2019IRS, Cao2020AirFL}.
The combined channel from the $u$-th user to the BS via the STAR-RIS can be written as
\begin{align}
\bar{h}_u = h_u +  \boldsymbol{\bar{r}}^{\rm H} \mathbf{\Theta}_u \boldsymbol{r}_u, \ \forall u \in \mathcal{U}.
\end{align}

\subsection{Training Process of Federated Learning}
Let $\mathcal{T} = \{1, 2, \ldots, T\}$ denote the set of time slots for training rounds (a.k.a., communication rounds).
Each training round is composed of four main steps including global model broadcast, local gradient calculation, local gradient upload, and global model update \cite{Yang2020TWC, Ren2020Scheduling}.
For example, at the $t$-th training round, the BS first broadcasts the latest global model $\boldsymbol{w}^{(t)} \in \mathbb{R}^Q$ to all AirFL users, and then they calculate the local gradients based on the their individual dataset.
Specifically, the $k$-th AirFL user obtains its local gradient as
\begin{align} \label{local_gradient}
\boldsymbol{g}_k^{(t)} = \nabla F_k ( \boldsymbol{w}^{(t)} ), \ \forall k \in \mathcal{K},
\end{align}
where $\nabla$ is the gradient operator, and $F_k (\boldsymbol{w}^{(t)})$ is the local loss function of the $k$-th user that can be one of the typical loss quantification methods, such as mean square error and cross entropy loss \cite{Chen2021Convergence, Dinh2021Federated}.

 Next, all AirFL users upload their local gradients $\{ \boldsymbol{g}_k^{(t)} \}$ to the BS for global synchronization.
 After receiving all local gradients, the global model is updated by
\begin{align} \label{synchronization}
\boldsymbol{w}^{(t+1)} = \boldsymbol{w}^{(t)} - \lambda \boldsymbol{g}^{(t)},
{\rm \ with \ }
\boldsymbol{g}^{(t)} = \frac{1}{K} \sum \nolimits_{k \in  \mathcal{K}} \boldsymbol{g}_k^{(t)},
\end{align}
where $\boldsymbol{g}^{(t)}$ is the global gradient and $\lambda$ is the learning rate (a.k.a. step size) adopted by the BS\footnote{In the proposed framework, AirFL users are allowed to train local models with different numbers of data samples, thus the sample differences among users are ignored in (\ref{local_gradient}). The global model aggregation rule defined in (\ref{synchronization}) holds for both cases of balanced and imbalanced samples. Additionally, the uplink communications of all users are assumed to be synchronized.}.
The above communication round continues until preset convergence accuracy is satisfied or the maximum number of iterations $T$ is reached.

\subsection{Concurrent Uplink Communication}
To perform concurrent uplink communication at the $t$-th training round, the local data $\boldsymbol{d}_n^{(t)}$ of the $n$-th NOMA user is encoded into the information-bearing signal (i.e., communication symbol $s_n^{(t)}$), the local gradient $\boldsymbol{g}_k^{(t)}$ of the $k$-th AirFL user is similarly transformed into the transmit signal (i.e., computation symbol $s_k^{(t)}$).
With the aid of the STAR-RIS, both NOMA users and AirFL users transmit simultaneously over the same time-frequency resource, and thus the superposition signal received at the BS is given by
\begin{align}
y^{(t)}
= \underbrace{\sum \nolimits_{n = 1}^{N} \bar{h}_n^{(t)} p_n^{(t)} s_n^{(t)}}_{\mathbf {NOMA~users}}
+ \underbrace{\sum \nolimits_{k = N+1}^{N+K} \bar{h}_k^{(t)} p_k^{(t)} s_k^{(t)}}_{\mathbf {AirFL~users}}
+ \underbrace{z_0^{(t)}}_{\mathbf {noise}},
\end{align}
where $p_n^{(t)}$ ($p_k^{(t)}$) is the power scalar at the $n$-th ($k$-th) user, and $z_{0} \sim \mathcal{CN}(0, \sigma^{2})$ is the additive white Gaussian noise (AWGN) received at the BS.

For ease of power control and without loss of generality, the symbols are assumed to  be statistically independent and are normalized to have unit variance \cite{Qi2020Integrated, Ni2021Over}, i.e., $\mathbb{E} [ | s_u^{(t)} | ^2 ] = 1, \forall u \in \mathcal{U}$ and $\mathbb{E} [ (s_u^{(t)})^{\rm H} s_{v}^{(t)} ] = 0, \forall u \ne v$.
Then, the transmit power of the $u$-th user at the $t$-th communication round is constrained by
\begin{align} \label{power_constraint}
\mathbb{E} \left[ | p_u^{(t)} s_u^{(t)} |^2 \right]  = \left| p_u^{(t)} \right| ^{2} \le P_{u}, \ \forall u \in \mathcal{U},
\end{align}
where $P_{u}$ is the maximum transmit budget at the $u$-th user.
Additionally, over the all $T$ communication rounds, an average power constraint $\bar{P}_{u}$ is imposed at the $u$-th user, and thus we should have
\begin{align} \label{average_power_constraint}
\frac{1}{T} \sum \limits_{t=1}^{T} \mathbb{E} \left[ | p_u^{(t)} s_u^{(t)} |^2 \right]  = \frac{1}{T} \sum \limits_{t=1}^{T} \left| p_u^{(t)} \right| ^{2} \le \bar{P}_{u}, \ \forall u \in \mathcal{U}.
\end{align}

Although the STAR-RIS can simultaneously provide concurrent communication for both the reflection signals and the transmission signals by allowing hybrid users share the same orthogonal (time-frequency) wireless resources, it also leads to severe co-channel interference, especially in the case of massive connectivity.
However, as mentioned before, the co-channel interference has different effects on the performance of the communication-centric NOMA users and the learning-centric AirFL users.
On the one hand, the co-channel interference is harmful to the NOMA users, whose individual signals should be decoded from the received superposed signal.
On the other hand, the co-channel interference is conducive to improving the aggregation performance of AirFL users by turning the wireless channel into a computer with the functionality of a weighted sum.
Hence, an efficient signal processing scheme is needed to coordinate the co-channel interference in heterogeneous fusion networks, which plays a pivotal role in enhancing the overall system performance in terms of communication throughput and learning behavior.

\begin{figure*} [t]
	\color{black}
	\centering
	\includegraphics[width=7 in]{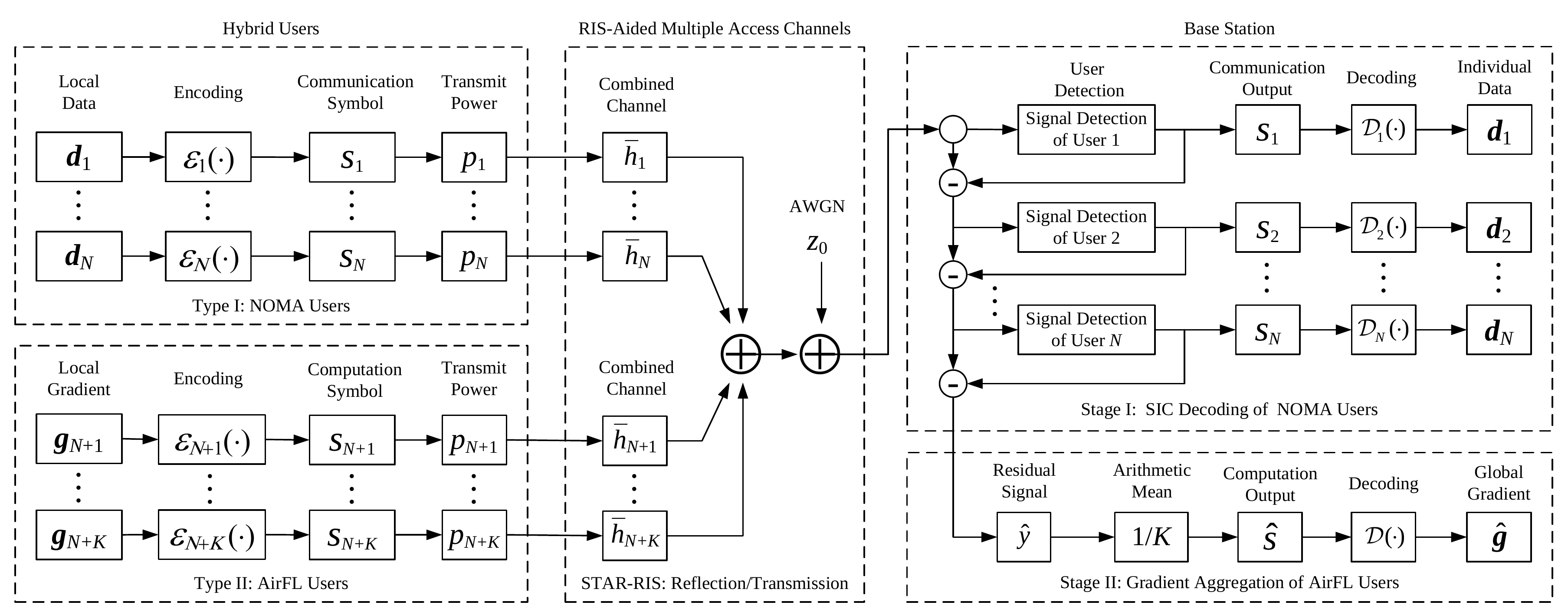}
	\caption{Block diagram of the designed signal processing scheme for the STAR-RIS integrated NOMA and AirFL framework.}
	\label{block_diagram}
\end{figure*}

\subsection{Successive Signal Processing}
By taking advantage of the SIC technique at the BS side, the signals from strong users can be decoded one by one to remove the co-channel interference for weak users \cite{Liu2017NOMA, Liu2022Evolution}.
In the sequel, the residual interference composed of signals from weak users can be harnessed for function computation \cite{Nazer2007Computation, Ni2020Federated, Sery2020Over}.
To actualize this design, we propose a successive signal processing (SSP) scheme taking full advantage of the superposition property of multiple access channels.
The block diagram of the proposed SSP scheme is shown in Fig. \ref{block_diagram}, where there are two key stages at the BS.
The first stage is devised to mitigate interference for NOMA users.
The second stage exploits interference for AirFL users.
The premise of this scheme is that all NOMA users are anticipated to be strong users with larger channel gains and all AirFL users are expected to become weak users having smaller channel gains.
Fortunately, by virtue of its flexible configuration ability, the STAR-RIS is capable of creating a controllable signal propagation environment over the full space.
Therefore, the STAR-RIS is particularly well suited for tackling the challenging issue of dynamically adjusting the channel conditions of hybrid users throughout the network.

To obtain the desired signal processing order\footnote{In this paper, the SIC decoding order of NOMA users is considered to be fixed for simplicity, which can be further optimized for different utilities such as throughput maximization, fairness guarantee or physical layer security.}, the following channel conditions should be met by jointly optimizing the mode switching and phase shifts of the STAR-RIS:
\begin{equation} \label{channel_condition}
\underbrace{\left| \bar{h}_{1}^{(t)} \right|^2 \ge \left| \bar{h}_{2}^{(t)} \right|^2 \ge \cdots \ge \left| \bar{h}_{N}^{(t)} \right|^2}_{\mathbf {strong~users}}
\ge
\underbrace{\left| \bar{h}_{k}^{(t)} \right|^2, \ \forall k \in \mathcal{K}}_{\mathbf {weak~users}}.
\end{equation}
Based on the SIC decoding order in (\ref{channel_condition}), if we have $v=n+1 \leq N$, the BS first decodes the $n$-th strong user's signal and then subtracts it from the superposed signal to decode the desired signal of the $v$-th strong user, and so on.
Therefore, the received signal-to-interference-plus-noise ratio (SINR) of the $n$-th NOMA user at the BS is given by
\begin{equation} \label{SINR}
\gamma_{n}^{(t)} = \frac{ \left| \bar{h}_{n}^{(t)} \right| ^{2} \left| p_{n}^{(t)} \right| ^{2} }{ \sum \nolimits_{u=n+1}^{N+K} \left| \bar{h}_{u}^{(t)} \right| ^{2} \left| p_{u}^{(t)} \right| ^{2} + \sigma^{2}}, \ \forall n \in \mathcal{N}.
\end{equation}
After successfully eliminating all signals from NOMA users, the residual signal is only composed of the co-channel interference from AirFL users and the wireless noise, which is expressed as
\begin{equation} \label{reconstructed_signal}
\hat{y}^{(t)} = \sum \nolimits_{k \in \mathcal{K}} \bar{h}_{k}^{(t)} p_{k}^{(t)} s_{k}^{(t)} + z_{0}^{(t)}.
\end{equation}

Upon reconstructing the signal $\hat{y}^{(t)}$ in (\ref{reconstructed_signal}), the BS applies an arithmetic mean to compute the estimation of $s^{(t)} = \frac{1}{K} \sum_{k \in \mathcal{K}} s_{k}^{(t)}$.
The average gradient message of interest received at the BS is given by
\begin{equation} \label{averaged_signal}
\hat{s}^{(t)} = \frac{\hat{y}^{(t)}}{K} = \frac{1}{K} \left( \sum \limits_{k \in \mathcal{K}} \bar{h}_{k}^{(t)} p_{k}^{(t)} s_{k}^{(t)} + z_{0}^{(t)} \right).
\end{equation}
The computation distortion (a.k.a., aggregation error) of the recovered average gradient message with respect to (w.r.t.) the ideal average message $s^{(t)}$ can be measured by the instantaneous MSE defined as ${\rm MSE}^{(t)} \triangleq \mathbb{E} [ |\hat{s}^{(t)} - s^{(t)} |^2 ] $, i.e.,
\begin{equation} \label{MSE_s}
\begin{aligned}
{\rm MSE}^{(t)}
= \frac{1}{K^{2}} \left( \sum \limits_{k \in \mathcal{K}} \left| \bar{h}_{k}^{(t)} p_{k}^{(t)} - 1 \right| ^2 + \sigma^2 \right).
\end{aligned}
\end{equation}

In light of the design above, through leveraging the STAR-RIS to tune channel conditions of hybrid users, one main advantage of our SSP scheme is its capability of separating the individual signal of NOMA users while preserving the aggregated signal of AirFL users.
As such, both NOMA users and AirFL users can be simultaneously served via concurrent communications in our framework.
However, it can be observed from (\ref{SINR}) that the achievable communication rate of NOMA users is interfered by the signal from AirFL users.
Furthermore, the aggregation error in (\ref{MSE_s}) may lead to a notable drop of AirFL users' prediction accuracy in the inference process \cite{Yang2020TWC, Zhu2020Broadband, Guo2021Analog}.
In this context, despite the fact that different effects of interference can be exploited by our SSP scheme, a more elaborate resource management solution is desired to enhance the overall system performance.
Therefore, we jointly design the power scalar of hybrid users and the configuration mode of STAR-RIS to improve the communication-learning efficiency with MSE tolerance for fast model aggregation and with QoS requirements for reliable data transmission.
Toward this end, our next sections are to theoretically confirm the convergence of our framework and answer two fundamental questions:
\rmnum{1}) how non-ideal wireless channels affect the convergence behavior of AirFL users,
and \rmnum{2}) how to accelerate convergence while satisfying minimum data rate demands of NOMA users.
 
\section{Convergence Analysis and Problem Formulation} \label{section_convergence_and_problem}
\subsection{Basic Assumption}
To theoretically characterize the convergence performance of AirFL, similar to the works in \cite{Li2020On, Zhu2021One, Cao2020AirFL, Wang2019Adaptive}, we make the following standard assumptions on the global loss function and gradient estimations.
\begin{assumption}[$L$-smooth] \label{assumption_1}
	\emph{The global loss function $F \left( \boldsymbol{w} \right)$ is $L$-smooth. Namely, for any model parameters $\boldsymbol{w}$ and $\boldsymbol{v}$, there exists a non-negative constant $L$, such that}
	\begin{equation} \label{assumption_L_smooth}
	F \left( \boldsymbol{w} \right) - F \left( \boldsymbol{v} \right) \le  \nabla F \left( \boldsymbol{v} \right)^{\top} \left( \boldsymbol{w}- \boldsymbol{v} \right) + \frac{L}{2} \left\| \boldsymbol{w}- \boldsymbol{v} \right\| _2^2. 
	\end{equation}
\end{assumption}
\begin{assumption}[$\mu$-strongly convex] \label{assumption_2}
	\emph{The global loss function $F \left( \boldsymbol{w} \right)$ is strongly convex with a positive parameter $\mu$ such that for any $\boldsymbol{w}$ and $\boldsymbol{v}$, we have}
	\begin{equation} \label{strongly_convex}
	F \left( \boldsymbol{w} \right) - F \left( \boldsymbol{v} \right)  \ge \nabla F \left( \boldsymbol{v} \right)^{\top} \left( \boldsymbol{w}- \boldsymbol{v} \right) + \frac{\mu}{2} \left\| \boldsymbol{w}- \boldsymbol{v} \right\| _2^2. 
	\end{equation}
\end{assumption}
\begin{assumption}[Variance bound] \label{assumption_3}
	\emph{The expectation of the local gradients $\boldsymbol{g}_k$ at the $k$-th AirFL user is assumed to be independent and an unbiased estimation of the global gradient $\boldsymbol{g} = \nabla F ( \boldsymbol{w} )$ with coordinate bounded variance, i.e.,
	\begin{eqnarray} \label{assumption_gradient_bound}
	&\mathbb{E} \left[ \boldsymbol{g}_k \right] = \boldsymbol{g}, \ \forall k \in \mathcal{K}, \\
	&\mathbb{E} \left[ \left( g_{k,i} - g_i \right) ^2 \right] \le \delta_i^2, \ \forall k,i,
	\end{eqnarray}
	where $g_{k,i}$ and $g_i$ represent the $i$-th element of $\boldsymbol{g}_k$ and $\boldsymbol{g}$, respectively, and $\boldsymbol{\delta} = \left[ \delta_1, \delta_2, \ldots, \delta_Q \right]$ is a non-negative vector.}
\end{assumption}

The above assumptions make the convergence analysis of AirFL tractable.
Assumption \ref{assumption_1} on the Lipschitz smoothness holds for a broad range of typical loss functions, such as the linear regression, cross entropy, logistic regression, and softmax classifier \cite{Li2020On, Zhu2021One, Cao2020AirFL, Sery2020Over, Wang2019Adaptive}.
Assumption \ref{assumption_2} on the strong convexity can be easily extended to the Polyak-Lojasiewicz (PL) condition given in Appendix \ref{proof_of_theorem_1}.
Similar to \cite{Guo2021Analog}, the specific values of parameters $L$ and $\mu$ in Assumptions \ref{assumption_1} and \ref{assumption_2} can be obtained by calculating the maximum and minimum eigenvalues of $\nabla^2 F (\boldsymbol{w})$.
Assumption \ref{assumption_3} on the bounded variance is standard in the stochastic optimization literature \cite{Zhu2018Natasha}.
As an aside, these assumptions are made for analytical tractability \cite{Sery2020Over}.
In practice, many more complicated learning models that do not satisfy these assumptions can also perform well in practice, as is shown by the experiments in Section \ref{section_simulation}.

\subsection{Convergence Analysis} \label{convergence_analysis}
According to the reconstructed signal obtained in (\ref{reconstructed_signal}), the BS can recover local gradients $\{ \boldsymbol{g}_{k}^{(t)} \}$ from wireless signals $\{ s_{k}^{(t)} \}$.
Then, the received signal vector of interest at the BS can be equivalently given by
\begin{equation} \label{aggregated_gradient}
\boldsymbol{\hat{y}}^{(t)} = \sum \nolimits_{k \in \mathcal{K}} \bar{h}_{k}^{(t)} p_{k}^{(t)} \boldsymbol{g}_{k}^{(t)} + \boldsymbol{z}_{0}^{(t)},
\end{equation}
where $\boldsymbol{z}_{0}^{(t)} \in \mathbb{R}^Q$ is the receiver noise vector distributed as $\boldsymbol{z}_{0}^{(t)} \sim \mathcal{CN}(0, \sigma^2 \boldsymbol{I})$.
Similarly, the aggregated global gradient defined in (\ref{averaged_signal}) is equivalent to
\begin{equation} \label{averaged_gradient}
\boldsymbol{\hat{g}}^{(t)} = \frac{\boldsymbol{\hat{y}}^{(t)}}{K}
= \frac{1}{K} \left( \sum \nolimits_{k \in \mathcal{K}} \bar{h}_{k}^{(t)} p_{k}^{(t)} \boldsymbol{g}_{k}^{(t)} + \boldsymbol{z}_{0}^{(t)} \right).
\end{equation}

The global gradient estimation in (\ref{averaged_gradient}) highly depends on the communication factors (including transmit power, STAR-RIS configuration, the wireless channel and noise).
To reveal their impact on the one-round convergence behavior of AirFL, we develop a concrete metric to capture the contribution of each training round on the convergence rate, as presented in the following lemma.
\begin{lemma}[One-round convergence] \label{one_round}
	\emph{Consider a federated learning task satisfying Assumptions \ref{assumption_1} and \ref{assumption_3}, and suppose that $F(\boldsymbol{w}^{*})$ is the minimum value of the global loss function.
	Then, the one-round convergence behavior at the $t$-th communication round is bounded by}
	\begin{align} \label{one_round_expectation}
	&\mathbb{E} \left[ F ( \boldsymbol{w}^{(t+1)} ) \right] - F ( \boldsymbol{w}^{*} ) \le F ( \boldsymbol{w}^{(t)} ) - F ( \boldsymbol{w}^{*} ) \nonumber \\
	& - \left[ \sum \nolimits_{k \in \mathcal{K}} \left(  \frac{\lambda}{K} \bar{h}_{k}^{(t)} p_{k}^{(t)} - \frac{L \lambda^2}{2 K^2} ( \bar{h}_{k}^{(t)} p_{k}^{(t)} ) ^2 \right) \right] \| \boldsymbol{g}^{(t)}  \| _2^2 \nonumber \\
	& + \frac{L \lambda^2}{2 K^2} \sum \nolimits_{k \in \mathcal{K}} \left( \bar{h}_{k}^{(t)} p_{k}^{(t)} \right) ^2 \left\| \boldsymbol{\delta} \right\| _2^2 + \frac{L Q \lambda^2 \sigma^2}{2 K^2}.
	\end{align}
\end{lemma}

\begin{IEEEproof}
	See Appendix \ref{proof_of_one_round}.
\end{IEEEproof}

The expected performance gap of $\mathbb{E} [ F ( \boldsymbol{w}^{(t+1)} ) ] - F ( \boldsymbol{w}^{*} )$ is affected by four terms.
The first term on the right-hand-side of (\ref{one_round_expectation}) characterizes the learning gap of the previous round.
The second term is negatively related to the squared norm of the global gradient.
The third term is determined by the bound on the norm of local gradients.
The last term comes from the noisy channel and includes learning-related parameters.
Note that the first and last items are independent of the transmit power (active beamforming) and STAR-RIS configuration (passive beamforming) so that they can be regarded as two constants.
The middle two terms depend on the active and passive beamforming schemes and thus need to be optimized.

Based on the result in Lemma \ref{one_round}, we now investigate the total optimality gap (a.k.a. convergence upper bound) of AirFL.
Specifically, we provide a theoretical analysis of the learning performance of AirFL after $T$ communication rounds in the following theorem.
\begin{theorem}[Optimality gap] \label{theorem_1}
	\emph{When Assumptions \ref{assumption_1}, \ref{assumption_2} and \ref{assumption_3} hold and the learning rate is fixed, in the case with arbitrary transmit power at AirFL users and random configuration design at the STAR-RIS, the total optimality gap for AirFL after $T$ communication rounds is given by
	\begin{align} \label{theorem_optimality_gap}
	&\mathbb{E} \left[ F ( \boldsymbol{w}^{(T+1)} ) \right] - F ( \boldsymbol{w}^{*} ) \le \prod \limits_{t=1}^{T} \Lambda_3^{(t)} \left( F ( \boldsymbol{w}^{(1)} ) - F ( \boldsymbol{w}^{*} ) \right) \nonumber \\
	&+ \sum \limits_{t=1}^{T-1} ( \prod \limits_{i=t+1}^{T} \Lambda_3^{(i)} ) \Lambda_4^{(t)} + \Lambda_4^{(T)}
	\triangleq \Upsilon \left(  \{ p_u^{(t)} \}, \{ \mathbf{\Theta}_u^{(t)} \} \right) ,
	\end{align}
	where}
	\begin{align}
	\label{Lambda_3}
	\Lambda_3^{(t)} & \triangleq  1 \! - \! \sum \limits_{k \in \mathcal{K}} \! \left[  \! \frac{2 \mu \lambda  \bar{h}_{k}^{(t)} p_{k}^{(t)}}{ K} \! - \! \frac{\mu L \lambda^2}{ K^2} \left( \bar{h}_{k}^{(t)} p_{k}^{(t)} \right) ^2 \right], \\
	\label{Lambda_4}
	\Lambda_4^{(t)} & \triangleq \frac{L \lambda^2}{2 K^2} \sum \limits_{k \in \mathcal{K}} \left( \bar{h}_{k}^{(t)} p_{k}^{(t)} \right) ^2 \left\| \boldsymbol{\delta} \right\| _2^2 + \frac{L Q \lambda^2 \sigma^2}{2 K^2}.
	\end{align}
\end{theorem}

\begin{IEEEproof}
	See Appendix \ref{proof_of_theorem_1}.
\end{IEEEproof}

The optimality gap in Theorem \ref{theorem_1} can guarantee the convergence for any learning rate $\lambda$ satisfying $\Lambda_3^{(t)} < 1$ such that $\lim \nolimits_{T \rightarrow \infty} \prod_{t=1}^{T} \Lambda_3^{(t)} = 0$.
The first term on the right-hand-side of (\ref{theorem_optimality_gap}) shows that the effect of initial optimality gap vanishes as training proceeds.
To explicitly evaluate the convergence performance with a decayed learning rate $\lambda^{(t)}$, we replace the constant learning rate with a diminishing one such that the optimality gap can be given as follows.
\begin{corollary}[Gap with a diminishing learning rate] \label{diminishing_rate}
	\emph{Suppose that the diminishing learning rate $\lambda^{(t)}$ is designed as $\lambda^{(t)} = \frac{\Gamma}{t + \nu} \le \frac{2K \sum_{k \in \mathcal{K}} \bar{h}_{k}^{(t)} p_{k}^{(t)} - K^2}{L \sum_{k \in \mathcal{K}} ( \bar{h}_{k}^{(t)} p_{k}^{(t)} ) ^2}$ where $\Gamma > 1/ \mu$ and $\nu > 0$.
	Then, the expected optimality gap at the $t$-th communication round is given by
	\begin{align} \label{gap_diminishing}
	\mathbb{E} \left[ F ( \boldsymbol{w}^{(t+1)} ) \right] - F (\! \boldsymbol{w}^{*} ) \le \frac{\xi_t}{t+ 1 + \nu},
	\end{align}
	where}
	\begin{subequations}
	\begin{align}
		\quad \xi_t &= \max \left\lbrace (t + \nu) \left( F (\boldsymbol{w}^{(t)}) - F(\boldsymbol{w}^{*}) \right) , \widetilde{Q} \right\rbrace, \\
		\widetilde{Q} &= \frac{ L \Gamma^2 \left( \sum_{k \in \mathcal{K}} ( \bar{h}_{k}^{(t)} p_{k}^{(t)} ) ^2 \| \boldsymbol{\delta} \| _2^2 + Q \sigma^2 \right) }{ 2K^2 (\mu \Gamma - 1) }.
	\end{align}
\end{subequations}
\end{corollary}

\begin{IEEEproof}
	See Appendix \ref{proof_of_corollary_1}.
\end{IEEEproof}

From Corollary \ref{diminishing_rate}, one can observe that when the learning rate is diminishing, the expected optimality gap declines linearly as the communication round increases.
Namely, the asymptotic convergence rate is $\mathcal{O} \left( 1/t \right)$.
Thus far, we conclude that the designed framework is guaranteed to converge.
In the following, we would like to formulate a resource allocation problem of training acceleration by jointly optimizing the active and passive beamforming schemes.

\begin{corollary}[Impact of gradient aggregation error] \label{analysis_aggregation_error}
	\emph{To analysis the impact of aggregation error on the learning performance of AirFL, we denote $ \boldsymbol{\varepsilon}^{(t)} =  \boldsymbol{\hat{g}}^{(t)} - \boldsymbol{g}^{(t)} $ as the gradient aggregation error.
		Let $\lambda \le \frac{1}{2+L}$ and $\tilde{\mu} = 1 - \lambda \mu$.
		Then, the optimality gap (\ref{theorem_optimality_gap}) is rewritten as}
		\begin{align} \label{gap_with_aggregation_error}
		& \mathbb{E} \left[  F ( \boldsymbol{w}^{(T+1)} ) \right] - F ( \boldsymbol{w}^{*} ) \nonumber \\
		& \le \tilde{\mu}^{T} \left(  F (  \boldsymbol{w}^{(1)} ) - F ( \boldsymbol{w}^{*} )  \right) 
		+ \frac{1}{2} \sum \limits_{t=1}^{T} \tilde{\mu}^{T-t} \| \underbrace{ \mathbb{E} [ \boldsymbol{\varepsilon}^{(t)} ] }_{\rm bias} \| _2^2 \nonumber \\
		& + \frac{L \lambda^2}{2} \sum \limits_{t=1}^{T} \tilde{\mu}^{T-t} \big( L \| \underbrace{ \mathbb{E} [ \boldsymbol{\varepsilon}^{(t)} ] }_{\rm bias} \| _2^2
		+ \left\| \boldsymbol{\delta} \right\| _2^2
		+ \underbrace{ \mathbb{E} [ \| \boldsymbol{\varepsilon}^{(t)} \| _2^2 ] }_{\rm gradient~MSE} \big).
		\end{align}
\end{corollary}

\begin{IEEEproof}
	See Appendix \ref{proof_of_corollary_2}.
\end{IEEEproof}

From Corollary \ref{analysis_aggregation_error}, it can be observed that, in the non-ideal case where the bias $\mathbb{E} [ \boldsymbol{\varepsilon}^{(t)} ]$ and the gradient MSE $\mathbb{E} [ \| \boldsymbol{\varepsilon}^{(t)} \| _2^2 ]$ are not zero, the aggregation error will have a significant impact on the learning behavior of AirFL.
Specifically,
1) for $\mathbb{E} [ \boldsymbol{\varepsilon}^{(t)} ] \ne 0$, the optimality gap in (\ref{gap_with_aggregation_error}) will not approach to zero as $T$ increases, even if under a sufficiently small learning rate ($\lambda \rightarrow 0$).
2) For $\mathbb{E} [ \| \boldsymbol{\varepsilon}^{(t)} \| _2^2 ] \ne 0$, the gradient aggregation error at the later communication rounds has more impact on the optimality gap than that at initial rounds, due to the monotonically increasing coefficient $\tilde{\mu}^{T-t}$.
3) In contrast, for the ideal case of $\mathbb{E} [ \boldsymbol{\varepsilon}^{(t)} ] = \mathbb{E} [ \| \boldsymbol{\varepsilon}^{(t)} \| _2^2 ] = 0$ with zero gradient variance, the considered AirFL model is able to reach the optimal performance without any gaps as long as $T$ is large enough.

\subsection{Problem Formulation}
Given the insights above, we are now interested in speeding up the convergence rate by minimizing the optimality gap obtained in Theorem \ref{theorem_1}.
To this end, we aim to jointly optimize the transmit power at users and the configuration design of STAR-RIS for convergence acceleration while guaranteeing communication quality.
Accordingly, subject to the transmit power constraints, the QoS requirements of NOMA users and the MSE tolerance of AirFL users in the considered network, the optimization problem towards minimizing (\ref{theorem_optimality_gap}) can be formulated as
{\allowdisplaybreaks[4]
\begin{subequations} \label{original_problem}
	\begin{align}
	\label{original_objective}
	\min \
	\ & \Upsilon \left(  \{ p_u^{(t)} \}, \{ \mathbf{\Theta}_u^{(t)} \} \right) \\
	{\rm s.t.} \
	\label{original_constraint_QoS}
	\ & \log_2 ( 1 + \gamma_{n}^{(t)} )  \ge R_{n}^{\min}, \ \forall n \in \mathcal{N}, \\
	\label{original_constraint_MSE}
	\ &{\rm MSE}^{(t)} \le \varepsilon_0, \\
	\label{original_constraint_RIS_mode}
	\ & \beta_{m}^{(t)} \in \{0, 1\}, \ \forall m \in \mathcal{M}, \\
	\label{original_constraint_RIS_reflection}
	\ & \theta_{m}^{(t)} \in [0, 2 \pi], \ \forall m \in \mathcal{M}, \\
	\label{original_constraint_RIS_transmission}
	\ & \phi_{m}^{(t)} \in [0, 2 \pi], \ \forall m \in \mathcal{M}, \\
	\ & {\rm (\ref{power_constraint}), \ (\ref{average_power_constraint}) \ and \ (\ref{channel_condition})},
	\end{align}
\end{subequations}
where
$R_{n}^{\min}$ is the QoS requirement at the $n$-th NOMA user,
and $\varepsilon_0 > 0$ is the maximum MSE error allowed by the AirFL users.}
The main challenge in solving problem (\ref{original_problem}) comes from the close coupling between the multiple continuous variables $\{  p_u, \theta_{m}, \phi_{m} \}$ and the discrete variables $\{ \beta_{m} \}$ in both objective function and constraints.

Note that problem (\ref{original_problem}) is a MINLP problem.
In general, no existing method is applicable for solving such a NP-hard problem optimally.
More specifically, due to the temporal coupling in the objective function (\ref{original_objective}), problem (\ref{original_problem}) is still difficult to solve even when we only consider the transmit power allocation or STAR-RIS configuration design.
Compared to the conventional case with reflecting-only RIS, more optimization variables are involved in (\ref{original_problem}) as the element mode can also be optimized and the reflection/transmission phase shifts will have a joint effect on the system performance.
Besides, the mode optimization results in integer variables, which is more challenging compared with the reflecting-only RIS.

To address this intractable issue, an alternating optimization technique is resorted to solve this problem in an efficient manner, i.e., first fix the configuration mode of the STAR-RIS and optimize the transmit power at users, then repeat this in turn until termination conditions are satisfied.
To be specific, we propose to decouple (\ref{original_problem}) into two subproblems:
\begin{enumerate}
	\item \emph{Subproblem of transmit power allocation}: Given the configuration mode of the STAR-RIS, our first subproblem is to minimize the optimality gap by dynamically controlling the transmit power at users.
	As a result, subject to the transmit power constraints, the QoS requirements of NOMA users, and the MSE tolerance of AirFL users, the non-convex and non-linear subproblem of the power allocation can be written as
	\begin{subequations} \label{problem_power_allocation}
		\begin{align}
		\label{problem_power_allocation_objective}
		\min 
		\ & \Upsilon \left(  \{ p_u^{(t)} \} \right) \\
		{\rm s.t.}
		\ & {\rm (\ref{power_constraint}), \ (\ref{average_power_constraint}), \ (\ref{original_constraint_QoS}) \ and \ (\ref{original_constraint_MSE})}.
		\end{align}
	\end{subequations}
	\item \emph{Subproblem of STAR-RIS configuration}: Given the transmit power at users, our second subproblem is to speed up the convergence rate by judiciously adjusting the configuration mode of the STAR-RIS.
	Therefore, subject to the channel quality order, the throughput- and distortion-related constraints, as well as the STAR-RIS characteristics, the mixed-integer programming subproblem of the configuration design is given by
	\begin{subequations} \label{problem_RIS_design}
		\begin{align}
		\label{problem_RIS_design_objective}
		\min 
		\ & \Upsilon \left( \{ \mathbf{\Theta}_u^{(t)} \} \right) \\
		{\rm s.t.} \
		\ & {\rm (\ref{channel_condition}) \ and \ (\ref{original_constraint_QoS}) - (\ref{original_constraint_RIS_transmission})}.
		\end{align}
	\end{subequations}
\end{enumerate}

Although the decoupled subproblems are simpler than (\ref{original_problem}), it is still extremely difficult to derive globally optimal solutions in the considered framework.
For example, the transmit power scalars are coupled over communication rounds and users in subproblem (\ref{problem_power_allocation}), which makes it highly non-convex.
Additionally, solving the configuration design subproblem (\ref{problem_RIS_design}) may result in exponential complexity due to the combinatorial feature of binary mode switching.
To obtain low-complexity solutions for the original MINLP problem (\ref{original_problem}), the subproblems are transformed to convex ones, as described next.

\section{Alternating Optimization} \label{section_optimization}

\subsection{Transmit Power Allocation} \label{section_power_allocation}
When the configuration mode of the STAR-RIS is given, we introduce an auxiliary variable $\tau \ge 0$ to rewrite the power allocation subproblem (\ref{problem_power_allocation}) as follows:
\begin{subequations} \label{problem_power_allocation_rewrite}
	\begin{align}
	\label{problem_power_allocation_rewrite_objective}
	\min \limits_{ \{ p_u^{(t)} \}, \tau}
	\ & \ \tau \\
	{\rm s.t.} \
	\label{gap_constraint}
	\ & \Upsilon (  \{ p_u^{(t)} \} ) \le \tau, \\
	\ & {\rm (\ref{power_constraint}), \ (\ref{average_power_constraint}), \ (\ref{original_constraint_QoS}) \ and \ (\ref{original_constraint_MSE})}.
	\end{align}
\end{subequations}
To overcome the non-convexity of constraints (\ref{original_constraint_QoS}) and  (\ref{original_constraint_MSE}), we introduce auxiliary variables $\rho_{u} = | p_{u} | ^{2}$ and $\eta_{k} = | \bar{h}_{k} p_{k} - 1 | ^2$.
Then we rewrite (\ref{original_constraint_QoS}) and  (\ref{original_constraint_MSE}) as
\begin{align}
\label{original_constraint_QoS_rewrite}
&
| \bar{h}_{n}^{(t)} | ^{2} \rho_{n}^{(t)} \!\ge\! \zeta_n \left( \sum \nolimits_{u=n+1}^{N+K} | \bar{h}_{u}^{(t)} | ^{2} \rho_{u}^{(t)} \!+\! \sigma^{2} \right)\!, \forall n \!\in\! \mathcal{N}\!, \\
\label{original_constraint_MSE_rewrite}
&
\sum \nolimits_{k \in \mathcal{K}} \eta_{k}^{(t)} + \sigma^2 \le \epsilon_0 K^2,
\end{align}
where $\zeta_n = 2^{R_n^{\min}} -1, \forall n \in \mathcal{N}$ are constant parameters.

The constraints in (\ref{original_constraint_QoS_rewrite}) and (\ref{original_constraint_MSE_rewrite}) are convex,
as are the other two power-related constraints in (\ref{power_constraint}) and (\ref{average_power_constraint}).
The main difficulty of solving problem (\ref{problem_power_allocation_rewrite}) lies in the non-convex constraint (\ref{gap_constraint}).
To handle this, we adopt the SCA method to approximate $\Upsilon ( \{ p_{u}^{(t)} \} )$ by applying the first-order Taylor expansion at a local point $\{ p_{u}^{(t)}[\ell] \}$ obtained in the $\ell$-th iteration.
The approximated constraint is thus given by
\begin{align} \label{approximated_gap}
& \Upsilon \left(  \{ p_u^{(t)} \} \right) \simeq \widetilde{\Upsilon} \left(  \{ p_u^{(t)} \} \right) \triangleq \Upsilon \left( \{ p_{u}^{(t)}[\ell] \} \right) \nonumber \\
& + \sum \nolimits_{u} \left( p_{u}^{(t)} - p_{u}^{(t)}[\ell] \right) \nabla \Upsilon \left( \{ p_{u}^{(t)}[\ell] \} \right), \ \forall u \in \mathcal{K},
\end{align}
where $\nabla \Upsilon ( \{ p_{u}^{(t)}[\ell] \} )$ denotes the first-order derivative at the local point $ \{ p_{u}^{(t)}[\ell] \}$, and is given in (\ref{exact_approximated_gap}).
The detailed derivations are given in Appendix \ref{proof_of_equation_33}.

\begin{figure*}[t]
	\begin{subequations}
	\label{exact_approximated_gap}
	\begin{eqnarray}
	& \nabla \Upsilon \left( p_u^{(t)} [\ell] \right)
	= & - \frac{ 2 \mu \lambda  \bar{h}_{u}^{(t)} \left( F^{(1)} - F^* \right) }{ K } \left( 1 - \frac{ \lambda  L\bar{h}_{u}^{(t)} p_{u}^{(t)} }{ K } \right) \prod \nolimits_{i \in \mathcal{T} \backslash \{t\}} \Lambda_3^{(i)} %
	+ \frac{ L  \lambda^2\left\| \boldsymbol{\delta} \right\| _2^2}{ K^2 } \left( \bar{h}_{u}^{(t)} \right) ^2 p_{u}^{(t)} \prod \nolimits_{i=t+1}^{T} \Lambda_3^{(i)} \nonumber \\ %
	&{}& - \frac{ 2 \mu \lambda  \bar{h}_{u}^{(t)} }{ K } \left( 1 - \frac{ \lambda L \bar{h}_{u}^{(t)} p_{u}^{(t)} }{K} \right) \sum_{j=1}^{t-1} \Lambda_4^{(j)} \frac{ \prod_{i=j+1}^{T} \Lambda_3^{(i)} }{\Lambda_3^{(t)}}, \ u \in \mathcal{K}, \ t \in \mathcal{T} \backslash \{1\}, \\ %
	& \nabla \Upsilon \left( p_u^{(1)} [\ell] \right)
	= & - \frac{ 2 \mu \lambda  \bar{h}_{u}^{(1)} \left( F^{(1)} - F^* \right) }{ K }  \left( 1 - \frac{ \lambda L \bar{h}_{u}^{(1)} p_{u}^{(1)} }{ K} \right) \prod \nolimits_{i=2}^{T} \Lambda_3^{(i)} %
	+ \frac{L \lambda^2 \left\| \boldsymbol{\delta} \right\| _2^2}{ K^2} \left( \bar{h}_{u}^{(1)} \right) ^2 p_{u}^{(1)} \prod \nolimits_{i=2}^{T} \Lambda_3^{(i)}.
	\end{eqnarray}
	\hrulefill
	\end{subequations}
\end{figure*}

To ensure that the linear approximation in (\ref{approximated_gap}) is accurate, a series of trust region constraints are constructed over iterations:
\begin{equation} \label{trust_region}
|  p_u^{(t)} - p_u^{(t)} [\ell] | \le r[\ell], \ \forall u \in \mathcal{U},
\end{equation}
where $ r[\ell] $ is the trust region radius at the $\ell$-th iteration.

Based on the above approximations, the transmit power allocation problem (\ref{problem_power_allocation_rewrite}) can be approximated by
\begin{subequations} \label{problem_power_allocation_rewrite_convex}
	\begin{align}
	\min \limits_{ \{ p_{u}^{(t)} \}, \{ \rho_{u}^{(t)} \}, \{ \eta_{k}^{(t)} \}, \tau }
	\ & \ \tau \\
	{\rm s.t.} \qquad \quad
	\ & \widetilde{\Upsilon} (  \{ p_u^{(t)} \} ) \le \tau, \ \forall u \in \mathcal{K}, \\
	\ & \rho_{u}^{(t)} \ge | p_{u}^{(t)} | ^{2}, \ \forall u \in \mathcal{U}, \\
	\ & \eta_{k}^{(t)} \ge | \bar{h}_{k}^{(t)} p_{k}^{(t)} - 1 | ^2, \ \forall k \in \mathcal{K}, \\
	\ & {\rm (\ref{power_constraint}), \ (\ref{average_power_constraint}), \ (\ref{original_constraint_QoS_rewrite}), \ (\ref{original_constraint_MSE_rewrite})  \ and \ (\ref{trust_region})},
	\end{align}
\end{subequations}
which is convex w.r.t. $\{ p_{u}^{(t)} \}, \{ \rho_{u}^{(t)} \}, \{ \eta_{k}^{(t)} \}$ and $\tau$. Thus, it can be directly solved using standard optimization toolbox such as CVX \cite{Grant2014CVX}.
The proposed trust region-based SCA algorithm for solving problem (\ref{problem_power_allocation}) is given in Algorithm \ref{algorithm_1}, where the radius of the trust region gradually decreases over iterations.
This algorithm terminates when $r[\ell]$ is lower than the preset threshold $\epsilon_1$ or the maximum iteration number $L_1$ is reached.

\begin{algorithm}[t]
	\caption{Trust Region-Based SCA Algorithm}
	\label{algorithm_1}
	\begin{algorithmic}[1]
		\renewcommand{\algorithmicrequire}{\textbf{Initialize}}
		\renewcommand{\algorithmicensure}{\textbf{Output}}
		\STATE \textbf{Initialize} $\{ p_u^{(t)}[0] \}$, $\{ \rho_u^{(t)}[0] \}$, $\{ \eta_k^{(t)}[0] \}$, $\tau[0]$, the trust region radius $r[0] = 1$, the threshold $\epsilon_1$, the maximum iteration number $L_1$, and set $\ell_1=0$.
		\REPEAT
		\STATE With given $\{ p_u^{(t)}[\ell_1] \}$, $\{ \rho_u^{(t)}[\ell_1] \}$, $\{ \eta_k^{(t)}[\ell_1] \}$ and $\tau[\ell_1]$, obtain $\{ p_u^{(t)}[\ell_1+1] \}$, $\{ \rho_u^{(t)}[\ell_1+1] \}$, $\{ \eta_k^{(t)}[\ell_1+1] \}$ and $\tau[\ell_1+1]$ by solving problem (\ref{problem_power_allocation_rewrite_convex}).
		\STATE Update $r[\ell_1+1] \leftarrow r[\ell_1] / 2$.
		\STATE Update $\ell_1 \leftarrow \ell_1 + 1$.
		\UNTIL $r[\ell_1]  \le \epsilon_1$ or $\ell_1 \ge L_1$.
		\STATE \textbf{Output} the converged transmit power solution.
	\end{algorithmic}
\end{algorithm}

\subsection{STAR-RIS Configuration} \label{section_RIS_configuration}
Given the transmit power at users, while replacing constraints (\ref{original_constraint_QoS}) and  (\ref{original_constraint_MSE}) with (\ref{original_constraint_QoS_rewrite}) and (\ref{original_constraint_MSE_rewrite}), the subproblem of STAR-RIS configuration can be rewritten as follows:
\begin{subequations} \label{subproblem_RIS_design}
	\begin{align}
	\label{subproblem_RIS_design_objective}
	\min 
	\ & \Upsilon \left( \{ \mathbf{\Theta}_u^{(t)} \} \right) \\
	{\rm s.t.}
	\label{subproblem_RIS_design_constraints}
	\ & {\rm (\ref{channel_condition}), \ (\ref{original_constraint_QoS_rewrite}) \ and \ (\ref{original_constraint_MSE_rewrite})}, \\
	\ & \boldsymbol{\Theta}_u^{(t)} \in \mathcal{Q}, \ \forall u \in \mathcal{U},
	\end{align}
\end{subequations}
where $\mathcal{Q} = \{ (\beta_m, \theta_m, \phi_m) \mid \beta_m \in \{0,1\}, \theta_m \in [0, 2 \pi], \phi_m \in [0, 2 \pi] \}$ denotes the feasible set for the mode switching and its corresponding phase shift per STAR-RIS element.

Let $\bar{\beta}_m = 1- \beta_m$. Then the STAR-RIS configuration vector is given by
\begin{equation}
	\boldsymbol{q}_u \!=\! \!\left\{\!
	\begin{array}{ll}
			\!\left[\! \beta_1 e^{j \theta _1}, \beta_2 e^{j \theta _2}, \!\ldots\!, \beta_M e^{j \theta _M} \right]^{\rm H}\!, \!\forall u \!\in\! \mathcal{N}_R \!\cup\! \mathcal{K}_R, \\
			\!\left[\! \bar{\beta}_1 e^{j \phi _1}, \bar{\beta}_2 e^{j \phi _2}, \!\ldots\!, \bar{\beta}_M e^{j \phi _M} \right]^{\rm H}\!, \!\forall u \!\in\! \mathcal{N}_T \!\cup\! \mathcal{K}_T. \\
	\end{array} \right.
\end{equation}

As such, we can obtain $\boldsymbol{\bar{r}}^{\rm H} \mathbf{\Theta}_u \boldsymbol{r}_u = \boldsymbol{q}_u^{\rm H} \boldsymbol{R}_u$ by changing variables from $\{ \mathbf{\Theta}_u \}$ to $\{ \boldsymbol{q}_u \}$, while $\boldsymbol{R}_u = {\rm diag} \{ \boldsymbol{\bar{r}}^{\rm H} \} \boldsymbol{r}_u$.
Next, the combined channel gain can be expressed as
\begin{align} \label{combined_channel}
\left| \bar{h}_u \right|^2
& = \left| h_u + \boldsymbol{\bar{r}}^{\rm H} \mathbf{\Theta}_u \boldsymbol{r}_u \right|^2 = \left| h_u + \boldsymbol{q}_u^{\rm H} \boldsymbol{R}_u \right|^2 \nonumber \\
& = \boldsymbol{q}_u^{\rm H} \boldsymbol{R}_u \boldsymbol{R}_u^{\rm H} \boldsymbol{q}_u + \boldsymbol{q}_u^{\rm H} \boldsymbol{R}_u h_u^{\rm H} + h_u \boldsymbol{R}_u^{\rm H} \boldsymbol{q}_u + \left| h_u \right|^2 \nonumber \\
& = \boldsymbol{\bar{q}}_u^{\rm H} \boldsymbol{\widetilde{R}}_u \boldsymbol{\bar{q}}_u + \left| h_u \right|^2,
\end{align}
where
\begin{equation}
\boldsymbol{\widetilde{R}}_u = \left[ \begin{array}{cc}
\boldsymbol{R}_u \boldsymbol{R}_u^{\rm H} & \boldsymbol{R}_u h_u^{\rm H} \\
h_u \boldsymbol{R}_u^{\rm H} & 0 
\end{array} 
\right ]
\text{and} \
\boldsymbol{\bar{q}}_u = \left[ \begin{array}{c}
\boldsymbol{q}_u \\
1
\end{array} 
\right ].
\end{equation}
Since $\boldsymbol{\bar{q}}_u^{\rm H} \boldsymbol{\widetilde{R}}_u \boldsymbol{\bar{q}}_u = {\rm tr} \left( \boldsymbol{\widetilde{R}}_u \boldsymbol{\bar{q}}_u \boldsymbol{\bar{q}}_u^{\rm H} \right)$, we have
\begin{equation} \label{combined_channel_gain}
\left| \bar{h}_u \right|^2
= {\rm tr} \left( \boldsymbol{\widetilde{R}}_u \boldsymbol{\bar{q}}_u \boldsymbol{\bar{q}}_u^{\rm H} \right) + \left| h_u \right|^2
= {\rm tr} \left( \boldsymbol{\widetilde{R}}_u \boldsymbol{Q}_u  \right) + \left| h_u \right|^2,
\end{equation}
where $\boldsymbol{Q}_u = \boldsymbol{\bar{q}}_u \boldsymbol{\bar{q}}_u^{\rm H}$.
The new introduced variables $\boldsymbol{Q}_u$ satisfy $\boldsymbol{Q}_u \succeq 0$, ${\rm rank} \left( \boldsymbol{Q}_u  \right) = 1$ and ${\rm Diag} \left( \boldsymbol{Q}_u  \right) = \boldsymbol{\beta}_u$.
Specifically, ${\rm Diag} \left( \boldsymbol{Q}_u  \right)$ denotes a vector whose elements are extracted from the main diagonal elements of matrix $\boldsymbol{Q}_u$, and $\boldsymbol{\beta}_u$ denotes the mode switching vector, i.e.,
\begin{equation}
\boldsymbol{\beta}_u = \left\{
\begin{array}{ll}
\left[ \beta_1, \beta_2, \ldots, \beta_M \right]^{\rm H}, \ \forall u \in \mathcal{N}_R \cup \mathcal{K}_R, \\
\left[ \bar{\beta}_1, \bar{\beta}_2, \ldots, \bar{\beta}_M \right]^{\rm H}, \ \forall u \in \mathcal{N}_T \cup \mathcal{K}_T. \\
\end{array} \right.
\end{equation}

Based on the matrix lifting technique adopted in (\ref{combined_channel_gain}), the constraints in (\ref{channel_condition}) and (\ref{original_constraint_QoS_rewrite}) can be rewritten as
{\allowdisplaybreaks[4]
\begin{align}
	\label{tr_Q_channel}
	& {\rm tr} \left( \boldsymbol{\widetilde{R}}_1^{(t)} \boldsymbol{Q}_1^{(t)}  \right) + \left| h_1^{(t)} \right|^2
	\ge \ldots \ge {\rm tr} \left( \boldsymbol{\widetilde{R}}_N^{(t)} \boldsymbol{Q}_N^{(t)}  \right) + \left| h_N^{(t)} \right|^2 \nonumber \\
	& \ge {\rm tr} \left( \boldsymbol{\widetilde{R}}_k^{(t)} \boldsymbol{Q}_k ^{(t)} \right) + \left| h_k^{(t)} \right|^2, \ \forall k \in \mathcal{K}, \\
	\label{tr_Q_QoS}
	 & \left[ {\rm tr} \left( \boldsymbol{\widetilde{R}}_n^{(t)} \boldsymbol{Q}_n^{(t)}  \right) + \left| h_n^{(t)} \right|^2 \right] \left| p_{n}^{(t)} \right| ^{2} \ge \zeta_n \sigma^2 \nonumber \\
	& + \zeta_n \sum \limits_{u=n+1}^{N+K} \left[ {\rm tr} \left( \boldsymbol{\widetilde{R}}_u^{(t)} \boldsymbol{Q}_u^{(t)}  \right) \!+\! \left| h_u^{(t)} \right|^2 \right] \left| p_{u}^{(t)} \right| ^{2}\!, \forall n \!\in\! \mathcal{N}.
\end{align}}

To transform the constraint (\ref{original_constraint_MSE_rewrite}) into a more tractable form, the following lemma is given to tackle its non-convexity.
\begin{lemma} \label{lemma_1}
	\emph{Let $\boldsymbol{\mathring{R}}_k = \boldsymbol{R}_k p_k$, $\widehat{h}_k = h_k p_k - 1$ and
	\begin{equation}
	\boldsymbol{\widehat{R}}_k = \left[ \begin{array}{cc}
	\boldsymbol{\mathring{R}}_k \boldsymbol{\mathring{R}}_k^{\rm H} & \boldsymbol{\mathring{R}}_k \widehat{h}_k^{\rm H} \\
	\widehat{h}_k \boldsymbol{\mathring{R}}_k^{\rm H} & 0 
	\end{array}
	\right ].
	\end{equation}
	Then we have
	\begin{equation} \label{constraint_MSE_transformed}
	\left| \bar{h}_{k} p_{k} - 1 \right| ^2 = {\rm tr} \left( \boldsymbol{\widehat{R}}_k \boldsymbol{Q}_k \right) + \left| \widehat{h}_k \right|^2.
	\end{equation}}
\end{lemma}

\begin{IEEEproof}
	See Appendix \ref{proof_of_lemma_1}.
\end{IEEEproof}

According to Lemma \ref{lemma_1}, the non-convex constraint in (\ref{original_constraint_MSE_rewrite}) can be rewritten as the following convex one:
\begin{equation} \label{constraint_MSE_rewritten}
\sum _{k \in \mathcal{K}} \left[ {\rm tr} \left( \boldsymbol{\widehat{R}}_k^{(t)} \boldsymbol{Q}_k^{(t)} \right) + \left| \widehat{h}_k^{(t)} \right|^2 \right] + \sigma^2 \le \epsilon_0 K^2.
\end{equation}

\begin{lemma} \label{lemma_2}
	\emph{Let $\boldsymbol{\mathring{R}}_k = \boldsymbol{R}_k p_k$, $\check{h}_k = \mathring{h}_k - \frac{K}{L \lambda}$ with $\mathring{h}_k = h_k p_k$,
	\begin{equation}
	\boldsymbol{\check{R}}_k = \left[ \begin{array}{cc}
	\boldsymbol{\mathring{R}}_k \boldsymbol{\mathring{R}}_k^{\rm H} & \boldsymbol{\mathring{R}}_k \check{h}_k^{\rm H} \\
	\check{h}_k \boldsymbol{\mathring{R}}_k^{\rm H} & 0 
	\end{array}
	\right ]\!,
	{\rm \ and \ }
	\boldsymbol{\bar{R}}_k = \left[ \begin{array}{cc}
	\boldsymbol{\mathring{R}}_k \boldsymbol{\mathring{R}}_k^{\rm H} & \boldsymbol{\mathring{R}}_k \mathring{h}_k^{\rm H} \\
	\mathring{h}_k \boldsymbol{\mathring{R}}_k^{\rm H} & 0 
	\end{array}
	\right ]\!. \nonumber
	\end{equation}
	Then, it holds that
	\begin{align}
		\nabla \Lambda_3 \left( \boldsymbol{Q}_k \right) & = \frac{ \mu L \lambda^2 }{ K^2 } \boldsymbol{\check{R}}_k^{\top}, \\
		\nabla \Lambda_4 \left( \boldsymbol{Q}_k \right) & = \frac{L \lambda^2 \left\| \boldsymbol{\delta} \right\| _2^2}{2 K^2} \boldsymbol{\bar{R}}_k^{\top}.
	\end{align}}
\end{lemma}

\begin{IEEEproof}
	See Appendix \ref{proof_of_lemma_2}.
\end{IEEEproof}

Based on Lemma \ref{lemma_2}, we can employ the first-order Taylor expansion to tackle the non-convexity of the objective function (\ref{subproblem_RIS_design_objective}).
For a given point $\{ \boldsymbol{Q}_u^{(t)}[\ell] \}$ in the $\ell$-th iteration of the SCA method, the objective (\ref{subproblem_RIS_design_objective}) can be approximated by
\begin{align}
\label{Upsilon_convex}
& \Upsilon \left( \{ \mathbf{\Theta}_u^{(t)} \} \right) \simeq
\widehat{\Upsilon} \left( \{ \boldsymbol{Q}_u^{(t)} \} \right) \triangleq \Upsilon \left( \{ \boldsymbol{Q}_u^{(t)}[\ell] \} \right) \nonumber \\
& + \left( \boldsymbol{Q}_u^{(t)} - \boldsymbol{Q}_u^{(t)}[\ell] \right) ^{\top} \nabla \Upsilon \left( \{ \boldsymbol{Q}_u^{(t)}[\ell] \} \right), \ u \in \mathcal{K},
\end{align}
where explicit expressions of $\nabla \Upsilon ( \{ \boldsymbol{Q}_u^{(t)}[\ell] \} )$ are given in (\ref{exact_Upsilon_derivative}).

\begin{algorithm}[t]
	\caption{Penalty-Based SDR Algorithm}
	\label{algorithm_2}
	\begin{algorithmic}[1]
		\renewcommand{\algorithmicrequire}{\textbf{Initialize}}
		\renewcommand{\algorithmicensure}{\textbf{Output}}
		\STATE \textbf{Initialize} $\{ \boldsymbol{Q}_u^{(t)}[0] \}$, $\{ \boldsymbol{\beta}_u^{(t)}[0] \}$, the penalty parameter $\chi$, the scaling factor $\varrho$, the threshold $\epsilon_p, \epsilon_c$, the maximum iteration number $L_2$, and set $\ell_2=0$.
		\REPEAT
		\STATE Let the iteration index $\ell_2 \leftarrow 0$.
		\REPEAT
		\STATE With given $\{ \boldsymbol{Q}_u^{(t)}[\ell_2] \}$ and $\{ \boldsymbol{\beta}_u^{(t)}[\ell_2] \}$, solve  (\ref{subproblem_RIS_design_convex}) to obtain $\{ \boldsymbol{Q}_u^{(t)}[\ell_2+1] \}$ and $\{ \boldsymbol{\beta}_u^{(t)}[\ell_2+1] \}$.
		\STATE Update $\ell_2 \leftarrow \ell_2 + 1$.
		\UNTIL $\left| \widetilde{\Psi}[\ell_2] - \widetilde{\Psi}[\ell_2-1] \right| \le \epsilon_p$ or $\ell_2 \ge L_2$.
		\STATE Update $\{ \boldsymbol{Q}_u^{(t)}[0] \}$ and $\{ \boldsymbol{\beta}_u^{(t)}[0] \}$ with the current solutions $\{ \boldsymbol{Q}_u^{(t)}[\ell_2] \}$ and $\{ \boldsymbol{\beta}_u^{(t)}[\ell_2] \}$.
		\STATE Update $\chi \leftarrow \varrho \chi$.
		\UNTIL the constraint violation is below $\epsilon_c$.
		\STATE \textbf{Output} the converged STAR-RIS configuration scheme.
	\end{algorithmic}
\end{algorithm}

\begin{figure*}[t]
	\begin{subequations}
		\label{exact_Upsilon_derivative}
		\begin{eqnarray}
		& \nabla \Upsilon ( \boldsymbol{Q}_u^{(t)} [\ell] )
		= & \frac{\mu L \lambda^2 \left( F^{(1)} - F^* \right) }{  K^2 } \left( \boldsymbol{\check{R}}_u^{(t)} \right)^{\top} \prod \nolimits_{i \in \mathcal{T} \backslash \{t\}} \Lambda_3^{(i)}
		+ \frac{L \lambda^2 \left\| \boldsymbol{\delta} \right\| _2^2}{2 K^2 } \left( \boldsymbol{\bar{R}}_u^{(t)} \right)^{\top} \prod \nolimits_{i=t+1}^{T} \Lambda_3^{(i)} \nonumber \\
		&{}& + \frac{ \mu L \lambda^2 }{ K^2 } \left( \boldsymbol{\check{R}}_u^{(t)} \right)^{\top} \sum_{j=1}^{t-1} \Lambda_4^{(j)} \frac{ \prod_{i=j+1}^{T} \Lambda_3^{(i)} }{\Lambda_3^{(t)}}, \ u \in \mathcal{K}, \ t \in \mathcal{T} \backslash \{1\}, \\ %
		& \nabla \Upsilon ( \boldsymbol{Q}_u^{(1)} [\ell] )
		= & \frac{ \mu L \lambda^2 \left( F^{(1)} - F^* \right) }{ K^2  } \left( \boldsymbol{\check{R}}_u^{(1)} \right)^{\top} \prod \nolimits_{i=2}^{T} \Lambda_3^{(i)}
		+  \frac{L \lambda^2 \left\| \boldsymbol{\delta} \right\| _2^2}{2 K^2} \left( \boldsymbol{\bar{R}}_u^{(1)} \right)^{\top} \prod \nolimits_{i=2}^{T} \Lambda_3^{(i)}, \ u \in \mathcal{K}.
		\end{eqnarray}
		\hrulefill
	\end{subequations}
\end{figure*}

As a result, by replacing the non-convex terms (\ref{subproblem_RIS_design_objective})-(\ref{subproblem_RIS_design_constraints}) with (\ref{tr_Q_channel}), (\ref{tr_Q_QoS}), (\ref{constraint_MSE_rewritten}) and (\ref{Upsilon_convex}), problem (\ref{subproblem_RIS_design}) can be transformed into the following one:
{\allowdisplaybreaks[4]
\begin{subequations} \label{subproblem_RIS_design_transformed}
	\begin{align}
	\label{subproblem_RIS_design_objective_transformed}
	\min \limits_{ \{ \boldsymbol{Q}_u^{(t)} \}, \{ \boldsymbol{\beta}_u^{(t)} \} }
	\ & \widehat{\Upsilon} \left( \{ \boldsymbol{Q}_u^{(t)} \} \right) \\
	{\rm s.t.} \quad \
	\ & {\rm (\ref{tr_Q_channel}), \ (\ref{tr_Q_QoS}) \ and \ (\ref{constraint_MSE_rewritten})}, \\
	\label{Diag_Q}
	\ & {\rm Diag} \left( \boldsymbol{Q}_u^{(t)}  \right) = \boldsymbol{\beta}_u^{(t)}, \ \forall u \in \mathcal{U}, \\
	\label{Q_0}
	\ & \boldsymbol{Q}_u^{(t)} \succeq 0, \ \forall u \in \mathcal{U}, \\
	\label{rank_Q}
	\ & {\rm rank} \left( \boldsymbol{Q}_u^{(t)}  \right) = 1, \ \forall u \in \mathcal{U}, \\
	\label{beta_01}
	\ & \beta_{m}^{(t)} \in \{0, 1\}, \ \forall m \in \mathcal{M}.
	\end{align}
\end{subequations}
At this point, the remaining non-convexity of problem (\ref{subproblem_RIS_design_transformed}) lies in the rank-one constraint (\ref{rank_Q}) and the binary constraint (\ref{beta_01}).}
To tackle these issues, the SDR method can be invoked to drop the non-convex rank-one constraint directly \cite{Wu2019IRS, Ni2021Resource}.
Then, the penalty method can be combined with the linear relaxation to solve the combinatorial optimization problem w.r.t. the binary mode switching.

Specifically, the binary constraint (\ref{beta_01}) can be equivalently rewritten as
\begin{equation} \label{binary_constraint}
\beta_{m}^{(t)} ( 1 - \beta_{m}^{(t)} ) = 0, \ \forall m \in \mathcal{M},
\end{equation}
where the binary variables are relaxed into the continuous ones, i.e., $\beta_{m}^{(t)} \in \{0, 1\} \rightarrow \beta_{m}^{(t)} \in [0, 1], \forall m \in \mathcal{M}$.
Then, by adding the equality (\ref{binary_constraint}) as a penalty term into the objective function (\ref{subproblem_RIS_design_objective_transformed}), it can be obtained as
\begin{equation} \label{penalty_objective}
\Psi \left( \{ \boldsymbol{Q}_u^{(t)} \}, \{ \boldsymbol{\beta}_u^{(t)} \} \right) = \widehat{\Upsilon} \left( \{ \boldsymbol{Q}_u^{(t)} \} \right) + \chi \sum \nolimits_{m=1}^{M} \beta_{m}^{(t)} ( 1 - \beta_{m}^{(t)}),
 \end{equation}
where $\chi > 0$ denotes the positive penalty parameter that penalizes the objective function if $\beta_{m}^{(t)} \in (0, 1)$.

Note that (\ref{penalty_objective}) is still non-convex w.r.t. $ \{ \boldsymbol{\beta}_u^{(t)} \}$.
To transform it into a convex form, the penalty-based objective function (\ref{penalty_objective}) is approximated by using the first-order Taylor expansion at a given point $\beta_{m}^{(t)} [\ell]$ in the $\ell$-th iteration of the SCA method, which is given by
\begin{align} \label{penalty_objective_convex}
& \Psi \left( \{ \boldsymbol{Q}_u^{(t)} \}, \{ \boldsymbol{\beta}_u^{(t)} \} \right)
\simeq \widetilde{\Psi} \left( \{ \boldsymbol{Q}_u^{(t)} \}, \{ \boldsymbol{\beta}_u^{(t)} \} \right)
\triangleq \widehat{\Upsilon} \left( \{ \boldsymbol{Q}_u^{(t)} \} \right)  \nonumber \\ 
& + \chi \sum \nolimits_{m=1}^{M} \left[ \beta_{m}^{(t)} ( 1 - 2 \beta_{m}^{(t)} [\ell] ) + ( \beta_{m}^{(t)} [\ell] )^2 \right].
\end{align}
Combing the above approximation with the SDR method, we replace the objective function (\ref{subproblem_RIS_design_objective_transformed}) with its approximation in (\ref{penalty_objective_convex}) and drop constraint (\ref{rank_Q}) directly.
Then, problem (\ref{subproblem_RIS_design_transformed}) can be reformulated as
\begin{subequations} \label{subproblem_RIS_design_convex}
	\begin{align}
	\min
	\ &\widetilde{\Psi} \left( \{ \boldsymbol{Q}_u^{(t)} \}, \{ \boldsymbol{\beta}_u^{(t)} \} \right) \\
	{\rm s.t.}
	\ & {\rm (\ref{tr_Q_channel}), (\ref{tr_Q_QoS}), (\ref{constraint_MSE_rewritten}), (\ref{Diag_Q}) \ and \ (\ref{Q_0})}, \\
	\ & \beta_{m}^{(t)} \in [0, 1], \ \forall m \in \mathcal{M}.
	\end{align}
\end{subequations}
Since problem (\ref{subproblem_RIS_design_convex}) is convex, it can be efficiently solved by existing optimization solvers such as CVX.
In summary, we can obtain a suboptimal solution to problem (\ref{subproblem_RIS_design}) by solving its approximated problem (\ref{subproblem_RIS_design_convex}) in the inner loop and gradually increasing $\chi$ in the outer loop.
If the obtained solution fails to be rank-one, the Gaussian randomization method can be adopted to construct a rank-one solution  \cite{Ni2021Resource}.
The penalty-based SDR algorithm for solving (\ref{subproblem_RIS_design}) is given in Algorithm \ref{algorithm_2}, where $\varrho$ is a scaling factor for the penalty parameter.

\begin{figure} [t!]
	\centering
	\includegraphics[width=3.5 in]{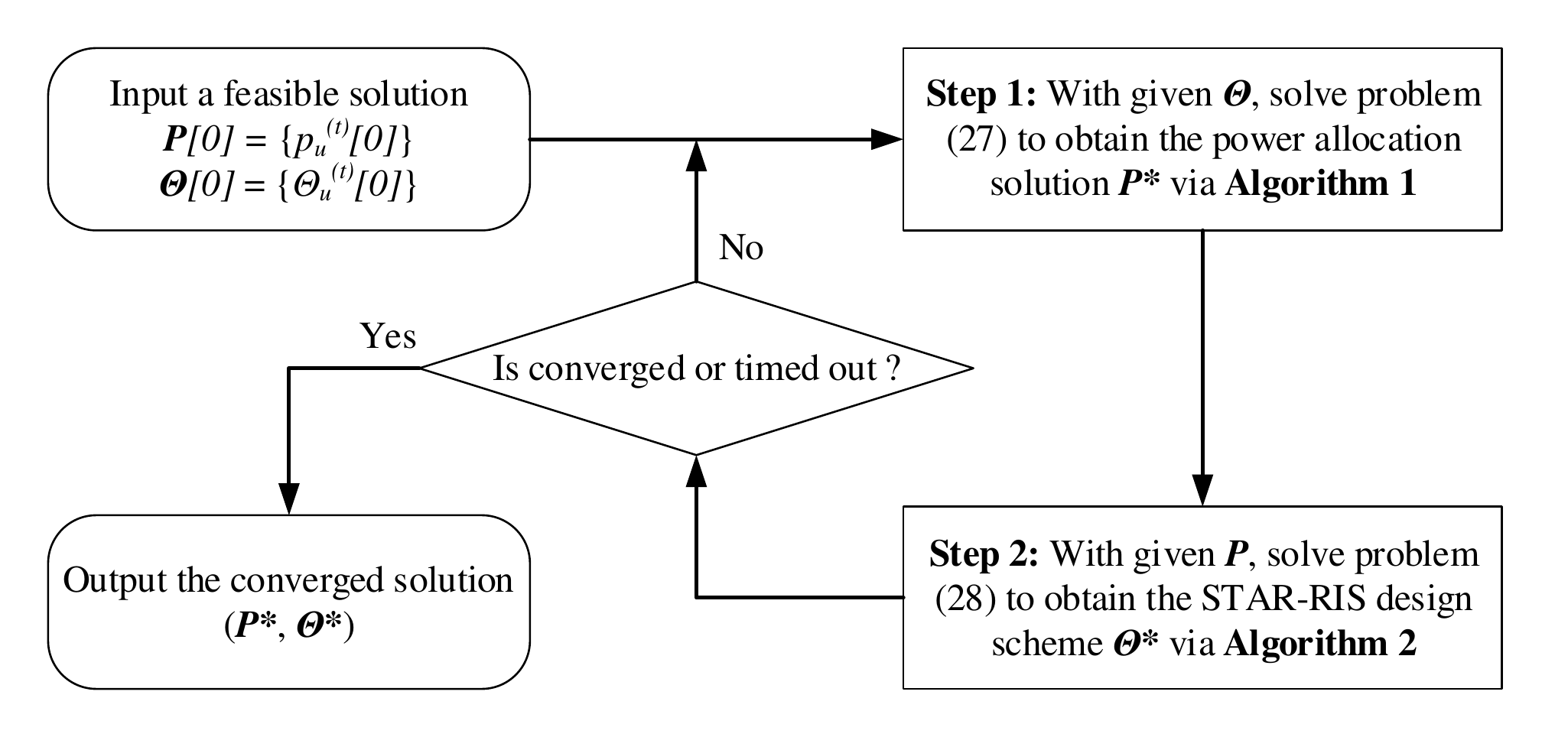}
	\caption{A flowchart of the proposed alternating optimization algorithm.}
	\label{flowchart}
\end{figure}

A final flowchart of the proposed alternating optimization algorithm is given in Fig. \ref{flowchart}.
Owing  to the iterative optimization in steps 1 and 2, the expected learning gap is non-increasing over iterations.
Since the achievable optimality gap is lower bounded by zero, the sequence $\{ \Upsilon [\ell] \}$ converges to a locally optimal solution as long as the number of iterations is sufficiently large.
Furthermore, if the interior-point method is considered in each optimization step, the main complexity of Algorithm \ref{algorithm_1} is given by $ \mathcal{O}\left( ( 2NT+3KT+T )^{3} \max \{ \log(1/\epsilon_1), L_1 \} \right)$.
Similarly, the complexity of Algorithm 2 is $ \mathcal{O}\left( (M^2T + MT)^{3} L_p \log(1/\epsilon_c) \right)$, where $L_p = \max \{ \log(1/\epsilon_p), L_2 \}$ denotes the number of inner iterations required for convergence and $\log(1/\epsilon_c)$ represents the outer iterations required for satisfying the preset precision.
Usually, we have $M>N$ and $M>K$, thus the main complexity of the proposed alternating optimization algorithm is given by
$ \mathcal{O} \left( L_{\rm a} L_p M^6 T^3 \log(1/\epsilon_c) \right)$ where $L_a$ denotes the maximum iteration number.

\subsection{Extension to MIMO Setup}
	The proposed framework in the multiple-input multiple-output (MIMO) setup differs from its single-antenna counterpart.
	Suppose that the BS has $N_r$ antennas and each user has $N_t$ antennas.
	The BS-User link, the RIS-User link, and the BS-RIS link are denoted by $\boldsymbol{H}_u \in \mathbb{C}^{N_r \times N_t}$, $\boldsymbol{R}_u \in \mathbb{C}^{M \times N_t}$, and $\boldsymbol{\bar{R}} \in \mathbb{C}^{M \times N_r}$, respectively.
	The multi-antenna BS applies the receive beamforming $\boldsymbol{m}_n \in \mathbb{C}^{N_r \times 1}$ to detect the individual signal of the $n$-th NOMA user and applies $\boldsymbol{a} \in \mathbb{C}^{N_r \times 1}$ to aggregate the superposed signal of AirFL users.	
	Let $\boldsymbol{p}_u \in \mathbb{C}^{N_t \times 1}, \forall u$ be the transmit beamforming at the $u$-th user.
	Then, the received signal at the BS for the $n$-th NOMA user is given by
	\begin{equation}
	\setlength{\abovedisplayskip}{3pt}
	\setlength{\belowdisplayskip}{3pt}
	{y}_{n}
	= \boldsymbol{m}_{n}^{\rm H} \left( 
	\sum \limits_{{n'}=1}^{N} \boldsymbol{\bar{H}}_{n'} \boldsymbol{p}_{n'} {s}_{n'}
	\!+\! \sum \limits_{k=1}^{K} \boldsymbol{\bar{H}}_k \boldsymbol{p}_k {s}_k
	\!+\! \boldsymbol{z}_0
	\right), \forall n \in \mathcal{N},
	\end{equation}
	where
	$\boldsymbol{\bar{H}}_u = \boldsymbol{H}_u  +  \boldsymbol{\bar{R}}^{\rm H} \mathbf{\Theta}_u \boldsymbol{R}_u, \forall u \in \mathcal{U} = \mathcal{N} \cup \mathcal{K}$ denote the combined channel coefficients,
	the receive beamforming satisfies $\left\| \boldsymbol{m}_{n} \right\| _2^2 = 1, \forall n \in \mathcal{N}$ and,
	$\boldsymbol{z}_0 \sim \mathcal{CN}(0, \sigma^{2} \boldsymbol{I})$ is the AWGN at the BS.
	
	Similar to the decoding order (\ref{channel_condition}) in the single antenna case, the combined channel coefficients in the MIMO setup can be ranked as
	\begin{equation}  \label{channel_condition_MIMO}
	\underbrace{\left\| \boldsymbol{\bar{H}}_{1} \right\|_F^2 \ge \left\| \boldsymbol{\bar{H}}_{2} \right\|_F^2 \ge \cdots \ge \left\| \boldsymbol{\bar{H}}_{N} \right\|_F^2}_{\mathbf {strong~users}}
	\ge
	\underbrace{\left\| \boldsymbol{\bar{H}}_{k} \right\|_F^2, \ \forall k \in \mathcal{K}}_{\mathbf {weak~users}}. 
	\end{equation}
	
	Using SIC, the achievable data rate (bps/Hz) of the $n$-th NOMA user is given by
	\begin{equation}  \label{SINR_MIMO}
	\Gamma_{n} = \log_2 \left( 1 + \frac{\left| \boldsymbol{m}_n^{\rm H} \boldsymbol{\bar{H}}_n \boldsymbol{p}_n \right| ^2 }{\sum \nolimits_{u=n+1}^{N+K} \left| \boldsymbol{m}_n^{\rm H} \boldsymbol{\bar{H}}_u \boldsymbol{p}_u \right| ^2 + \sigma^{2} }  \right)  , \ \forall n \in \mathcal{N}.
	\end{equation}
	
	After successfully removing NOMA communication symbols $\{ {s}_n \}$ from the superposition signal, we have $\boldsymbol{\hat{y}} = \sum \nolimits_{k = N+1}^{N+K} \boldsymbol{\bar{H}}_k \boldsymbol{p}_k {s}_k + \boldsymbol{z}_0$.
	Then, the computation output is given by $\hat{s} = \frac{1}{K} \boldsymbol{a}^{\rm H} \boldsymbol{\hat{y}}$.
	Next, the MSE of $\hat{s}$ with respect to ${s} = \frac{1}{K} \sum_{k \in \mathcal{K}} {s}_k$ can be written as
	\begin{align} \label{MSE_MIMO}
	{\rm \widetilde{MSE}}
	= \frac{1}{K^{2}} \left( \sum \nolimits_{k \in \mathcal{K}} \left| \boldsymbol{a}^{\rm H} \boldsymbol{\bar{H}}_k \boldsymbol{p}_k - 1 \right| ^2
	+ \sigma^2 \left\| \boldsymbol{a}^{\rm H} \right\| _2^2 \right).
	\end{align}
	
	Substituting the above new definitions and expressions into problem (\ref{original_problem}), we can obtain the optimization problem for the MIMO setup.
	Compared to problem (\ref{original_problem}) in the single-antenna case, we need to optimize extra variables (i.e., the receive beamforming $\{\boldsymbol{m}_n^{(t)}\}$ for NOMA users and $\{\boldsymbol{a}^{(t)}\}$ for AirFL users) in the MIMO setup.
	By using the problem decomposition method of Section \Rmnum{3}, the optimization problem formulated for the MIMO setup can be decomposed into three subproblems: transmit beamforming, STAR-RIS configuration, and receiving beamforming.
	For the transmit beamforming subproblem, the zero-forcing precoding and matrix lifting-based SDR technique can be adopted to solve it heuristically \cite{Zhu2019AirComp}.
	For the STAR-RIS configuration subproblem, the proposed penalty-based method can be extended to solve it efficiently, due to the similar constraint structure as problem (\ref{original_problem}).
	For the receiving beamforming subproblem, the minimum mean square error (MMSE) receivers can be employed to balance system performance and design complexity \cite{Qi2020Integrated}.
	To avoid redundancy, the details of solving the optimization problem for the MIMO setup are omitted here for brevity.

\section{Simulation Settings and Results} \label{section_simulation}

\subsection{Experimental Setup}
\begin{figure} [t!]
	\color{black}
	\centering
	\includegraphics[width=3.2 in]{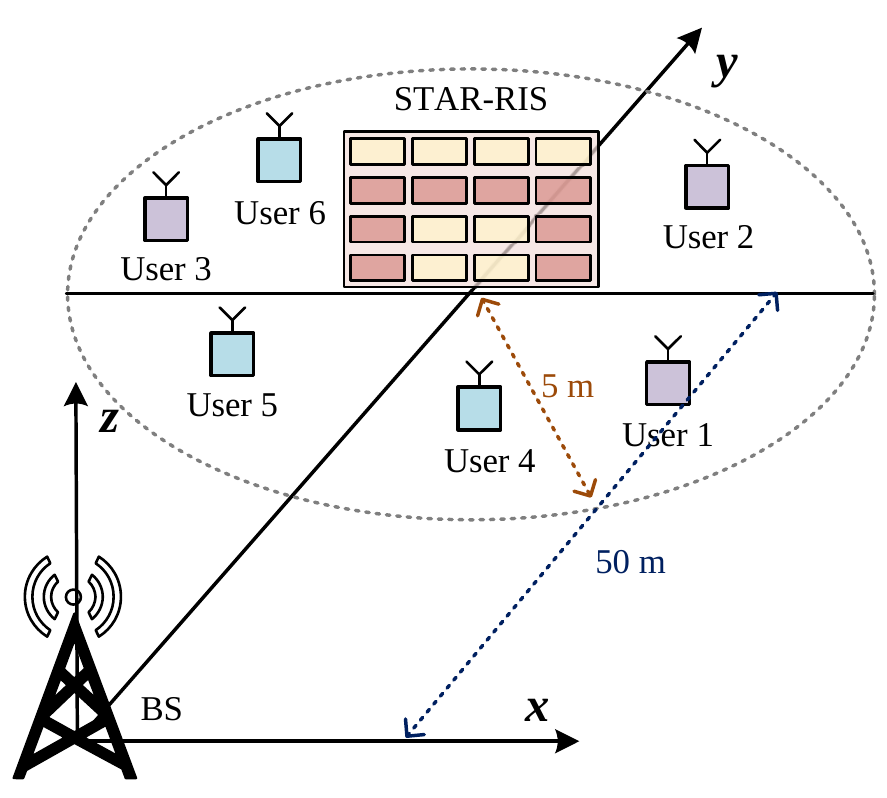}
	\caption{Topology setup for simulation (3D view).}
	\label{simulation_setup}
\end{figure}

\textbf{Network and topology settings:}
The topology setup for the simulations is shown in Fig. \ref{simulation_setup}.
We consider that there are $U=6$ users and one BS in the STAR-RIS assisted heterogeneous network, where $N=3$ NOMA users and $K=3$ AirFL users are randomly and uniformly distributed in a circle centered at the STAR-RIS with a radius of ${\rm  5 \ m}$.
In the three-dimensional (3D) Cartesian coordinates, the BS and the STAR-RIS are located at $(0, 0, 0)$ and $(0, 50, 0)$, respectively.
The settings of channel model are similar to \cite{Ni2021Resource}, where the reference path loss is set as $\varsigma_0 = -30 {\rm \ dBm}$,
the large-scale path loss exponent is $\alpha=2.2$,
and the Rician factor is $\kappa = 2$.
The power budget of the $u$-th user is set as $P_{u}=23 {\rm \ dBm}$ and $\bar{P}_{u} = 20{\rm \ dBm}$, the noise power is set to be $\sigma^{2} = -80 {\rm \ dBm}$.
The minimum data rate requirement of the $n$-th NOMA user is assumed to be $R_{n}^{\min} = 1 {\rm \ bps/Hz}$, and the aggregation error tolerance of the AirFL users is $\varepsilon_0 = 0.01$.
The number of reflecting elements is set as $M=20$, unless otherwise stated.

\textbf{Datasets and learning tasks:}
To evaluate the effectiveness of our proposed algorithms for the STAR-RIS assisted heterogeneous network, we use one synthetic dataset and two real datasets in our experiments.
To be specific, for the synthetic dataset, we examine the proposed algorithm with one simple regression task, where the optimality gap over the training process is adopted as the performance metric by training a linear regression model with MSE loss function \cite{Guo2021Analog, Dinh2021Federated}.
For the real datasets, AirFL users are required to execute classification tasks where they need to collaboratively train (machine learning) ML models for image classification on the MINST and CIFAR-10 datasets \cite{Chen2021Convergence,Chen2021PNAS,Yang2020TWC ,Zhu2020Broadband, Guo2021Analog, Zhu2021One, Zhang2021Gradient}.
Specifically, all datasets are divided randomly into a training set with $75\%$ samples and a testing set with $25\%$ ones.
The learning task is to train a 6-layered convolutional neural network (CNN) for handwritten digit identification on the MNIST dataset, and a 50-layered residual network (ResNet) for object recognition on the CIFAR-10 dataset.
The cross-entropy error is adopted as the loss function, unless otherwise indicated,
the constant learning rate is set to be $\lambda = 10^{-4}$,
and the maximum communication round is set as $T=200$.
At the end of each round, we assess the learning performance of the aggregated model in terms of the training loss and test accuracy.

\textbf{Benchmark schemes:}
For comparison, we consider the following benchmark schemes in the experiments.
\begin{itemize}
	\item[\rmnum{1}.] Noise-free FL: This is an ideal case where there is no communication noise in the considered wireless networks and the signal from NOMA users can be perfectly decoded and removed, i.e., the BS and users can interact the accurate information without any distortion. This scheme can be viewed as the optimal performance of AirFL users.
	\item[\rmnum{2}.] Conventional RIS: In this case, all users in full space are served by one reflecting-only RIS and one transmitting-only RIS. For fairness, the two conventional RISs are placed at the same location as the STAR-RIS, but each one is equipped with $M/2$ elements. Therefore, this baseline can be treated as a special case of the considered STAR-RIS where the mode of all elements is fixed.
	\item[\rmnum{3}.] Random STAR-RIS: In this case, we assume that the phase shifts of all elements are randomly selected from $[0, 2 \pi]$ and only the mode switching indicators of the STAR-RIS need to be optimized by the proposed Algorithm \ref{algorithm_2} at each round. This scheme degenerates subproblem (\ref{problem_RIS_design}) into an element allocation problem.
	\item[\rmnum{4}.] Equal power allocation: This is also a simplified baseline case where the transmit power of all users keeps the same during different communication rounds under the constraint of average power budget, i.e., $p_u^{(t)} = \bar{P}_{u}^{(t)} / T, \forall u, t$, and the STAR-RIS is configured by Algorithm \ref{algorithm_2}.
\end{itemize}

\subsection{Performance Evaluation}

\begin{figure} [t!]
	\color{black}
	\centering
	\subfloat[]{\label{Linear_reg_Opt_gap}
		\begin{minipage}[t]{0.5 \textwidth}
			\centering
			\includegraphics[width= 3.5 in]{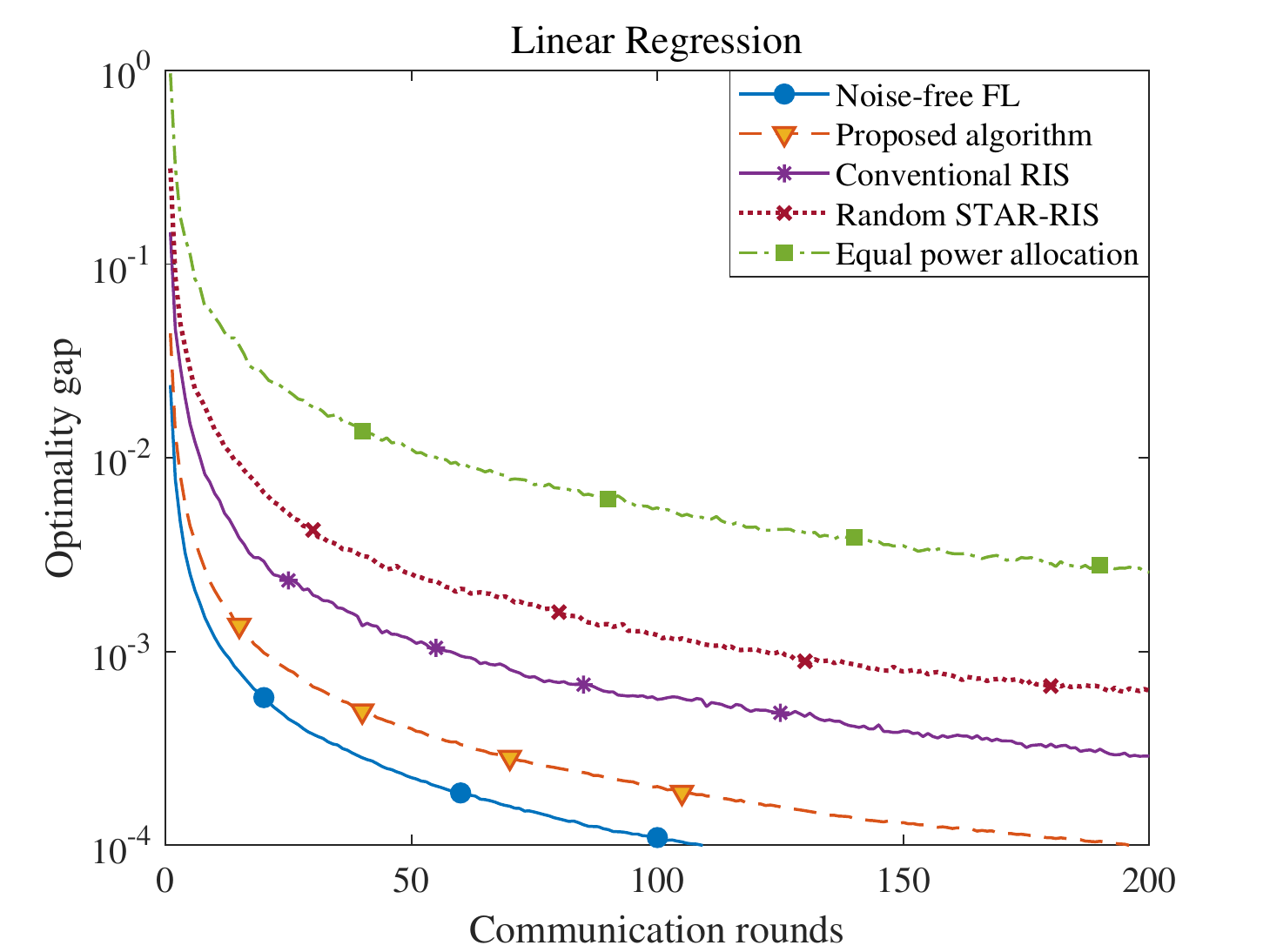}
		\end{minipage}
	} \\  % [-3 pt]
	\subfloat[]{\label{Linear_reg_vs_M}
		\begin{minipage}[t]{0.5 \textwidth}
			\centering
			\includegraphics[width= 3.5 in]{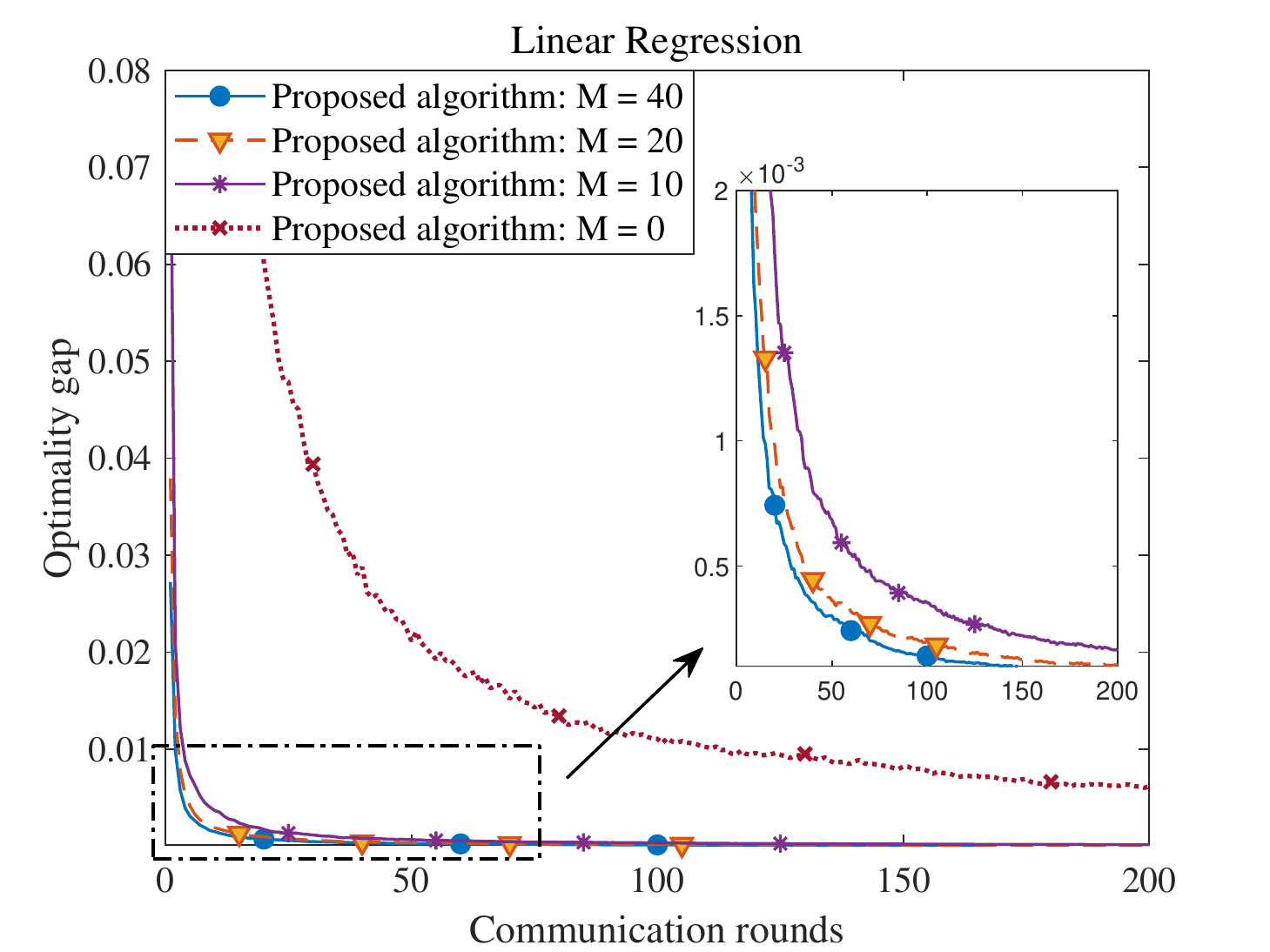}
		\end{minipage}
	}
	\caption{Learning performance of training a linear regression model on the synthetic dataset: (a) Optimality gap under different schemes. (b) Optimality gap under different $M$.}
	\label{optimality_gap}
	%	\vspace{-3 mm}
\end{figure}

In Fig. \ref{optimality_gap}, we compare the optimality gap of all the above schemes versus the communication rounds in the training process of the linear regression task.
To prepare the synthetic dataset, we generate a total of $4 \times 10^4$ data pairs including $3 \times 10^4$ samples for training and $10^4$ ones for testing.
Following the independent and identically distributed (IID) setting, the training set is distributed evenly to $3$ AirFL users.
In each training round, only $50$ samples are used to train the local model at the local users, and another $50$ samples are used to test the aggregated model at the BS.
The generated data pairs $(\mathbf{x}, y)$ follow the function $y = \mathbf{c}^{\top} \mathbf{x} + 0.5 n_0$ where the input vector $\mathbf{x} \in \mathbb{R}^{10}$ follows the IID Gaussian distribution as $\mathcal{N}(0, \boldsymbol{I})$ and the observation noise $n_0$ obeys $\mathcal{N}(0, 1)$ as well as $\mathbf{c} = [1,2,\ldots,10]^{\top}$.
From Fig. \ref{Linear_reg_Opt_gap}, we observe that the convergence rate of the proposed algorithm is closest to that of the ideal noise-free scheme compared to other benchmark schemes.
Moreover, it can be seen that the curve of all schemes has a fast linear convergence rate when the communication round $t$ is small.
However, as $t$ grows larger, the optimality gap decreases slower than that in the initial training process.
This is because the learning performance is suffered from the limited size of the local datasets as the training goes on.
Fig. \ref{Linear_reg_vs_M} demonstrates the optimality gap under different numbers of STAR-RIS elements during the training process.
From this figure, one can see that increasing the number of STAR-RIS elements is beneficial to achieve lower optimality gap though the gains are not so significant.
However, if the STAR-RIS is removed from the integrated system (i.e., $M=0$), the learning performance of AirFL users degrades significantly due to the strong interference from NOMA users.
This validates the importance of interference management of non-orthogonal transmissions, and also reveals that using STAR-RIS to maintain the decoding order (\ref{channel_condition}) is an effective design.

\begin{figure} [t]
	\centering
	\includegraphics[width=3.5 in]{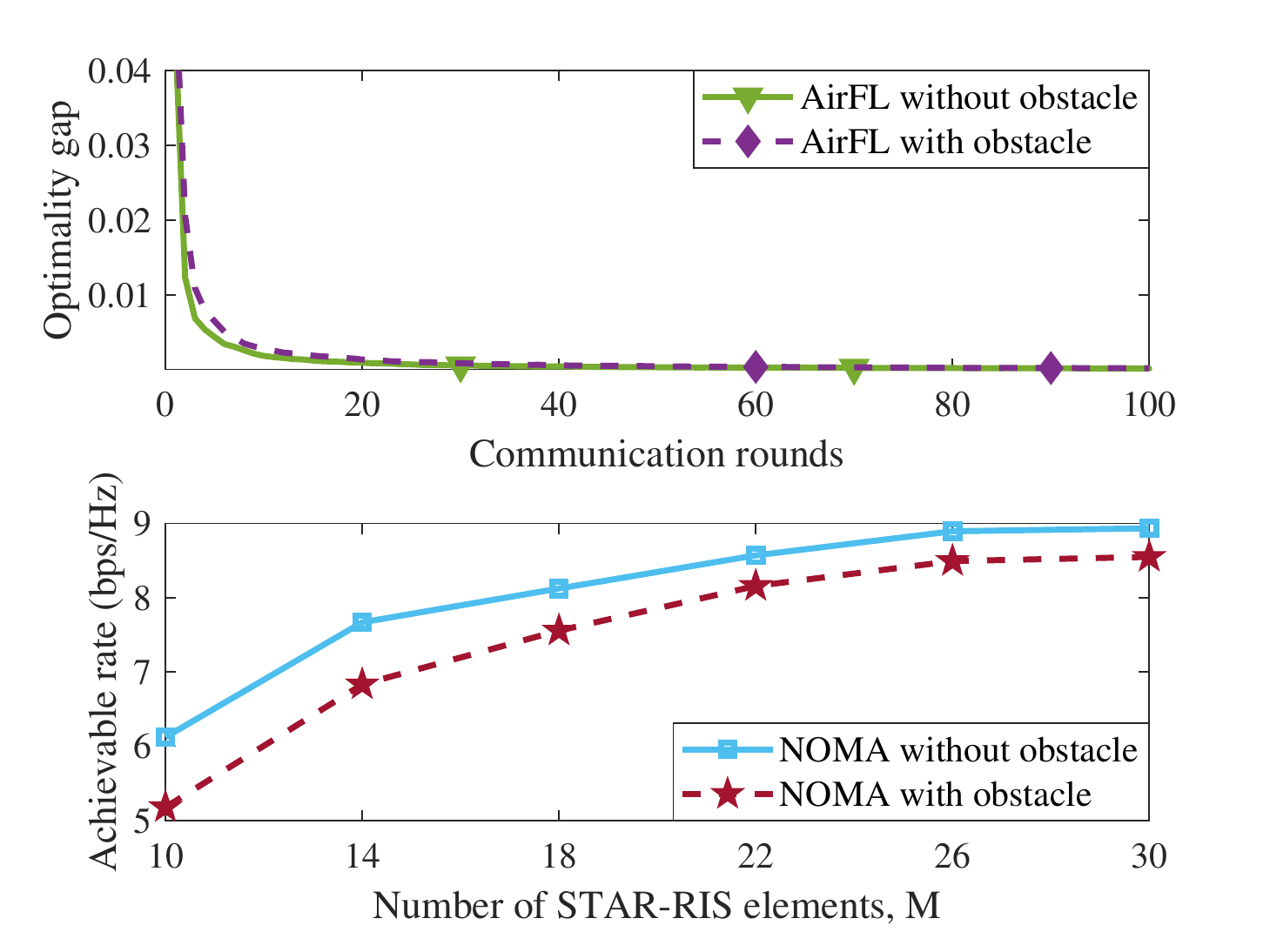}
	\caption{The impact of the obstacle on the optimality gap of AirFL users and the achievable rate of NOMA users.}
	\label{gap_and_rate}
\end{figure}

In Fig. \ref{gap_and_rate}, we demonstrate the impact of the obstacle on the optimality gap of AirFL users and the achievable rate of NOMA users.
Specifically, when the obstacle is considered in the integrated network, the direct links between the BS and all users are assumed to be blocked.
Otherwise, the LoS link is available to all NOMA and AirFL users.
From this figure, we can observe two phenomena in terms of the convergence behavior and data transmission rate.
First, compared to the blocking case (i.e., AirFL with obstacle), lower optimality gap can be achieved by the unblocking case (i.e., AirFL without obstacle) at the early communication rounds if the LoS links between the BS and AirFL users are available.
However, as the number of communication rounds increases, the performance gap between the above two schemes is negligible.
Second, the achievable uplink rate of NOMA users would be compromised to some extent when the LoS links between the BS and AirFL users are blocked.
Furthermore, it can be notice that the negative effects of link blocking are not be eliminated as the number of RIS elements increases
On the whole, the blocking issue experienced by the considered network affects the communication rate of NOMA users more than the learning performance of AirFL users.

\begin{figure} [t]
	\color{black}
	\centering
	\subfloat[]{\label{achievable_rate_vs_RIS_location}
		\begin{minipage}[t]{0.5 \textwidth}
			\centering
			\includegraphics[width= 3.5 in]{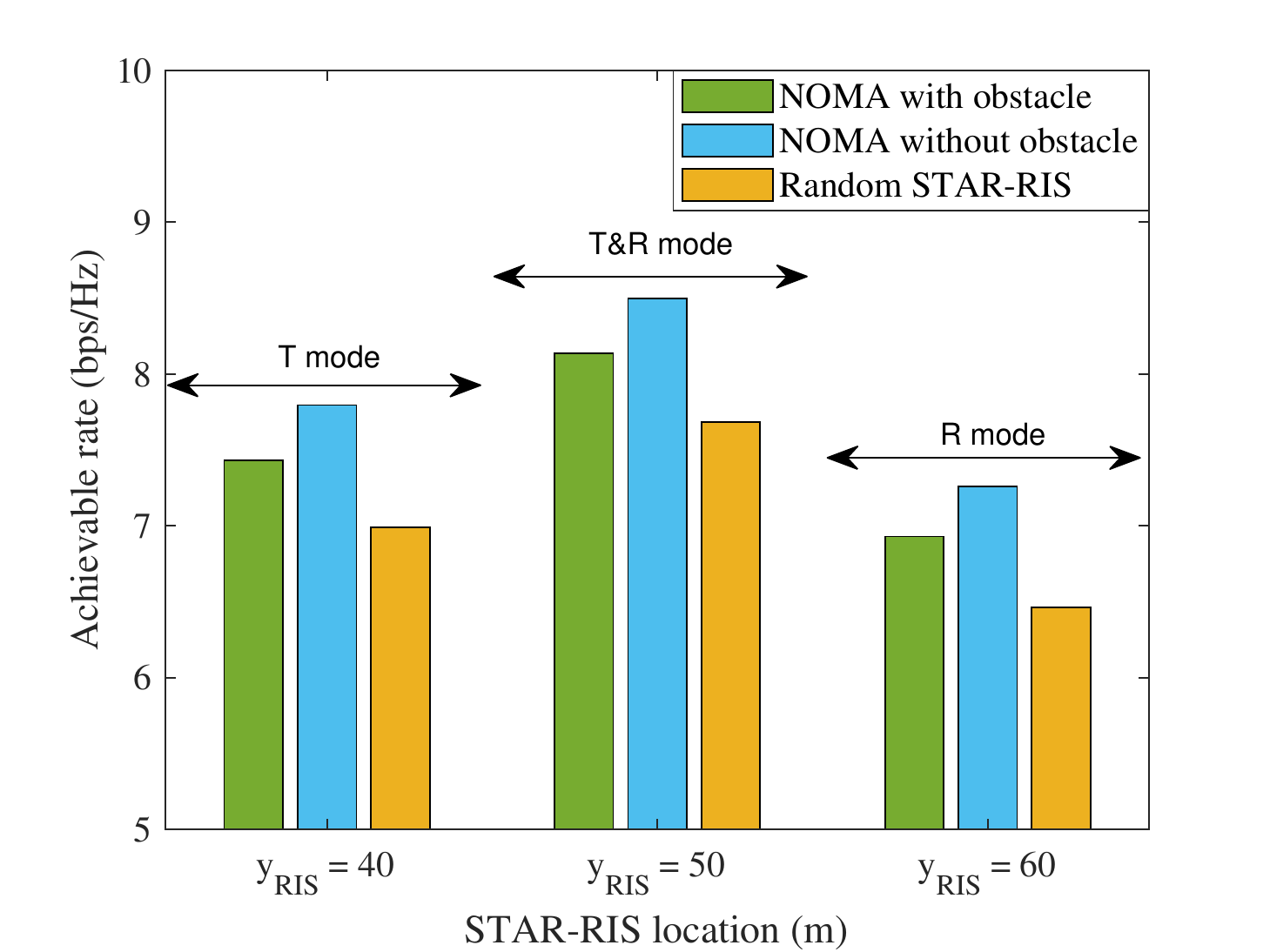}
		\end{minipage}
	} \\  % [-3 pt]
	\subfloat[]{\label{achievable_rate_vs_M}
		\begin{minipage}[t]{0.5 \textwidth}
			\centering
			\includegraphics[width= 3.5 in]{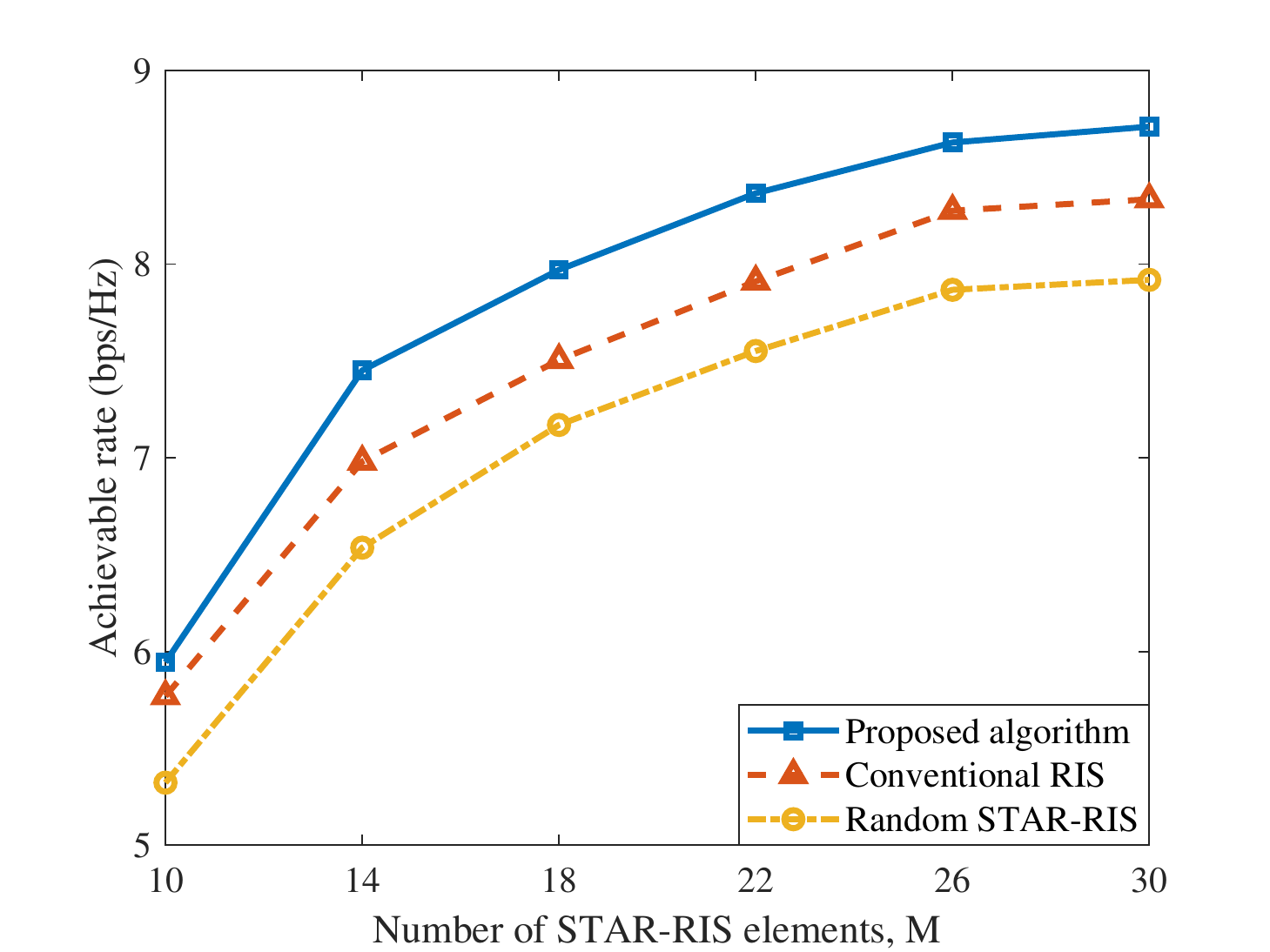}
		\end{minipage}
	}
	\caption{Achievable uplink rate of NOMA users: (a) Impact of the location of STAR-RIS. (b) Impact of the number of STAR-RIS elements.}
	\label{achievable_rate}
%	\vspace{-3 mm}
\end{figure}

\begin{figure*} [t!]
	\color{black}
	\centering
	\subfloat[]{\label{training_loss_MNIST_IID}
		\begin{minipage}[t]{0.24 \textwidth}
			\centering
			\includegraphics[width=1.89 in]{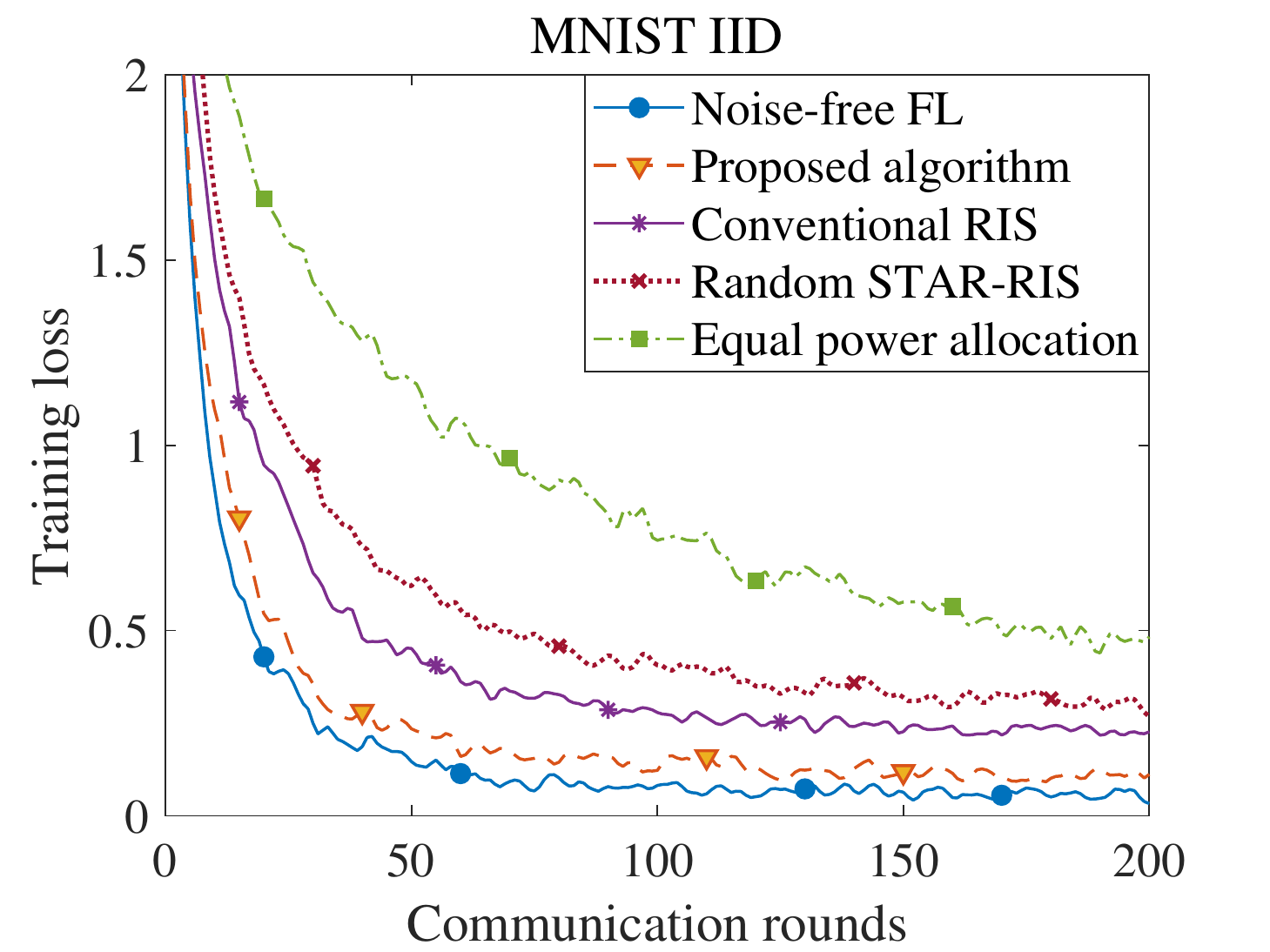}
		\end{minipage}
	}
	\subfloat[]{\label{test_accuracy_MNIST_IID}
		\begin{minipage}[t]{0.24 \textwidth}
			\centering
			\includegraphics[width=1.89 in]{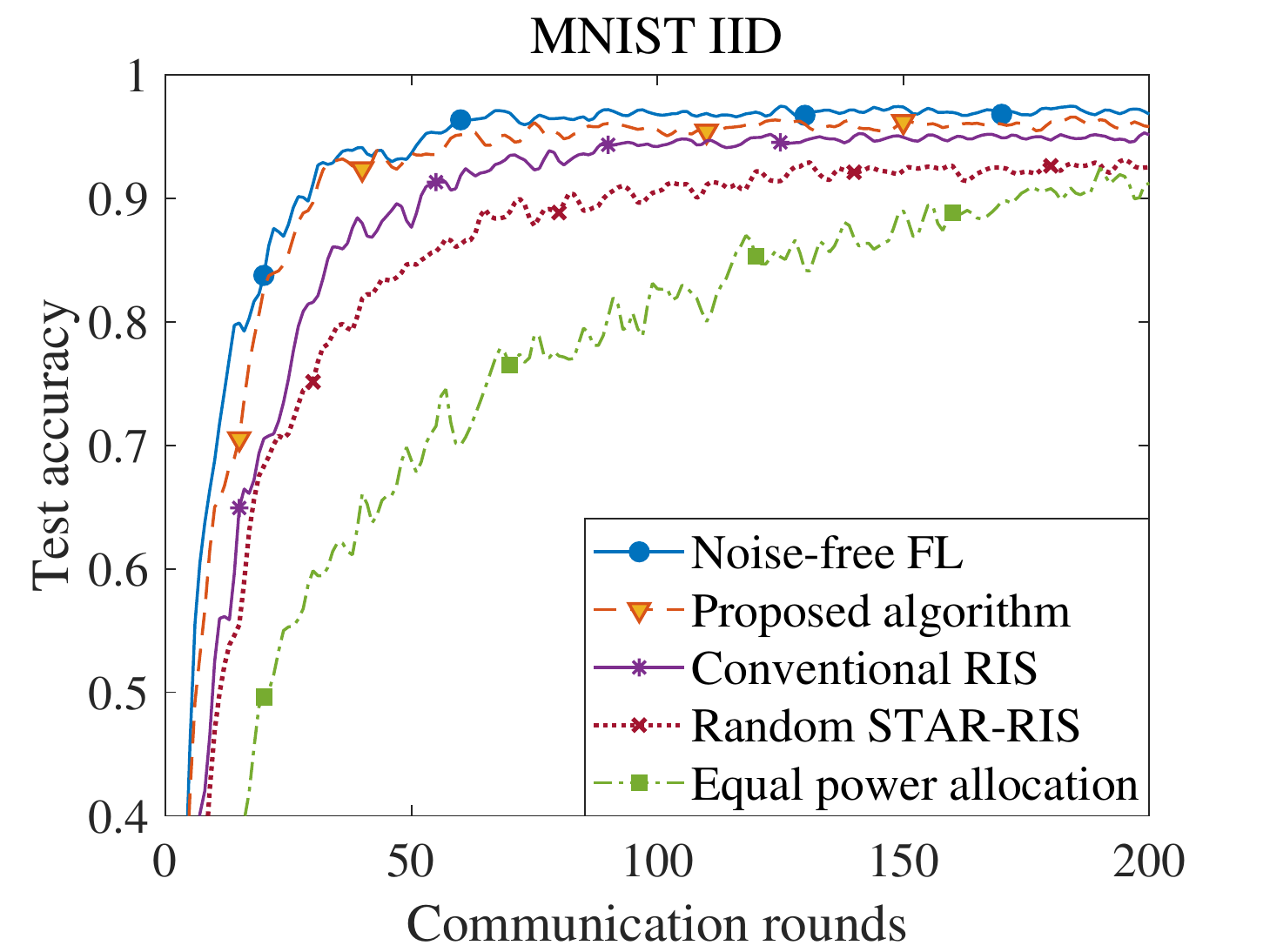}
		\end{minipage}
	}
	\subfloat[]{\label{training_loss_MNIST_Non_IID}
		\begin{minipage}[t]{0.24 \textwidth}
			\centering
			\includegraphics[width=1.89 in]{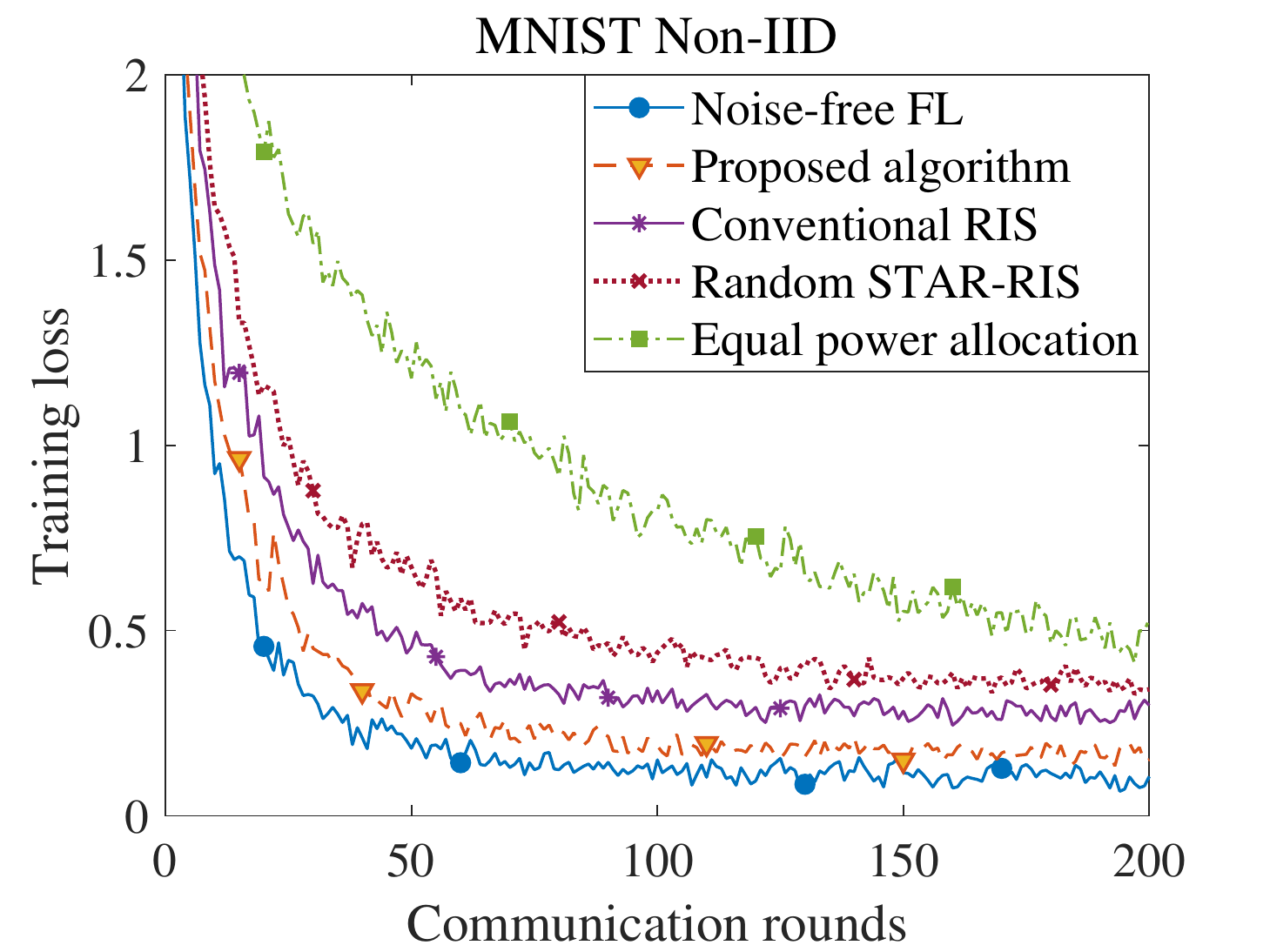}
		\end{minipage}
	}
	\subfloat[]{\label{test_accuracy_MNIST_Non_IID}
		\begin{minipage}[t]{0.24 \textwidth}
			\centering
			\includegraphics[width=1.89 in]{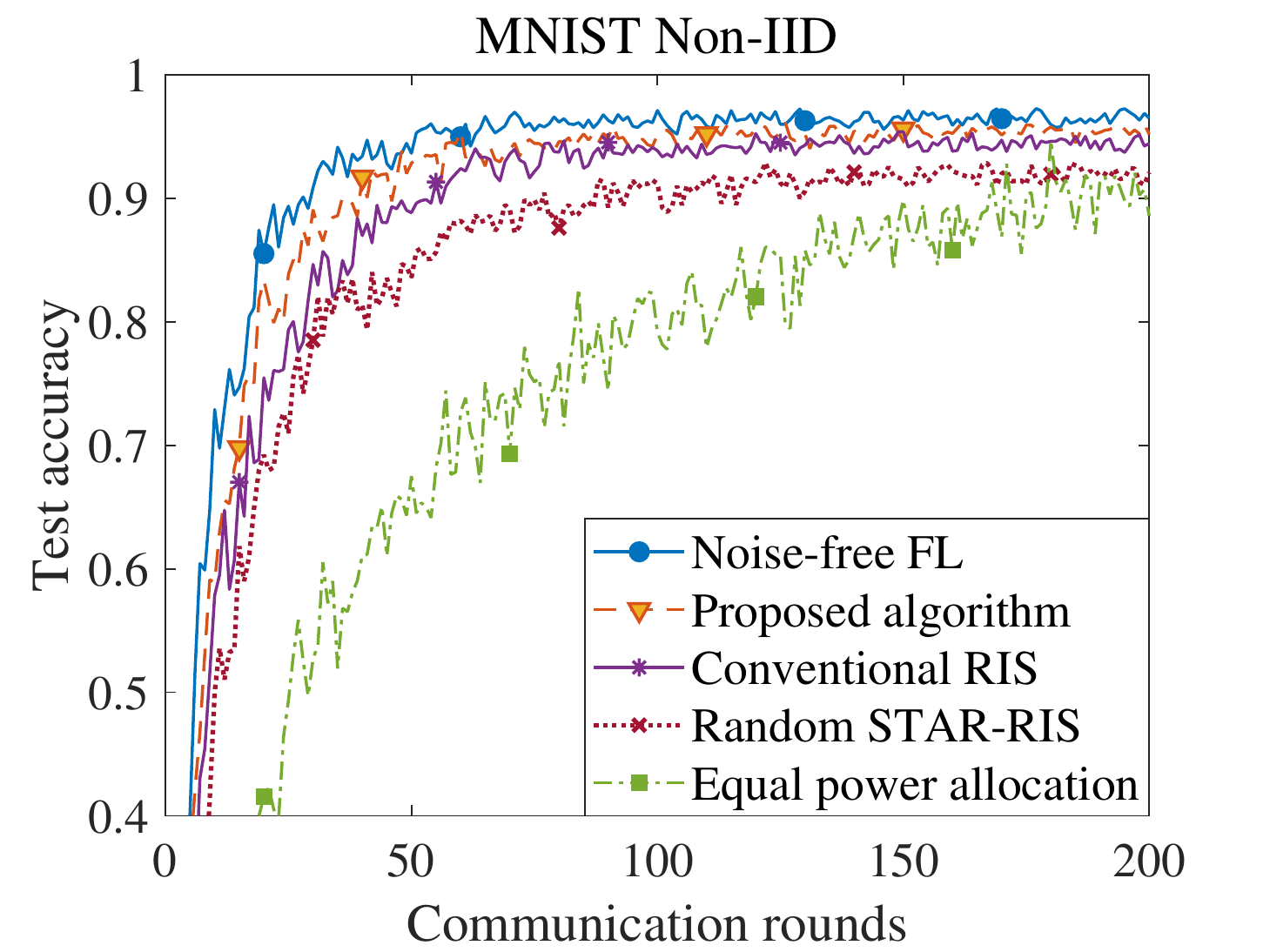}
		\end{minipage}
	}
	\caption{Learning performance of training a CNN on the MNIST dataset with IID and non-IID settings.}
	\label{result_MNIST}
\end{figure*}

\begin{figure*} [t!]
	\color{black}
	\centering
	\subfloat[]{\label{training_loss_CIFAR_IID}
		\begin{minipage}[t]{0.24 \textwidth}
			\centering
			\includegraphics[width=1.89 in]{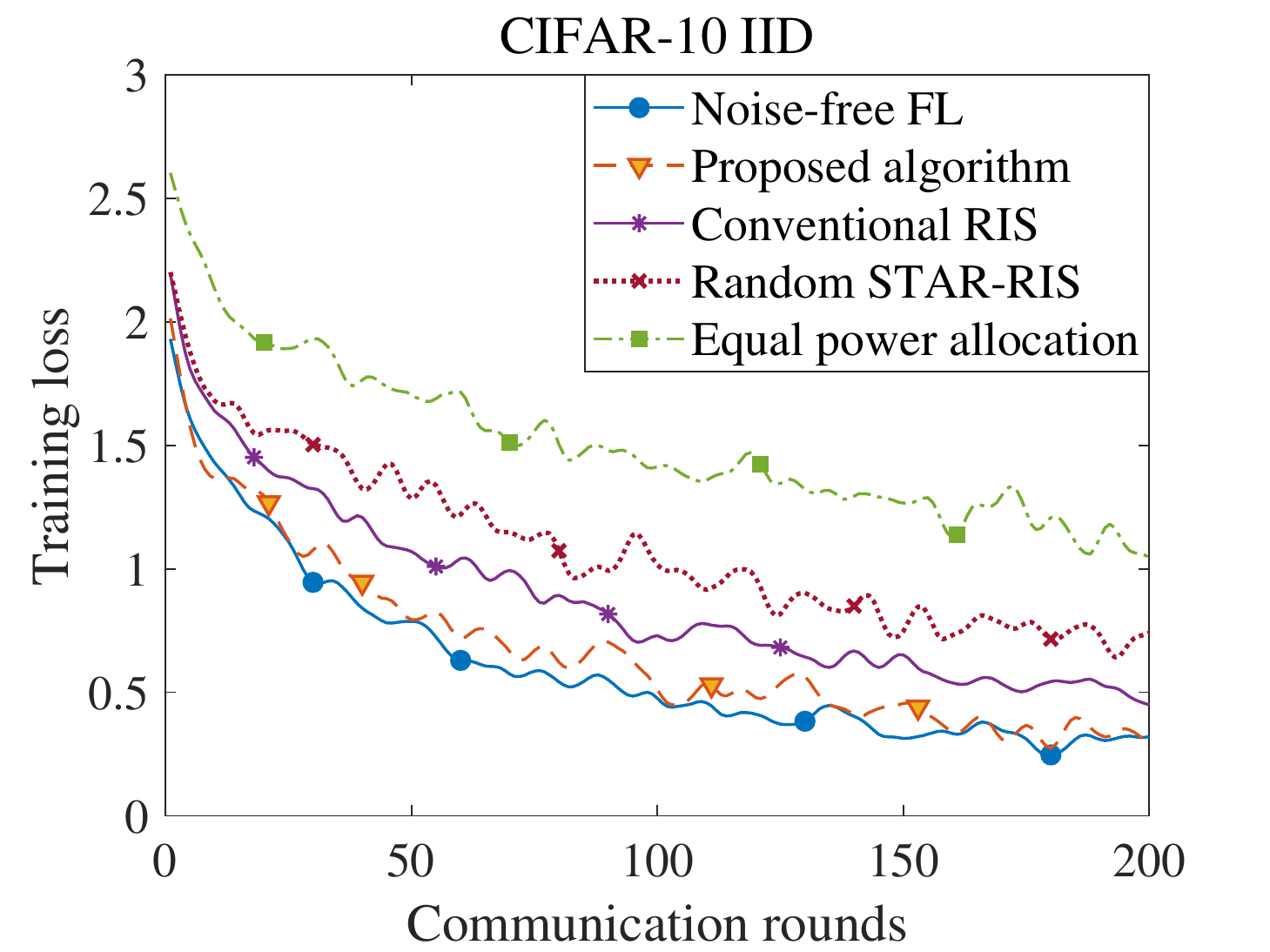}
		\end{minipage}
	}
	\subfloat[]{\label{test_accuracy_CIFAR_IID}
		\begin{minipage}[t]{0.24 \textwidth}
			\centering
			\includegraphics[width=1.89 in]{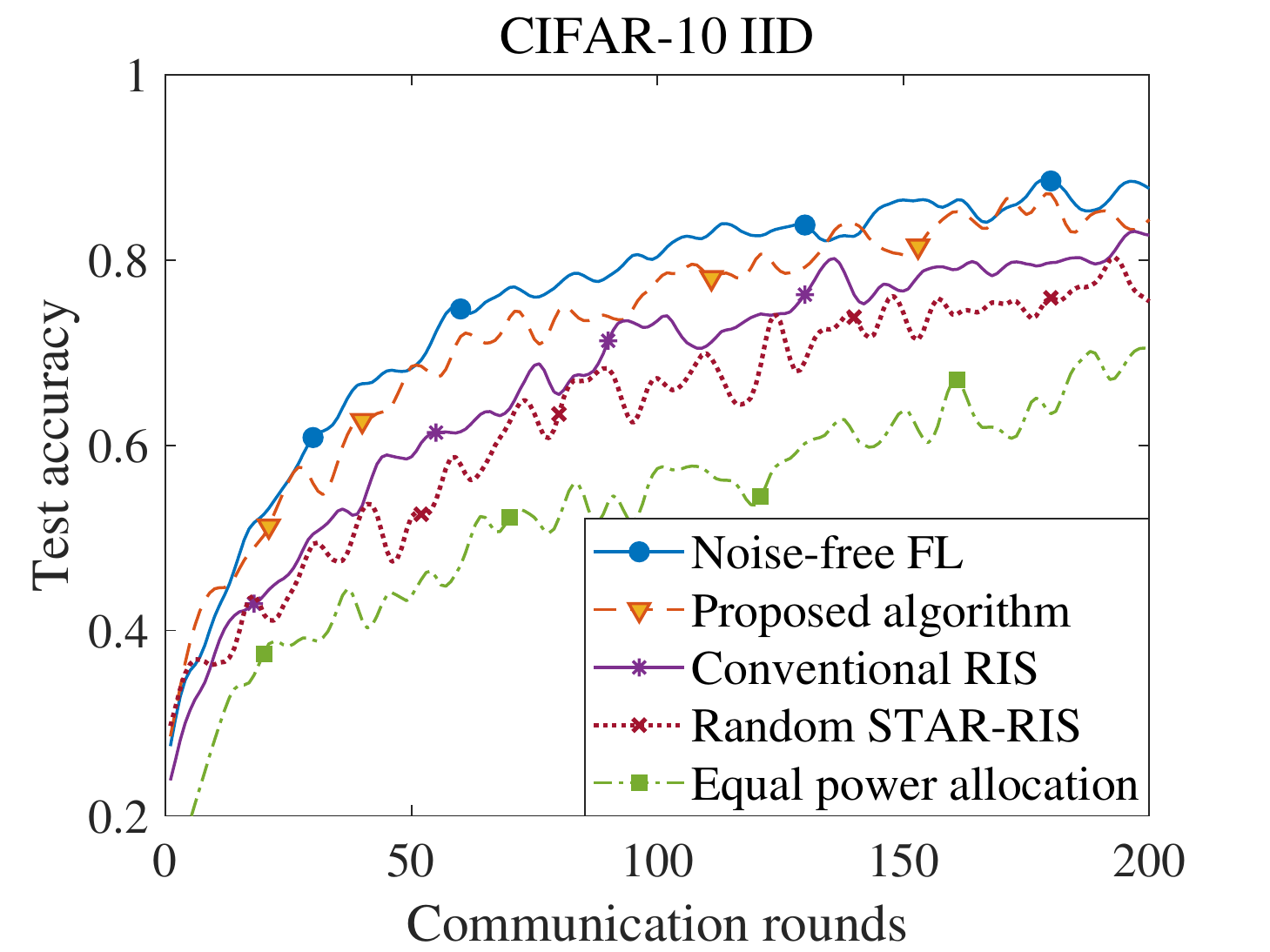}
		\end{minipage}
	}
	\subfloat[]{\label{training_loss_CIFAR_Non_IID}
		\begin{minipage}[t]{0.24 \textwidth}
			\centering
			\includegraphics[width=1.89 in]{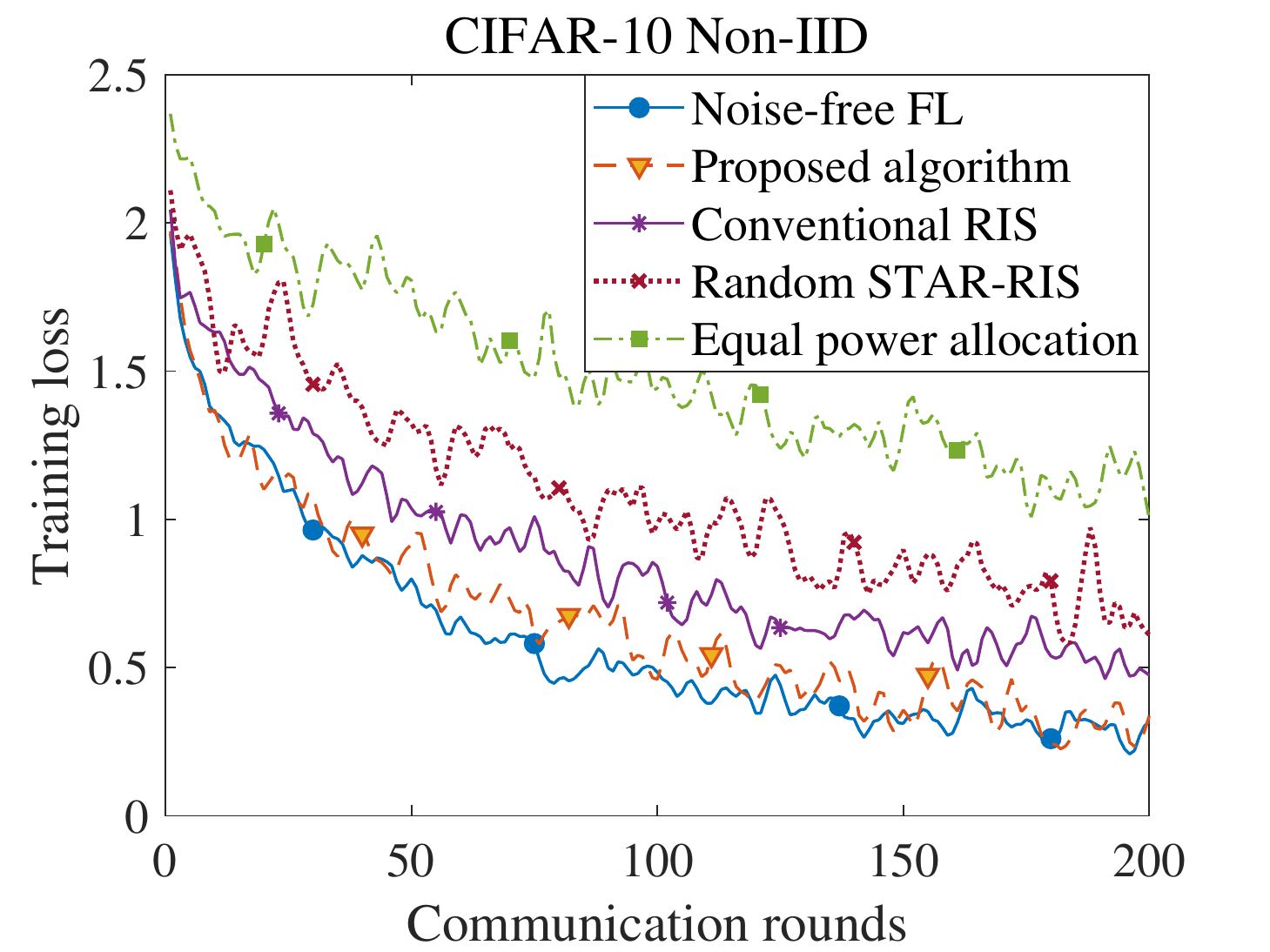}
		\end{minipage}
	}
	\subfloat[]{\label{test_accuracy_CIFAR_Non_IID}
		\begin{minipage}[t]{0.24 \textwidth}
			\centering
			\includegraphics[width=1.89 in]{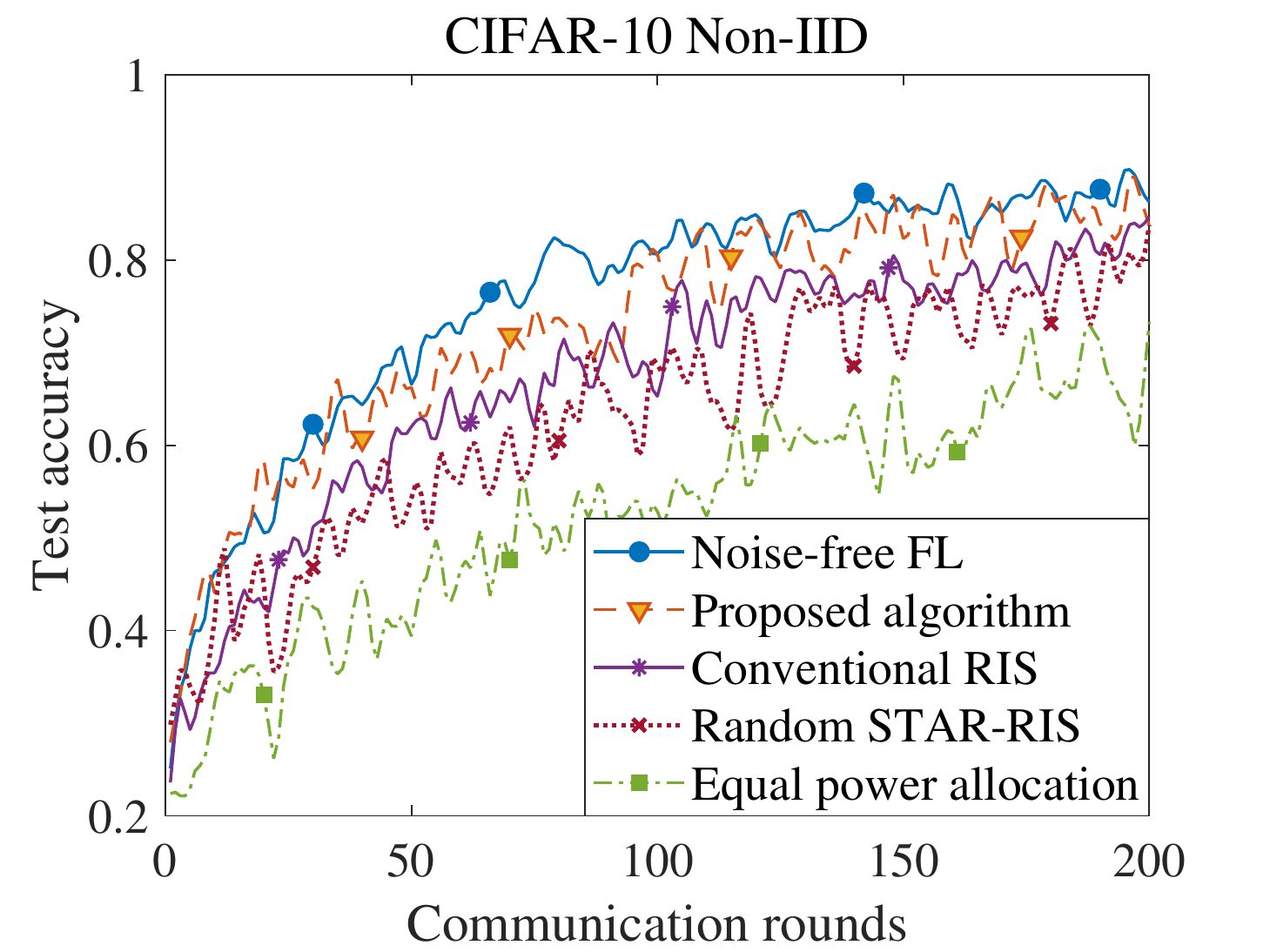}
		\end{minipage}
	}
	\caption{Learning performance of training a ResNet on the CIFAR-10 dataset with IID and non-IID settings.}
	\label{result_CIFAR10}
	\vspace{-1 mm}
\end{figure*}

In Fig. \ref{achievable_rate}, we show the impact of the STAR-RIS deployment on the achievable communication rate of NOMA users when $y_{\rm RIS} \in \{40, 50, 60\}$ and $M \in [10 \ 30]$, respectively.
In this figure, the achievable sum rate of NOMA users is used as the performance metric, given by
$R_{\rm NOMA} = \sum \nolimits_{n=1}^{N} \log_2 \left( 1 + \gamma_{n} \right)$.
For the cases of $y_{\rm RIS} = 40$ and $60 {\rm \ m}$ in Fig. \ref{achievable_rate_vs_RIS_location}, the STAR-RIS can only work in one mode (i.e., T mode for $y_{\rm RIS} = 40 {\rm \ m}$ and R mode for $y_{\rm RIS} = 60 {\rm \ m}$) to assist the communication from users to the BS.
From Fig. \ref{achievable_rate_vs_RIS_location}, it is observed that the achievable rate of all schemes first increases and then decreases while achieving their maximum spectrum efficiency at the case of $y_{\rm RIS} = 50 {\rm \ m}$.
Namely, the system throughput can be greatly improved when the STAR-RIS is deployed very close to the users.
Furthermore, it is verified that when the direct links from users to the BS are blocked by obstacles (this is equivalent to the direct links experiencing rich scattering fading with unfavorable propagation conditions), the performance gain brought about by the STAR-IRS is deteriorated since only the signal transmitted/reflected by the STAR-IRS can be received at the BS compared to the case without obstacles.
In Fig. \ref{achievable_rate_vs_M}, we compare the achievable rate versus the number of STAR-RIS elements.
By comparing the proposed algorithm with conventional RIS, we verify the performance gain of the mode switching design.
Comparing our algorithm with random STAR-RIS, one can verify the performance gain of the transmitting/reflecting phase shifts design.
It is clear that the uplink throughput for all schemes increases with the number of passive elements at the STAR-IRS, which reveals the effectiveness of deploying more elements at the STAR-IRS before the performance gain reaches a saturation point.
It is evident from Fig. \ref{achievable_rate_vs_M} that the proposed algorithm outperforms the other two benchmarks in term of spectrum efficiency, which validates the necessity for the element-wise optimization of mode switching and phase shifts at the STAR-RIS.

In Fig. \ref{result_MNIST}, we present the learning performance of training a CNN on the MNIST dataset.
To conduct simulations under different data distributions, we divide the MNIST dataset into two types of data partitions:
a) IID setting, where the full training samples are shuffled and then each local dataset is evenly assigned with $2 \times 10^4$ data samples covering all ten labels;
b) Non-IID setting: where all data samples are sorted by digit label and then each local dataset is randomly assigned with $3$ or $4$ labels.
For the latter setting, each AirFL user only has images labeled by part of digital numbers.
It is the unbalanced partition that guarantees the non-IID distribution of samples.
Regardless of whether it is an IID setting or a non-IID setting in Fig. \ref{result_MNIST}, the noise-free FL scheme obtains the optimal learning performance after $T$ communication rounds are completed, while the equal power allocation scheme performs worst with a remarkable gap.
Among all benchmark schemes in Fig. \ref{result_MNIST}, the proposed algorithm is capable of achieving the learning performance closest to the ideal noise-free FL scheme in terms of training loss and test accuracy.
This verifies again the effectiveness of joint optimization of the transmit power at the users and the configuration design at the STAR-RIS.
Moreover, we notice that if the phase shifts are not properly tuned (i.e., random STAR-RIS), conventional RIS achieves a better learning performance.
The result implies that when the number of reflecting and transmitting elements is limited, optimizing the phase shifts dominates the performance.
This observation is also applicable to the results in Fig. \ref{optimality_gap} and \ref{achievable_rate}.
Compared with the IID settings, one can see that the learning curves for all schemes in the non-IID settings become noisier and suffer from poorer convergence behavior due to the unbalanced data distribution.
Additionally, Fig. \ref{result_CIFAR10} illustrates the learning performance of training a ResNet on the CIFAR-10 dataset with the IID and non-IID settings.
%The data partition of CIFAR-10 is the same as MNIST dataset.
Note that the overall trends of all curves in Fig. \ref{result_CIFAR10} are similar to that in Fig. \ref{result_MNIST}.
Thus, the discussions for Fig. \ref{result_CIFAR10} are omitted here for brevity.
Further comprehensive evaluation with more common network topologies, more diverse data distributions, and more realistic user mobility remain as future work.

\section{Conclusions and Future Work} \label{section_conclusion}
In this paper, we advocated a STAR-RIS integrated NOMA and AirFL framework to overcome the spectrum scarcity and support heterogeneous services (e.g., communication- and learning-centric tasks).
Based on the SIC decoding technique, a new design of successive signal processing for concurrent uplink communication was developed to coordinate interference among hybrid users and thus enhance system performance.
Then, we provided a closed-form expression for the optimality gap to characterize the impact of unfavorable wireless communication on the convergence rate of AirFL.
Next, we investigated a non-convex resource allocation problem in the considered heterogeneous network by jointly optimizing the transmit power at users and the configuration design at the STAR-RIS to minimize the optimality gap while guaranteeing QoS requirements.
Due to the non-convex nature of this problem, we developed an alternating optimization algorithm by devising a trust region-based SCA approach and a penalty-based SDR method to solve the decoupled subproblems iteratively.
Finally, experiments using synthetic and real datasets were conducted both to corroborate our theoretical analysis and to demonstrate the effectiveness of the proposed solution.
Simulation results also verified that our algorithms achieve better performance in terms of learning behavior and communication capacity than alternative benchmarks.

%This work is an initial step in forming a universal framework for fulfilling 6G's vision of ubiquitous connectivity and pervasive intelligence.
%A number of interesting key problems and several possible extensions in this active and ongoing area are worthy of further investigation.
%One important future direction is to study channel acquisition methods for the STAR-RIS and design robust active and passive beamforming schemes under imperfect CSI, as well as analyze the impact of channel estimation error on the convergence rate of AirFL users and on the achievable throughput of NOMA users.
%Another promising direction is to consider the theoretical analysis of imperfect SIC decoding, non-IID data and non-convex loss functions on the system performance w.r.t. throughput and convergence.
%Finally, extending the two-layer single-cell framework shown in Fig. \ref{system_model} to a multi-layer and/or multi-cell computing network is also of interest.
%In addition, research problems such as hierarchical/personalized federated learning and adaptive gradient coding in the integrated networks are also important topics yet to be explored.

This work is an initial step in forming a universal framework for fulfilling 6G's vision of ubiquitous connectivity and pervasive intelligence.
A number of interesting key problems and several possible extensions in this active and ongoing area are worthy of further investigation.
We hereby conclude this paper by discussing a few promising research directions, and highlighting several open problems:
\begin{itemize}
	\item
	\textbf{Privacy and security issue:}
	For one thing, in the wireless network, there may exist eavesdroppers which attempt to obtain the raw data uploaded by NOMA users or the model parameters updated by AirFL users.
	For another, some malicious NOMA users are likely to inject fake data, while some untrustable AirFL users may launch model inference attacks to obtain private information of other users.
	Therefore, it is essential to enhance the privacy and security of the proposed framework by tackling the issues such as illegal eavesdroppers, bogus data injection, and model inference attacks.
	
	\item
	\textbf{Resilient and robust design:}
	 User mobility brings challenging problems such as time-varying channels, dynamic node dropout/join, and inaccurate CSI.
	 As a result, one important future direction is to consider the impact of user mobility on the achievable performance of the proposed framework.
	 For example, it is critical but remain unsolved to develop a resilient solution that allows asynchronous model aggregation to adapt to the dynamic node dropout/join of high-mobility users.
	 In addition, some pending problems are to study the efficient channel acquisition method for the STAR-RIS-aided system and design robust joint active and passive beamforming schemes under time-varying channels and/or imperfect CSI, as well as analyze the impact of channel estimation error on the convergence rate of AirFL users and on the achievable throughput of NOMA users.
	 
	 \item
	 \textbf{Other open problems:}
	 Other valuable problems are to investigate the hybrid learning method at the BS by using the raw data and local updates transmitted by these local users, and to derive the theoretical analysis of practical scenarios such as non-convex loss functions and noisy data labels.
	 Also, extending the two-layer single-cell framework shown in Fig. \ref{system_model} to a multi-layer and/or multi-cell computing network is also of interest.
	 Besides, research problems such as hierarchical/personalized federated learning and adaptive gradient coding in the integrated networks are also important topics yet to be explored.
\end{itemize}

\appendices
\section{Proof of Lemma \ref{theorem_1}} \label{proof_of_one_round}
When synchronization is performed $t$ rounds, the loss difference between two consecutive rounds is given by $\Delta_t = F(\boldsymbol{w}^{(t+1)}) - F (\boldsymbol{w}^{(t)})$.
Let $\boldsymbol{w} = \boldsymbol{w}^{(t+1)}$ and $\boldsymbol{v} = \boldsymbol{w}^{(t)}$.
From Assumption \ref{assumption_1}, it follows that
\begin{equation} \label{Delta_t_1}
\Delta_t
\!\le\! \nabla F ( \boldsymbol{w}^{(t)} )^{\top} \!( \boldsymbol{w}^{(t+1)} \!-\! \boldsymbol{w}^{(t)} \!)
\!+\! \frac{L}{2} \| \boldsymbol{w}^{(t+1)} \!-\! \boldsymbol{w}^{(t)} \| _2^2.
\end{equation}

According to the model update rule defined in (\ref{synchronization}), the global model in the $t$-th communication round is given by
\begin{equation} \label{model_update}
\boldsymbol{w}^{(t+1)} = \boldsymbol{w}^{(t)} - \lambda \boldsymbol{\hat{g}}^{(t)},
\end{equation}
where $\boldsymbol{\hat{g}}^{(t)}$ is the aggregated global gradient obtained in (\ref{averaged_gradient}).
Denoting $\boldsymbol{g}^{(t)} = \nabla F ( \boldsymbol{w}^{(t)} )$, while plugging (\ref{model_update}) and (\ref{averaged_gradient}) into (\ref{Delta_t_1}), we have
{\allowdisplaybreaks[4]
\begin{align} \label{Delta_t_2}
\Delta_t
& \le ( \boldsymbol{g}^{(t)} ) ^{\top} \left( - \lambda \boldsymbol{\hat{g}}^{(t)} \right)
+ \frac{L}{2} \left\| - \lambda \boldsymbol{\hat{g}}^{(t)} \right\| _2^2 \nonumber \\
& = - \underbrace{ \frac{\lambda}{K} ( \boldsymbol{g}^{(t)} ) ^{\top} \left( \sum \nolimits_{k \in \mathcal{K}} \bar{h}_{k}^{(t)} p_{k}^{(t)} \boldsymbol{g}_{k}^{(t)} + \boldsymbol{z}_{0}^{(t)} \right) }_{ \triangleq \Lambda_1^{(t)} } \nonumber \\
& \quad + \underbrace{ \frac{L \lambda^2}{2 K^2} \left\| \sum \nolimits_{k \in \mathcal{K}} \bar{h}_{k}^{(t)} p_{k}^{(t)} \boldsymbol{g}_{k}^{(t)} + \boldsymbol{z}_{0}^{(t)} \right\| _2^2 }_{ \triangleq \Lambda_2^{(t)} }.
\end{align}
}

Let $\boldsymbol{w}^*$ be the optimal global learning model and subtract $F ( \boldsymbol{w}^{*} )$ from both sides of (\ref{Delta_t_2}).
Then,
\begin{equation} \label{Delta_t_3}
F ( \boldsymbol{w}^{(t+1)} ) \!-\! F ( \boldsymbol{w}^{*} ) \le F ( \boldsymbol{w}^{(t)} ) \!-\! F ( \boldsymbol{w}^{*} ) \!-\! \Lambda_1^{(t)} \!+\! \Lambda_2^{(t)},
\end{equation}
where the last two terms are defined in (\ref{Delta_t_2}).

Now, the expectation of $\Lambda_1^{(t)}$ is given by
{\allowdisplaybreaks[4]
\begin{subequations} \label{expectation_Lambda_1}
\begin{align}
\mathbb{E} [ \Lambda_1^{(t)} ]
& = \mathbb{E} \left[ \frac{\lambda}{K} ( \boldsymbol{g}^{(t)} ) ^{\top} \left( \sum \limits_{k \in \mathcal{K}} \bar{h}_{k}^{(t)} p_{k}^{(t)} \boldsymbol{g}_{k}^{(t)} + \boldsymbol{z}_{0}^{(t)} \right) \right] \\
& \overset{(a)}{=} \frac{\lambda}{K} ( \boldsymbol{g}^{(t)} ) ^{\top} \sum \nolimits_{k \in \mathcal{K}} \bar{h}_{k}^{(t)} p_{k}^{(t)} \mathbb{E} \left[ \boldsymbol{g}_{k}^{(t)} \right] \\
& \overset{(b)}{=} \frac{\lambda}{K} \sum \nolimits_{k \in \mathcal{K}} \bar{h}_{k}^{(t)} p_{k}^{(t)} \| \boldsymbol{g}^{(t)} \| _2^2,
\end{align}
\end{subequations}
where (a) follows from $\mathbb{E} [ \boldsymbol{z}_{0}^{(t)} ] = 0$ and (b) is because $\mathbb{E} [ \boldsymbol{g}_k^{(t)} ] = \boldsymbol{g}^{(t)}, \forall k \in \mathcal{K}$, as defined in Assumption \ref{assumption_3}.}
Next, the expectation of $\Lambda_2^{(t)}$ can be expressed as
{\allowdisplaybreaks[4]
\begin{subequations} \label{expectation_Lambda_2}
\begin{align}
\mathbb{E} [ \Lambda_2^{(t)} ]
& \!=\! \mathbb{E} \left[ \frac{L \lambda^2}{2 K^2} \left\| \sum \nolimits_{k \in \mathcal{K}} \bar{h}_{k}^{(t)} p_{k}^{(t)} \boldsymbol{g}_{k}^{(t)} + \boldsymbol{z}_{0}^{(t)} \right\| _2^2 \right] \\
& \!\overset{(c)}{=}\! \frac{L \lambda^2}{2 K^2} \sum \limits_{k \in \mathcal{K}} ( \bar{h}_{k}^{(t)} p_{k}^{(t)} )^2 \mathbb{E} \left[ \| \boldsymbol{g}_{k}^{(t)} \| _2^2 \right] + \frac{L Q \lambda^2 \sigma^2}{2 K^2} \\
& \!\overset{(d)}{\le}\! \frac{L \lambda^2}{2 K^2} \sum \limits_{k \in \mathcal{K}} ( \bar{h}_{k}^{(t)} p_{k}^{(t)} )^2 \left( \| \boldsymbol{g}^{(t)} \| _2^2 + \left\| \boldsymbol{\delta} \right\| _2^2 \right) \!+\! \widetilde{Z},
\end{align}
\end{subequations}
where (c) comes from $\mathbb{E} [ \| \boldsymbol{z}_{0}^{(t)} \| _2^2 ] = \frac{L Q \lambda^2 \sigma^2}{2 K^2} = \widetilde{Z}$
and (d) is because $\mathbb{E} [ \| \boldsymbol{g}_{k}^{(t)} \| _2^2 ] = ( \mathbb{E} [ \boldsymbol{g}_{k}^{(t)} ] ) ^2 +  \mathbb{E} [ \| \boldsymbol{g}_{k}^{(t)} - \boldsymbol{g}^{(t)} \|_2^2 ]$.}

Based on (\ref{expectation_Lambda_1}) and (\ref{expectation_Lambda_2}), by taking expectation at both sides of (\ref{Delta_t_3}) to average out the randomness, it holds that
{\allowdisplaybreaks[4]
\begin{align} \label{expectation}
&\mathbb{E} \left[ F ( \boldsymbol{w}^{(t+1)} ) \right] - F ( \boldsymbol{w}^{*} ) \nonumber \\
&\le \mathbb{E} \left[ F ( \boldsymbol{w}^{(t)} ) \right] - F ( \boldsymbol{w}^{*} ) - \mathbb{E} \left[ \Lambda_1^{(t)} \right] + \mathbb{E} \left[ \Lambda_2^{(t)} \right] \nonumber \\
& \le \mathbb{E} \left[ F ( \boldsymbol{w}^{(t)} ) \right] - F ( \boldsymbol{w}^{*} ) - \frac{\lambda}{K} \sum \nolimits_{k \in \mathcal{K}} \bar{h}_{k}^{(t)} p_{k}^{(t)} \| \boldsymbol{g}^{(t)} \| _2^2 \nonumber \\
& \quad + \frac{L \lambda^2}{2 K^2} \sum \nolimits_{k \in \mathcal{K}} \left( \bar{h}_{k}^{(t)} p_{k}^{(t)} \right) ^2 \left( \| \boldsymbol{g}^{(t)} \| _2^2 + \left\| \boldsymbol{\delta} \right\| _2^2 \right) + \widetilde{Z},
\end{align}
which completes the proof of the one-round convergence rate.}

\section{Proof of Theorem \ref{theorem_1}} \label{proof_of_theorem_1}
Since $F ( \boldsymbol{w})$ is a $\mu$-strongly convex function, we have
\begin{equation} \label{mu_convex}
\Delta_t \ge ( \boldsymbol{g}^{(t)} )^{\top} ( \boldsymbol{w}^{(t+1)} - \boldsymbol{w}^{(t)} )
+ \frac{\mu}{2} \| \boldsymbol{w}^{(t+1)} -\boldsymbol{w}^{(t)} \| _2^2.
\end{equation}

By minimizing both sides of (\ref{mu_convex}) w.r.t. $\boldsymbol{w} = \boldsymbol{w}^{(t+1)}$, it follows that
\begin{align} \label{min_mu_convex}
& \min \limits_{\boldsymbol{w}} \left[ F(\boldsymbol{w}^{(t+1)}) - F (\boldsymbol{w}^{(t)}) \right] \nonumber \\
& \ge \min \limits_{\boldsymbol{w}} \left[ (\boldsymbol{g}^{(t)} )^{\top} ( \boldsymbol{w}^{(t+1)} - \boldsymbol{w}^{(t)} )
+ \frac{\mu}{2} \| \boldsymbol{w}^{(t+1)} -\boldsymbol{w}^{(t)} \| _2^2 \right].
\end{align}
Notice that the left-hand side of (\ref{min_mu_convex}) is minimized at $\boldsymbol{w}^{(t+1)} = \boldsymbol{w}^{*}$ while the right-hand side of (\ref{min_mu_convex}) is achieved when $\boldsymbol{w}^{(t+1)} = \boldsymbol{w}^{(t)} - \frac{1}{\mu} \boldsymbol{g}^{(t)}$.
Thus, the inequality in (\ref{min_mu_convex}) becomes the following Polyak-Lojasiewicz (PL) condition:
\begin{equation} \label{assumption_PL}
2 \mu \left( F (\boldsymbol{w}^{(t)}) - F(\boldsymbol{w}^{*}) \right) \le \| \boldsymbol{g}^{(t)} \| _2^2.
\end{equation}

Combining (\ref{assumption_PL}) with the one-round convergence in (\ref{one_round_expectation}), the expected performance gap in the $t$-th communication round is bounded by
\begin{equation} \label{one_round_T}
\mathbb{E} [ F ( \boldsymbol{w}^{(t+1)} ) ] \!-\! F ( \boldsymbol{w}^{*} ) \!\le\! \Lambda_3^{(t)} \!\left( F ( \boldsymbol{w}^{(t)} ) \!-\! F ( \boldsymbol{w}^{*} ) \right) \!+\! \Lambda_4^{(t)},
\end{equation}
where $\Lambda_3^{(t)}$ and $\Lambda_4^{(t)}$ are defined in (\ref{Lambda_3}) and (\ref{Lambda_4}), respectively.
By recursively applying (\ref{one_round_T}), the cumulative optimality gap after $T$ communication rounds can be derived as
{\allowdisplaybreaks[4]
\begin{align}
&\mathbb{E} [ F ( \boldsymbol{w}^{(T+1)} ) ] - F ( \boldsymbol{w}^{*} ) \nonumber \\
&\le \Lambda_3^{(T)} ( F ( \boldsymbol{w}^{(T)} ) - F ( \boldsymbol{w}^{*} ) ) + \Lambda_4^{(T)} \nonumber \\
&\le \Lambda_3^{(T)} [  \Lambda_3^{(T-1)} ( F ( \boldsymbol{w}^{(T-1)} ) \!-\! F ( \boldsymbol{w}^{*} ) ) \!+\! \Lambda_4^{(T-1)} ] \!+\! \Lambda_4^{(T)} \nonumber \\
&\le \cdots \le \prod \limits_{t=1}^{T} \Lambda_3^{(t)} \Delta_1^* + \sum \limits_{t=1}^{T-1} ( \prod \limits_{i=t+1}^{T} \Lambda_3^{(i)} ) \Lambda_4^{(t)} + \Lambda_4^{(T)},
\end{align}
where $\Delta_1^* = F(\boldsymbol{w}^{(1)}) - F (\boldsymbol{w}^{*})$, completing the proof.}

\section{Proof of Corollary \ref{diminishing_rate}} \label{proof_of_corollary_1}
By replacing the constant learning rate with diminishing ones and denoting $\varpi_{k}^{(t)} = \bar{h}_{k}^{(t)} p_{k}^{(t)}$, the one-round convergence rate in Lemma \ref{one_round} can be rewritten as
\begin{align} \label{one_round_diminishing_1}
&\mathbb{E} \left[ F ( \boldsymbol{w}^{(t+1)} ) \right] - F ( \boldsymbol{w}^{(t)} ) \nonumber \\
&\le - \left[ \sum \limits_{k \in \mathcal{K}} \left(  \frac{ \lambda^{(t)} \varpi_{k}^{(t)} }{K} - \frac{L}{2 K^2} ( \lambda^{(t)} \varpi_{k}^{(t)} ) ^2 \right) \right] \| \boldsymbol{g}^{(t)}  \| _2^2 + \Lambda_4^{(t)} \nonumber \\
&= - \frac{\lambda^{(t)}}{2} \left[ \sum \limits_{k \in \mathcal{K}} \left(  \frac{2}{K} \varpi_{k}^{(t)} - \frac{L \lambda^{(t)}}{K^2} ( \varpi_{k}^{(t)} ) ^2 \right) \right] \| \boldsymbol{g}^{(t)}  \| _2^2 + \Lambda_4^{(t)} \nonumber \\
&\overset{(e)}{\le} - \frac{1}{2} \lambda^{(t)} \| \boldsymbol{g}^{(t)}  \| _2^2 + \Lambda_4^{(t)},
\end{align}
where (e) is because $\sum_{k \in \mathcal{K}} (  \frac{2}{K} \varpi_{k}^{(t)} - \frac{L \lambda^{(t)}}{K^2} ( \varpi_{k}^{(t)} ) ^2 ) \ge 1$ when $0 \le \lambda^{(t)} \le \frac{2K \sum_{k \in \mathcal{K}} \varpi_{k}^{(t)} - K^2}{L \sum_{k \in \mathcal{K}} ( \varpi_{k}^{(t)} ) ^2}$.

Substituting the PL-condition (\ref{assumption_PL}) into (\ref{one_round_diminishing_1}), we have
\begin{equation} \label{one_round_diminishing_2}
\mathbb{E} [ F ( \boldsymbol{w}^{(t+1)} ) ] \!-\! F ( \boldsymbol{w}^{(t)} ) \!\le \!-\! \lambda^{(t)} \mu \!\left(\! F (\boldsymbol{w}^{(t)}) \!-\! F(\boldsymbol{w}^{*}) \!\right) \!+\! \Lambda_4^{(t)}.
\end{equation}
Subtracting $F ( \boldsymbol{w}^{*} )$ at both sides of (\ref{one_round_diminishing_2}), it yields
\begin{align} \label{one_round_diminishing_3}
\mathbb{E} [ F ( \boldsymbol{w}^{(t+1)} ) ] \!-\! F (\! \boldsymbol{w}^{*} \!)\! \le\! (1\!-\! \lambda^{(t)} \mu ) \!\left(\! F (\boldsymbol{w}^{(t)}) \!-\! F (\! \boldsymbol{w}^{*} \!) \!\right)\! +\! \Lambda_4^{(t)}.
\end{align}
As defined in Corollary \ref{diminishing_rate}, $\xi_t = \max \{ (t + \nu) ( F (\boldsymbol{w}^{(t)}) - F(\boldsymbol{w}^{*}) ), [ L \Gamma^2 ( \sum_{k \in \mathcal{K}} ( \varpi_{k}^{(t)} ) ^2 \| \boldsymbol{\delta} \| _2^2 + Q \sigma^2) ] / [2K^2 (\mu \Gamma - 1)] \}$.
Then,
\begin{equation} \label{one_round_diminishing_4}
F (\boldsymbol{w}^{(t)}) - F(\boldsymbol{w}^{*}) \le \frac{\xi_t}{t + \nu}.
\end{equation}
 
 Substituting the diminishing learning rate $\lambda^{(t)} = [\Gamma/(t + \nu)]$ and the inequality (\ref{one_round_diminishing_4}) into (\ref{one_round_diminishing_3}), it holds that
 {\allowdisplaybreaks[4]
\begin{align} \label{one_round_diminishing_5}
& \mathbb{E} [ F ( \boldsymbol{w}^{(t+1)} ) ] - F (\! \boldsymbol{w}^{*} ) \le \left( 1 - \frac{\mu \Gamma}{t + \nu}\right) \frac{\xi_t}{t + \nu} \nonumber \\
& + \frac{L}{2 K^2} \left( \frac{\Gamma}{t + \nu}\right)^2 \sum_{k \in \mathcal{K}} ( \varpi_{k}^{(t)} ) ^2 \| \boldsymbol{\delta} \| _2^2 + \frac{L Q \sigma^2}{2 K^2} \left( \frac{\Gamma}{t + \nu}\right)^2 \nonumber \\
& = \frac{t + \nu - 1}{(t + \nu) ^2} \xi_t - \frac{\mu \Gamma - 1}{(t + \nu) ^2} \left( \xi_t -  \widetilde{Q} \right),
\end{align}
where $\widetilde{Q} = \frac{ L \Gamma^2 \left( \sum_{k \in \mathcal{K}} ( \varpi_{k}^{(t)} ) ^2 \| \boldsymbol{\delta} \| _2^2 + Q \sigma^2 \right) }{ 2K^2 (\mu \Gamma - 1) } \ge 0$ due to the fact that $\Gamma > 1 / \mu$.}
Thus,
\begin{equation}
\mathbb{E} [ F ( \boldsymbol{w}^{(t+1)} ) ] - F (\! \boldsymbol{w}^{*} )
\le \frac{t + \nu - 1}{(t + \nu) ^2} \xi_t \le \frac{\xi_t}{t + 1 + \nu},
\end{equation}
competing the proof.
 
 %%%%%%%%%%%%%%%%%%%%%%%%%%%%%
 \section{Proof of Corollary \ref{analysis_aggregation_error}} \label{proof_of_corollary_2}
 Denote $\boldsymbol{\varepsilon}^{(t)} = \boldsymbol{\hat{g}}^{(t)} - \boldsymbol{g}^{(t)}$ as the aggregation error.
 Based on (\ref{Delta_t_2}), we have
 {\allowdisplaybreaks[4]
 \begin{align} \label{Delta_t_4}
 \Delta_t
 & \le ( \boldsymbol{g}^{(t)} ) ^{\top} \left( - \lambda \boldsymbol{\hat{g}}^{(t)} \right)
 + \frac{L}{2} \left\| - \lambda \boldsymbol{\hat{g}}^{(t)} \right\| _2^2 \nonumber \\
 & = - \lambda ( \boldsymbol{g}^{(t)} ) ^{\top} \left( \boldsymbol{g}^{(t)} + \boldsymbol{\varepsilon}^{(t)} \right)
 + \frac{L \lambda^2}{2} \left\| \boldsymbol{g}^{(t)} + \boldsymbol{\varepsilon}^{(t)} \right\| _2^2 \nonumber \\
 & = - \lambda \left\| \boldsymbol{g}^{(t)} \right\| _2^2
 \underbrace{ - \lambda ( \boldsymbol{g}^{(t)} ) ^{\top} \boldsymbol{\varepsilon}^{(t)} }_{ \triangleq \Lambda_5^{(t)} }
 \underbrace{ + \frac{L \lambda^2}{2} \left\| \boldsymbol{g}^{(t)} \right\| _2^2 }_{ \triangleq \Lambda_6^{(t)} } \nonumber \\
 & \quad \underbrace{ + L \lambda^2 ( \boldsymbol{g}^{(t)} ) ^{\top} \boldsymbol{\varepsilon}^{(t)} }_{ \triangleq \Lambda_7^{(t)} }
 + \frac{L \lambda^2}{2} \left\| \boldsymbol{\varepsilon}^{(t)} \right\| _2^2.
 \end{align}
}
 
 By taking expectation of $\Lambda_5^{(t)}$, $\Lambda_6^{(t)}$ and $\Lambda_7^{(t)}$, we have
 \begin{align}
 \mathbb{E} [ \Lambda_5^{(t)} ] &= - \lambda ( \boldsymbol{g}^{(t)} ) ^{\top} \mathbb{E} [ \boldsymbol{\varepsilon}^{(t)} ] 
 \overset{(f)}{\le} \frac{ \lambda ^2}{2} \| \boldsymbol{g}^{(t)} \| _2^2 + \frac{1}{2} \| \mathbb{E} [ \boldsymbol{\varepsilon}^{(t)} ] \| _2^2, \nonumber \\
 \mathbb{E} [ \Lambda_6^{(t)} ] &= \frac{L \lambda^2}{2} \mathbb{E} [ \| \boldsymbol{g}^{(t)} \| _2^2 ]
 \overset{(g)}{\le} \frac{L \lambda^2}{2} ( \| \boldsymbol{g}^{(t)} \| _2^2 + \| \boldsymbol{\delta} \| _2^2 ),  \\
 \mathbb{E} [ \Lambda_7^{(t)} ] &= L \lambda^2 ( \boldsymbol{g}^{(t)} ) ^{\top} \mathbb{E} [ \boldsymbol{\varepsilon}^{(t)} ] \!
 \overset{(h)}{\le} \! \frac{\lambda^2}{2} \| \boldsymbol{g}^{(t)} \| _2^2 \!+\! \frac{L^2 \lambda^2}{2} \| \mathbb{E} [ \boldsymbol{\varepsilon}^{(t)} ] \| _2^2, \nonumber
 \end{align}
 where
 (f) comes from $- \boldsymbol{a}^{\top} \boldsymbol{b} \le \frac{1}{2} \| \boldsymbol{a} \| _2^2 + \frac{1}{2} \| \boldsymbol{b} \| _2^2$,
 (g) is because $ \mathbb{E} [ \| \boldsymbol{g}^{(t)} \| _2^2 ] = \mathbb{E} [ \| \frac{1}{K} \sum_{k \in \mathcal{K}} \boldsymbol{g}_{k}^{(t)} \| _2^2 ] = \frac{1}{K} \mathbb{E} [ \| \boldsymbol{g}_{k}^{(t)} \| _2^2 ] = \frac{1}{K} \{ ( \mathbb{E} [ \boldsymbol{g}_{k}^{(t)} ] ) ^2 +  \mathbb{E} [ \| \boldsymbol{g}_{k}^{(t)} - \boldsymbol{g}^{(t)} \|_2^2 ] \} \le \| \boldsymbol{g}^{(t)} \| _2^2 + \| \boldsymbol{\delta} \| _2^2$,
 and (h) is due to the Cauchy-Schwartz inequality.
 
Now, by taking expectation at both sides of (\ref{Delta_t_4}), we have
{\allowdisplaybreaks[4]
\begin{align} \label{Delta_t_5}
& \mathbb{E} \left[ F ( \boldsymbol{w}^{(t+1)} ) \right] - F ( \boldsymbol{w}^{(t)} ) \nonumber \\
& \le - \lambda \left\| \boldsymbol{g}^{(t)} \right\| _2^2 \!+\! \mathbb{E} [ \Lambda_5^{(t)} ] \!+\! \mathbb{E} [ \Lambda_6^{(t)} ] \!+\! \mathbb{E} [ \Lambda_7^{(t)} ]
\!+\! \frac{L \lambda^2}{2} \mathbb{E} [ \| \boldsymbol{\varepsilon}^{(t)} \| _2^2 ] \nonumber \\
& \le - \lambda \left\| \boldsymbol{g}^{(t)} \right\| _2^2 + \lambda ^2 \left\| \boldsymbol{g}^{(t)} \right\| _2^2 + \frac{1}{2} \left\| \mathbb{E} [ \boldsymbol{\varepsilon}^{(t)} ] \right\| _2^2
+ \frac{L \lambda^2}{2} \left\| \boldsymbol{g}^{(t)} \right\| _2^2 \nonumber \\
& \quad + \frac{L \lambda^2}{2} \left\| \boldsymbol{\delta} \right\| _2^2
+ \frac{L^2 \lambda^2}{2} \left\| \mathbb{E} [ \boldsymbol{\varepsilon}^{(t)} ] \right\| _2^2
+ \frac{L \lambda^2}{2} \mathbb{E} \left[ \left\| \boldsymbol{\varepsilon}^{(t)} \right\| _2^2 \right] \nonumber \\
& = - \lambda \left( 1- \lambda - \frac{L \lambda}{2} \right) \left\| \boldsymbol{g}^{(t)} \right\| _2^2 + \varXi^{(t)} \nonumber \\
& \overset{(i)}{\le} - \frac{\lambda}{2} \left\| \boldsymbol{g}^{(t)} \right\| _2^2 + \varXi^{(t)},
\end{align}
where
$\varXi^{(t)} = \frac{1 + L^2 \lambda^2}{2} \| \mathbb{E} [ \boldsymbol{\varepsilon}^{(t)} ] \| _2^2 
+ \frac{L \lambda^2}{2} \| \boldsymbol{\delta} \| _2^2
+ \frac{L \lambda^2}{2} \mathbb{E} [ \| \boldsymbol{\varepsilon}^{(t)} \| _2^2 ]$
and inequality (i) holds because $\lambda \le 1/(2+L)$.
%It can be noticed from  (\ref{Delta_t_5}) that, in the non-ideal case where the bias $\mathbb{E} [ \boldsymbol{\varepsilon}^{(t)} ]$ and the gradient MSE $\mathbb{E} [ \| \boldsymbol{\varepsilon}^{(t)} \| _2^2 ]$ are not zero, the aggregation error will have a significant impact on the learning behavior of AirFL.
}

Then, substituting (\ref{assumption_PL}) into (\ref{Delta_t_5}) and subtracting $F ( \boldsymbol{w}^{*} )$ at both sides, we obtain
\begin{align} \label{F_t_star}
& \mathbb{E} \left[ F ( \boldsymbol{w}^{(t+1)} ) \right] - F ( \boldsymbol{w}^{*} )
\le \tilde{\mu} \left( F (\boldsymbol{w}^{(t)}) - F(\boldsymbol{w}^{*}) \right) + \varXi^{(t)},
\end{align}
where $\tilde{\mu} = 1 - \lambda \mu$.
Finally, similar to (\ref{one_round_T}), the optimality gap in (\ref{gap_with_aggregation_error}) can be obtained by recursively applying (\ref{F_t_star}) $T$ rounds.

\section{Proof of Equation (\ref{exact_approximated_gap})} \label{proof_of_equation_33}
For $t \ne 1$,  by taking the first-order derivative of $\Upsilon (  \{ p_k^{(t)} \} )$ w.r.t. $p_k^{(t)}$, we have
{\allowdisplaybreaks[4]
\begin{align} \label{Upsilon_gradient}
\nabla \Upsilon \left(  \{ p_k^{(t)} \} \right)
& = \left( \prod \nolimits_{i=1, i \ne t}^{T} \Lambda_3^{(i)} \right) \nabla \Lambda_3^{(t)} ( F ( \boldsymbol{w}^{(1)} ) - F ( \boldsymbol{w}^{*} ) ) \nonumber \\
&  + \Lambda_4^{(1)} \left( \prod \nolimits_{i=2, i \ne t}^{T} \Lambda_3^{(i)} \right) \nabla \Lambda_3^{(t)} \nonumber \\
&  + \Lambda_4^{(2)} \left( \prod \nolimits_{i=3, i \ne t}^{T} \Lambda_3^{(i)} \right) \nabla \Lambda_3^{(t)} \nonumber \\
& \cdots \nonumber \\
&  + \Lambda_4^{(t-2)} \left( \prod \nolimits_{i=t-1, i \ne t}^{T} \Lambda_3^{(i)} \right) \nabla \Lambda_3^{(t)} \nonumber \\
&  + \Lambda_4^{(t-1)} \left( \prod \nolimits_{i=t, i \ne t}^{T} \Lambda_3^{(i)} \right) \nabla \Lambda_3^{(t)} \nonumber \\
&  + \nabla \Lambda_4^{(t)} \left( \prod \nolimits_{i=t+1}^{T} \Lambda_3^{(i)} \right).
\end{align}
}

Then, (\ref{Upsilon_gradient}) can be equivalently rewritten as
\begin{align} \label{Upsilon_t}
\nabla \Upsilon \left(  \{ p_k^{(t)} \} \right)
& = \left( \prod \nolimits_{i=1, i \ne t}^{T} \Lambda_3^{(i)} \right) \nabla \Lambda_3^{(t)} ( F ( \boldsymbol{w}^{(1)} ) - F ( \boldsymbol{w}^{*} ) ) \nonumber \\
& + \nabla \Lambda_3^{(t)} \sum \nolimits_{j=1}^{t-1} \Lambda_4^{(j)} \left( \prod \nolimits_{i=j+1, i \ne t}^{T} \Lambda_3^{(i)} \right) \nonumber \\
& + \nabla \Lambda_4^{(t)} \left( \prod \nolimits_{i=t+1}^{T} \Lambda_3^{(i)} \right).
\end{align}

Similarly, for $t=1$, we have
\begin{align} \label{Upsilon_1}
\nabla \Upsilon \left(  \{ p_k^{(1)} \} \right)
& = \left( \prod \nolimits_{i=2}^{T} \Lambda_3^{(i)} \right) \nabla \Lambda_3^{(1)} ( F ( \boldsymbol{w}^{(1)} ) - F ( \boldsymbol{w}^{*} ) ) \nonumber \\
& + \nabla \Lambda_4^{(1)} \left( \prod \nolimits_{i=2}^{T} \Lambda_3^{(i)} \right).
\end{align}

Finally, substituting
$\nabla \Lambda_3^{(t)} = - \frac{2 \mu \lambda  \bar{h}_{k}^{(t)}}{ K} ( 1 - \frac{L \lambda \bar{h}_{k}^{(t)} p_{k}^{(t)}}{K} )$
and
$\nabla \Lambda_4^{(t)} = \frac{L \lambda^2 \left\| \boldsymbol{\delta} \right\| _2^2}{ K^2 } ( \bar{h}_{k}^{(t)} ) ^2 p_{k}^{(t)}$
into (\ref{Upsilon_t}) and (\ref{Upsilon_1}), we can obtain (\ref{exact_approximated_gap}).
 
\section{Proof of Lemma \ref{lemma_1}} \label{proof_of_lemma_1}
According to the definition in (\ref{combined_channel}), it holds that
\begin{align} \label{eqn_74}
\left| \bar{h}_{k} p_{k} \!-\! 1 \right| ^{2} 
= \left| \left( h_k + \boldsymbol{q}_k^{\rm H} \boldsymbol{R}_k \right) p_k \!-\! 1 \right| ^{2}
= \left| \widehat{h}_k + \boldsymbol{q}_k^{\rm H} \boldsymbol{\mathring{R}}_k \right| ^{2},
\end{align}
where $\boldsymbol{\mathring{R}}_k = \boldsymbol{R}_k p_k$ and $\widehat{h}_k = h_k p_k - 1$, and we ignore the superscript $(t)$ in the following proof for brevity.
The expression (\ref{eqn_74}) can be written as 
\begin{align}
\left| \bar{h}_{k} p_{k} \!-\! 1 \right| ^{2} 
& = \boldsymbol{q}_k^{\rm H} \boldsymbol{\mathring{R}}_k \boldsymbol{\mathring{R}}_k^{\rm H} \boldsymbol{q}_k
+ \boldsymbol{q}_k^{\rm H} \boldsymbol{\mathring{R}}_k \widehat{h}_k^{\rm H}
\!+\! \widehat{h}_k \boldsymbol{\mathring{R}}_k^{\rm H} \boldsymbol{q}_k + \left| \widehat{h}_k \right| ^{2} \nonumber \\
& = \boldsymbol{\bar{q}}_k^{\rm H} \boldsymbol{\widehat{R}}_k \boldsymbol{\bar{q}}_k + \left| \widehat{h}_k \right| ^{2},
\end{align}
where
\begin{equation}
\boldsymbol{\widehat{R}}_k = \left[ \begin{array}{cc}
\boldsymbol{\mathring{R}}_k \boldsymbol{\mathring{R}}_k^{\rm H} & \boldsymbol{\mathring{R}}_k \widehat{h}_k^{\rm H} \\
\widehat{h}_k \boldsymbol{\mathring{R}}_k^{\rm H} & 0 
\end{array}
\right ]
\text{and} \
\boldsymbol{\bar{q}}_k = \left[ \begin{array}{c}
\boldsymbol{q}_k \\
1
\end{array} 
\right ].
\end{equation}
Since $\boldsymbol{\bar{q}}_k^{\rm H} \boldsymbol{\widehat{R}}_k \boldsymbol{\bar{q}}_k = {\rm tr} \left( \boldsymbol{\widehat{R}}_k \boldsymbol{\bar{q}}_k \boldsymbol{\bar{q}}_k^{\rm H} \right)$, we have
\begin{align}
\left| \bar{h}_{k} p_{k} \!-\! 1 \right| ^{2} 
= {\rm tr} \left( \boldsymbol{\widehat{R}}_k \boldsymbol{\bar{q}}_k \boldsymbol{\bar{q}}_k^{\rm H} \right) \!+\! \left| \widehat{h}_k \right|^2
\!=\! {\rm tr} \left( \boldsymbol{\widehat{R}}_k \boldsymbol{Q}_k \right) \!+\! \left| \widehat{h}_k \right|^2,
\end{align}
where $\boldsymbol{Q}_k = \boldsymbol{\bar{q}}_k \boldsymbol{\bar{q}}_k^{\rm H}$ satisfying $\boldsymbol{Q}_k \succeq 0$, ${\rm rank} \left( \boldsymbol{Q}_k  \right) = 1$ and ${\rm diag}^{-1} \left( \boldsymbol{Q}_k  \right) = \boldsymbol{\beta}_k, \forall k \in \mathcal{K}$.

\section{Proof of Lemma \ref{lemma_2}} \label{proof_of_lemma_2}
To prove Lemma \ref{lemma_2}, we first define
\begin{align}
	a = \frac{ 2 \mu \lambda }{ K } { \rm \quad and \quad } b = \frac{ \mu L \lambda^2 }{ K^2 }.
\end{align}

Then, $\Lambda_3$ in (\ref{Lambda_3}) can be rewritten as
\begin{align} \label{Lemma_2_Lambda_3}
\Lambda_3 
& = 1 - \sum \nolimits_{k \in \mathcal{K}} \left[ a  \bar{h}_{k} p_{k} - b \left( \bar{h}_{k} p_{k} \right) ^2 \right] \nonumber \\
& = 1 + b \sum \nolimits_{k \in \mathcal{K}} \left[ \left( \bar{h}_{k} p_{k} - \frac{a}{2b} \right) ^2 - \frac{a^2}{4b^2} \right].
\end{align}

From (\ref{averaged_gradient}), it can be noticed that $\bar{h}_k$ should be a real number due to the real-valued scalar $p_k$.
Then, we have $( \bar{h}_{k} p_{k} - \frac{a}{2b} ) ^2 = |  \bar{h}_{k} p_{k} - \frac{a}{2b} | ^2$.
Combined with (\ref{combined_channel}), it follows that
\begin{align} \label{check_h}
\left| \bar{h}_{k} p_{k} - \frac{a}{2b} \right| ^2
& = \left| \left( h_k + \boldsymbol{q}_k^{\rm H} \boldsymbol{R}_k \right)  p_{k} - \frac{a}{2b} \right| ^2
    = \left| \check{h}_k + \boldsymbol{q}_k^{\rm H} \boldsymbol{\mathring{R}}_k \right| ^2 \nonumber \\
& = \boldsymbol{q}_k^{\rm H} \boldsymbol{\mathring{R}}_k \boldsymbol{\mathring{R}}_k^{\rm H} \boldsymbol{q}_k
\!+\! \boldsymbol{q}_k^{\rm H} \boldsymbol{\mathring{R}}_k \check{h}_k^{\rm H}
\!+\! \check{h}_k \boldsymbol{\mathring{R}}_k^{\rm H} \boldsymbol{q}_k \!+\! \left| \check{h}_k \right| ^{2} \nonumber \\
& = \boldsymbol{\bar{q}}_k^{\rm H} \boldsymbol{\check{R}}_k \boldsymbol{\bar{q}}_k \!+\! \left| \check{h}_k \right| ^{2}
    = {\rm tr} \!\left(\! \boldsymbol{\check{R}}_k \boldsymbol{Q}_k \!\right) \!+\! \left| \check{h}_k \right|^2 \!,\!
\end{align}
where 
\begin{equation}
\boldsymbol{\check{R}}_k = \left[ \begin{array}{cc}
\boldsymbol{\mathring{R}}_k \boldsymbol{\mathring{R}}_k^{\rm H} & \boldsymbol{\mathring{R}}_k \check{h}_k^{\rm H} \\
\check{h}_k \boldsymbol{\mathring{R}}_k^{\rm H} & 0 
\end{array}
\right ]
\text{and} \
\check{h}_k = h_k p_k - \frac{a}{2b}.
\end{equation}
Plugging (\ref{check_h}) into (\ref{Lemma_2_Lambda_3}), we have
\begin{align} \label{Lambda_3_Q}
\Lambda_3 
= 1 + b \sum \limits_{k \in \mathcal{K}} \left[ {\rm tr} \left( \boldsymbol{\check{R}}_k \boldsymbol{Q}_k \right) + \left| \check{h}_k \right|^2 - \frac{a^2}{4b^2} \right].
\end{align}
By taking the derivative of (\ref{Lambda_3_Q}) w.r.t. $\boldsymbol{Q}_k$, we can obtain
\begin{align}
\nabla \Lambda_3 \left( \boldsymbol{Q}_k \right)  = \frac{ \mu L \lambda^2 }{ K^2 } \boldsymbol{\check{R}}_k ^{\top}.
\end{align}

Similarly, $\Lambda_4$ in (\ref{Lambda_4}) is given by
\begin{align} \label{Lemma_2_Lambda_4}
\Lambda_4 
= \frac{L \lambda^2 \left\| \boldsymbol{\delta} \right\| _2^2}{2 K^2} \sum \limits_{k \in \mathcal{K}} \!\left[\! {\rm tr} \left( \boldsymbol{\bar{R}}_k \boldsymbol{Q}_k  \right) \!+\! \left| \mathring{h}_k \right|^2 \!\right] \!+\! \frac{L Q \lambda^2 \delta^2}{2 K^2},
\end{align}
where 
\begin{equation}
\boldsymbol{\bar{R}}_k = \left[ \begin{array}{cc}
\boldsymbol{\mathring{R}}_k \boldsymbol{\mathring{R}}_k^{\rm H} & \boldsymbol{\mathring{R}}_k \mathring{h}_k^{\rm H} \\
\mathring{h}_k \boldsymbol{\mathring{R}}_k^{\rm H} & 0 
\end{array}
\right ]
\text{and} \
\mathring{h}_k = h_k p_k.
\end{equation}
By taking the derivative of (\ref{Lemma_2_Lambda_4}) w.r.t. $\boldsymbol{Q}_k$, it holds that
\begin{align}
\nabla \Lambda_4 \left( \boldsymbol{Q}_k \right)  = \frac{L \lambda^2 \left\| \boldsymbol{\delta} \right\| _2^2}{2 K^2}  \boldsymbol{\bar{R}}_k ^{\top},
\end{align}
which completes the proof.

\bibliographystyle{IEEEtran}
\bibliography{IEEEabrv,ref}

\end{document}